\documentclass[aps,prb,showpacs,twocolumn,amsmath,amssymb,floatfix,superscriptaddress]{revtex4-1}

\usepackage{hyperref}
\usepackage{graphicx}
\usepackage{tabularx}
\usepackage{bm}
\usepackage{xcolor}
\usepackage[applemac]{inputenc}
\usepackage{amssymb}
\usepackage{float}

\begin{document}

\title{Effects of the electrostatic environment on superlattice Majorana nanowires}

\author{Samuel D. Escribano}
\affiliation{Departamento de F{\'i}sica de la Materia Condensada C3 and}
\author{Alfredo Levy Yeyati}
\affiliation{Departamento de F{\'i}sica Te{\'o}rica de la Materia Condensada C5,
Condensed Matter Physics Center (IFIMAC) and Instituto Nicol\'as  Cabrera,
 Universidad Aut{\'o}noma de Madrid, E-28049 Madrid, Spain}
\author{Yuval Oreg}
\affiliation{Department of Condensed Matter Physics, Weizmann Institute of Science, Rehovot, Israel 7610}
\author{Elsa Prada}
 \email[Corresponding author: ]{elsa.prada@uam.es}
 \affiliation{Departamento de F{\'i}sica de la Materia Condensada C3 and}

\date{\today}

\begin{abstract}
Finding ways of creating, measuring and manipulating Majorana bound states (MBSs) in superconducting-semiconducting nanowires is a highly pursued goal in condensed matter physics. It was recently proposed that a periodic covering of the semiconducting nanowire with superconductor fingers would allow both gating and tuning the system into a topological phase while leaving room for a local detection of the MBS wavefunction. We perform a detailed, self-consistent numerical study of a three-dimensional (3D) model for a finite-length nanowire with a superconductor superlattice including the effect of the surrounding electrostatic environment, and taking into account the surface charge created at the semiconductor surface. We consider different experimental scenarios where the superlattice is on top or at the bottom of the nanowire with respect to a back gate. The analysis of the 3D electrostatic profile, the charge density, the low energy spectrum and the formation of MBSs reveals a rich phenomenology that depends on the nanowire parameters as well as on the superlattice dimensions and the external back gate potential. The 3D environment turns out to be essential to correctly capture and understand the phase diagram of the system and the parameter regions where topological superconductivity is established.
\end{abstract}

\maketitle

\section{Introduction}
\label{Introduction}
The appearance of Majorana bound states (MBSs) at the edges of topological superconductors in solid-state devices has attracted a great deal of attention both from theorists and experimentalists~\cite{Hasan:RMP10, Alicea:RPP12, Beenakker:arxiv11, Sato:JPSJ16, Aguado:rnc17, Lutchyn:NRM18}. These non-Abelian mid-gap zero energy modes are intriguing from a fundamental point of view and germane to topologically protected quantum computing applications \cite{Nayak:RMP08, Aasen:PRX16, Das:NPJ15}.
Due to their relative simplicity, most of the scrutiny has fallen onto one-dimensional (1D) proposals such as hybrid superconducting-semiconducting nanowires with strong spin-orbit coupling \cite{Lutchyn:NRM18} and ferromagnetic atomic chains on a superconductor (SC) \cite{Nadj-Perge:Science14, Ruby:PRL15, Feldman:17, Pawlak:17}. Tuning the system to appropriate conditions, experimentalists are able to find zero energy modes compatible with the existence of MBSs in the form of zero bias peaks in tunnelling spectroscopy experiments \cite{Mourik:Science12, Deng:Science16, Zhang:Nat17a, Chen:Science17, Deng:PRB18, Vaitiekenas:PRL18, Gul:NNano18, Grivnin:arxiv18, Vaitiekenas:arxiv18}.

However, due to the possibility of alternative explanations for the observed zero bias peak, the actual nature of these low-energy states has been brought into question \cite{Setiawan:PRB17, Liu:PRB17, Reeg:PRB18b, Moore:PRB18, Avila:arxiv18, Vuik:arxiv18}. A complementary measurement that could disperse the doubts would be to measure the actual zero mode probability density along the wire or chain, which should show for Majoranas an exponential decay from the edge towards its center with the Majorana localization length \cite{Klinovaja:PRB12}. Attempts in this direction, including simultaneous tunneling measurement at the the end and the bulk of the wire, were performed in Ref.~\onlinecite{Grivnin:arxiv18}.

The zero mode probability profile could in principle be accessed with the help of a scanning tunneling microscope (STM) that explores the local density of states at a certain energy along the wire \cite{Ben-Sach:PRB15}. STM measurements of this type have been carried out in iron chains on lead \cite{Nadj-Perge:Science14, Ruby:PRL15}, but in this case it is difficult to control the parameters of the system as these are fixed by material properties. In contrast, the parameters and topological phase transition of semiconducting wires can be manipulated by external magnetic and electric fields \cite{Lutchyn:NRM18}. This is one of the reasons making the semiconducting wire platforms so popular in the attempts to engineer topological superconductivity and to pursue MBSs. In these wires the induced pairing is achieved by proximity to a SC that can be either deposited or grown epitaxially over the wire \cite{Chang:Nnano15}. In the last case, hard superconducting gaps have been reported in InAs \cite{Chang:Nnano15} and InSb \cite{Gul:NNano17} wires with epitaxial Al layers.

These hybrid wires are subjected to an external in-plane magnetic field $B$ that generates a Zeeman energy for the electrons in the wire, $V_{\rm{Z}}=g\mu_{\rm{B}}B/2$, given in terms of the wire's $g$-factor and the Bohr magneton $\mu_{\rm{B}}$. According to simple 1D effective models \cite{Lutchyn:PRL10,Oreg:PRL10}, these wires experience  a phase transition to a topological state at Zeeman fields greater than $V_{\rm c}\equiv\sqrt{\Delta^2+\mu^2}$, where $\Delta$ is the induced gap and $\mu$ the wire's chemical potential. The charge density inside the wire, and thus $\mu$, can in principle be controlled by the voltage applied to a back gate, $V_{\rm gate}$. Due to their tunability, it would be ideal to perform STM experiments on these wires, a task that can be carried out nowadays \cite{Beidenkopf:private}.

\begin{figure}
\includegraphics{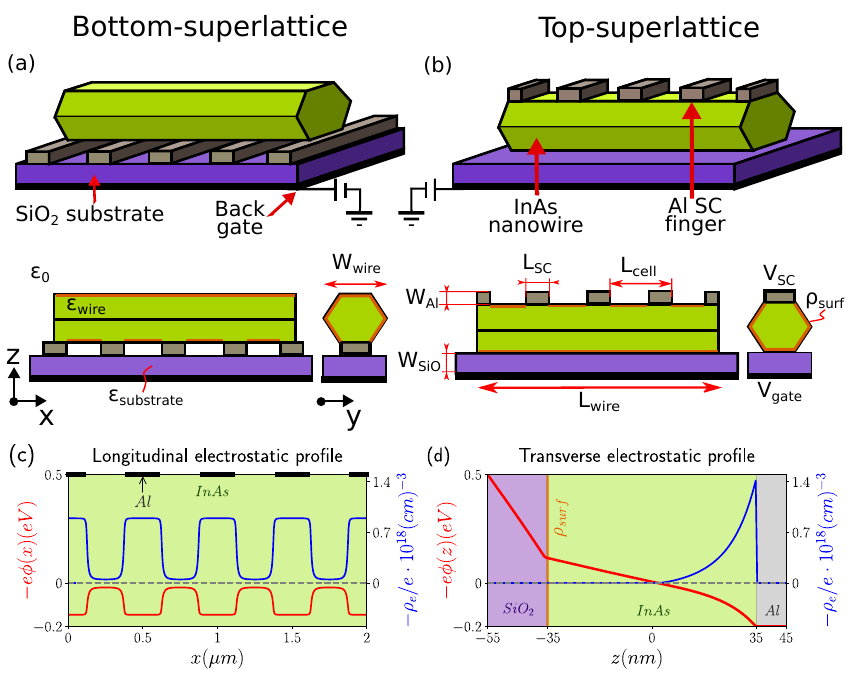}
\caption{(Colour online) Schematic 3D (top) and lateral view (bottom) representations of the two types of superlattice Majorana nanowires analysed in the text: the bottom-superlattice where the SC fingers are below (a) and the top-superlattice where they are on top of the nanowire (b). The nanowire is depicted in green, the SC superlattice in grey, the dielectric substrate in purple and the back gate in black. We choose the $x$-axis along the nanowire and the $z$-axis as the direction perpendicular to the back gate's surface. Different materials have different dielectric constants and dimensions. $V_{\rm SC}$ is the wire's conduction band offset to the metal Fermi level at the interface with the SC fingers, $\rho_{\rm surf}$ is the positive surface charge at the rest of the wire's facets and $V_{\rm gate}$ is the back gate's voltage. (c) and (d): examples of the self consistent solution of the Poisson-Schr\"odinger equations in the Thomas-Fermi approximation. The electrostatic potential energy profile (in red) and the charge density profile (in blue) are shown along the wire ($x$-direction at $z=30$nm) in (c) and across the wire ($z$-direction at $x=1\mu$m) in (d), for $V_{\rm gate}=-0.5$V, $V_{\rm SC}=0.2$V and $\rho_{\rm surf}=2\times 10^{17}e/cm^3$ in a surface layer of thickness $1nm$. Geometric parameters are: $L_{\rm wire}=2\mu$m, $L_{\rm cell}=500$nm, $L_{\rm SC}=250$nm, $W_{\rm Al}=10$nm, $W_{\rm SiO}=20$nm, $W_{\rm wire}=80$nm. Other parameters are given in Table \ref{Table_parameters}.}
\label{Fig1}
\end{figure}

Looking for an appropriate device to conduct such an experiment, Levine \textit{et al.} \cite{Levine:PRB17} recently showed that it is possible to find topological superconductivity in these wires when the superconductor (SC) is deposited periodically, forming a superlattice structure instead of covering continuously the length of the wire. A configuration with a superlattice of SC fingers at the bottom enables the STM tip to approach the wire from above, where it is free of any metal, and to drive a current between the tip and each of the SC fingers. Due to the metal free regions between the fingers, the back gate is capable of changing the charge density inside the wire due to the reduced screening by the finite size fingers. In this case \cite{Levine:PRB17}, the topological phase diagram becomes more complex than for the uniformly covered nanowire (due to the presence of longitudinal minibands created by the periodicity of the system), and extends over a wider region in parameter space (to lower Zeeman fields and higher values of the chemical potential).

Levine et al. \cite{Levine:PRB17} considered a minimal 1D model for the nanowire superstructure, in a similar fashion to other previous studies \cite{Sau:PRL12, Malard:PRB16, Hoffman:PRB16, Lu:PRB16} with related periodic structures. However, in the last couple of years it has been shown that the electrostatic environment and the three-dimensionality of these wires play an important role in all aspects concerning the trivial/topological phases and the appearance of MBSs. For instance, the electrostatic profile is not homogeneous along (and across) the wire, which creates a position-dependent chemical potential \cite{Prada:PRB12, Kells:PRB12, Liu:PRB17, Setiawan:PRB17, Moore:PRB18, Liu:PRB18} that has consequences for the topological phase transition and the shape and the overlap of MBSs \cite{Dominguez:NPJ17, Escribano:BJN18, Penaranda:PRB18, Fleckenstein:PRB18}. It also creates a position-dependent Rashba spin-orbit coupling \cite{Wojcik:PRB18, Moor:NJP18, Bommer:arxiv18}. Moreover, the charge density is not distributed uniformly across the wire and its location depends strongly on the external gate voltage \cite{Vuik:NJP16, Moor:NJP18}. This has consequences for the induced proximity effect \cite{Antipov:PRX18, Mikkelsen:PRX18} and the appearance of orbital magnetic effects \cite{Nijholt:PRB16, Kiczek:JoP17}. All these aspects influence the topological phase diagram and the topological protection of the Majorana zero modes \cite{Winkler:arxiv18}.

Motivated by the new possibilities afforded by the superlattice structures and the necessity to include electrostatic effects when analysing the performance of a particular device design, here we perform a detailed study of the systems shown in Fig. \ref{Fig1}.
 We consider two types of generic superlattice Majorana nanowires, one with the superconducting fingers at the bottom, between the nanowire and the back gate used to control the wire's charge density, see Fig. \ref{Fig1}(a), and the other with the fingers on top, further away from the back gate, see Fig. \ref{Fig1}(b). In this last case, the fingers themselves can play the role of local probes along the wire \cite{Grivnin:arxiv18,Huang:PRB18}. In both scenarios we assume that the fingers are connected to a macroscopic SC or grounded, so that we can neglect charging effects. Note that there are other works\cite{Sau:NatCom12, Fulga:NJP13, Lu:PRB16, Stenger:PRB18}  with periodic structures in the form of coupled quantum dots where the charging effect could be essential.

The physics of the setups analysed here is primarily affected by the periodic structure along the wire that creates, among other things, a periodic potential profile for the electrons, see Fig. \ref{Fig1}(c), a periodic spin-orbit coupling and a periodic induced pairing potential. These quantities are further dependent on the transverse coordinates, see Fig. \ref{Fig1}(d), which are in turn conditioned by the wire's boundary conditions (discussed in the next section).
All this gives rise to a rich phenomenology that has consequences for the topological phase diagram and the spectral properties of the wires. Fundamental parameters characterizing this phenomenology are the superlattice cell length $L_{\rm cell}$ and the SC coverage ratio $r_{\rm SC}=L_{\rm SC}/L_{\rm cell}$, where  $L_{\rm SC}$ is the size along the wire of the SC fingers. Since the geometry and the resulting 3D electrostatic profile in each setup are different, we find notable differences between both of them with advantages and disadvantages.

The bottom superlattice setup can be easily accessed from the top, for example with an STM tip as mentioned before, while its charge density is still controllable with the back gate thanks to the metal-free regions between the SC fingers. Nevertheless, the screening effect of the fingers is strong due to their vicinity to the gate, which produces sizeable potential oscillations for the electrons inside the wire. This in turn has negative consequences for the stability of the topological phase due to the appearance of localized states on top of the SC fingers that interact with the MBSs when they are present. Furthermore, the spin-orbit coupling changes sign along the wire with the periodicity of the superlattice, averaging to a small value. In contrast, in the top-superlattice device the charge density is more easily varied without the need of large back gate potentials and the topological phase is more readily accessible. The potential oscillations are thus softer and the spin-orbit coupling doesn't change sign and averages to a larger value. In turn, there is less nanowire surface exposed to open air and it is in principle more difficult to access.

In both setups the SC doesn't cover continuously the wire and consequently there is less induced superconductivity than in a uniformly covered one. We find that this leads to a reduced topological protection, manifested in a smaller topological minigap (energy difference between the Majorana zero energy mode and the continuum of states for $V_{\rm Z}>V_c$) and in a larger overlap between Majoranas at opposite ends of the wire (as measured by the Majorana charge\cite{Ben-Sach:PRB15}). Interestingly, the Majorana localization length is not only dependent on the SC coherence length, Fermi wavelength and spin-orbit length, as in the uniform hybrid wire, but also on the superlattice length.

To enhance the topological protection, at the end of the paper we propose an alternative configuration that combines a conventional hybrid Majorana nanowire (with one of its facets covered uniformly by a thin SC layer) and a superlattice of (normal or superconducting) fingers. This setup benefits from the advantages of the superlattice configuration while displaying a topological minigap and Majorana charge comparable to the uniform wire.

The structure of the paper is the following. In Sec. \ref{Methods} we describe the superlattice setups and the methodology employed to analyse them (further details on the numerical methods can be found in App. \ref{SI1}). We use a numerical approach that combines the effect of the electrostatic environment through the Poisson's equation and the wire's charge density through the Schr\"odinger's equation in a self-consistent manner. As in previous works where the electrostatic environment was considered \cite{Antipov:PRX18, Mikkelsen:PRX18, Vuik:NJP16, Winkler:arxiv18, Woods:PRB18}, our calculations are very demanding computationally, more so here since we have a superlattice structure. For this reason, we perform a series of approximations. For instance, we treat the proximity effect by the SC superlattice as a rigid boundary condition on the nanowire, effectively integrating out other SC degrees of freedom. We also ignore the orbital effects of the magnetic field. As we argue later on, this approximation will be justified at low densities and when the electron's wave function is pushed towards the SCs by the effect of the back gate.

It is important to note that in these systems there are many parameters as well as many length scales playing a role. Thus, we analyse different aspects separately in the first sections. In Subsec. \ref{Potential} we inspect the electrostatic potential profile along and across the wire for the two setups (further details in App. \ref{SI3}). In Subsec. \ref{Rashba} we analyse their inhomogeneous Rashba couplings. In Sec. \ref{Impact} we examine the impact of the superlattice on the nanowire spectral properties. We consider separately the effect of the inhomogeneous electrochemical potential, Subsec. \ref{Impact-mu}, the role of the wire's intrinsic doping, Subsec. \ref{Impact-Eint}, and the impact of the inhomogeneous induced pairing, Subsec. \ref{Impact-SC}. In Subsec. \ref{Optimal_parameters} we present a diagram in superlattice parameter space where we summarize the different features having a role in the stability of the topological phase analysed in the previous sections.

Finally, in Sec. \ref{3D_results} we consider all the previous ingredients together and analyse the behaviour of both  setups for realistic superlattice nanowire parameters. In particular, we find the spectrum over an extended range of external gate's voltages. We then focus on a particular longitudinal subband where the wire is topological and analyse the appearance of Majorana oscillations, the size of the topological minigap as well as the spatial profile of MBSs. An alternative configuration that enhances the topological protection is discussed in Subsec. \ref{Alternative}. For these calculations we solve the Schr\"{o}dinger-Poisson equation in the Thomas-Fermi approximation. To check its accuracy, we compare it with the full Schr\"{o}dinger-Poisson problem for some specific values of back gate's potential in App. \ref{SI2}. Finally, we conclude in Sec. \ref{Conclusions}.

\section{Setup and methodology}
\label{Methods}

Our aim is to study equilibrium properties of the superlattice Majorana nanowires of Figs. \ref{Fig1}(a,b) taking into account their electrostatic environment. To that end, we first compute the electrostatic potential by solving the Poisson's equation along and across the wire, taking into account its 3D geometry and the electrostatic parameters of the different materials. Then, we introduce this potential into the system's Bogoliubov-de Gennes Hamiltonian and diagonalize it to find its eigenvalues and eigenvectors (both for infinite and finite-length wires) as a function of external parameters such as the voltage applied to the back gate or the external magnetic field. Since the potential profile depends on the wire's charge density according to the Poisson's equation, and the charge density is calculated by diagonalizing the system's Hamiltonian, to solve the full Poisson-Schr\"odinger problem one needs to iterate the two in a self-consistent manner until convergence. In order to simplify this procedure we will employ the Thomas-Fermi approximation to calculate the wire's charge density, as explained below in this section. In doing so, and similarly to previous works \cite{Mikkelsen:PRX18, Winkler:arxiv18}, we assume that the potential is independent of the  magnetic field (calculated at $B=0$). This is justified since the charge density only depends slightly on $B$ for the $B$ values considered in this work, as we prove in App. \ref{SI2}.

A fully realistic calculation of the three-dimensional (3D) device would require to include the SC superlattice in the Hamiltonian at the same level as the nanowire itself. This is an involved problem that has been tackled in Refs. \onlinecite{Reeg:PRB17, Reeg:PRB18, Antipov:PRX18, Mikkelsen:PRX18, Winkler:arxiv18}. In general, it can be seen that the SC induces by proximity effect a renormalization of the wire's parameters such as $\mu$, $\alpha$ or $g$. When this renormalization is strong, called a metallization of the wire \cite{Reeg:PRB18}, it is detrimental for the appearance of a topological phase. Concerning the induced pairing, it is possible to find parameters (including the width of the SC layer \cite{Reeg:PRB17,Mikkelsen:PRX18}) where it is good, but it is in general necessary to assume a certain degree of disorder \cite{Winkler:arxiv18} in the SC to obtain a hard induced gap in the nanowire that is close to the parent's one. Here, and due to the complexity already introduced by the superlattice, we will treat the SC as a rigid boundary. Nonetheless, the SC superlattice width $W_{\rm Al}$ and its infinite dielectric constant will be taken into account when solving the electrostatic problem. We will assume good proximity effect described by a constant pairing amplitude $\Delta_0$, comparable to that of the SC bulk gap, at the sites in contact to the SC fingers (determined by the superlattice parameters $L_{\rm cell}$ and $L_{\rm SC}$). Good proximity in such superlattice devices could be achieved, for example, by using molecular beam epitaxy, either by shadowing techniques or by etching half-shell coated wires \cite{Nygard:private}.

We model the superlattice Majorana nanowire generalizing the 1D Hamiltonian of Refs. \onlinecite{Lutchyn:PRL10, Oreg:PRL10} to 3D space
\begin{eqnarray}
\label{Hamiltonian}
H= \frac{1}{2}\int \psi^\dagger(\vec{r})\hat{H}(\vec{r})\psi(\vec{r})d\vec{r}, \nonumber \\
\hat{H}(\vec{r})= \left[\frac{\hbar^2 k^2}{2m^*}-e\phi(\vec{r})-E_{\rm F}\right]\hat{\sigma}_0\hat{\tau}_z - \nonumber \\
-\frac{i}{2}\hat{\vec{\sigma}}\cdot\left[\vec{k}\times\vec{\alpha}_{\rm R}(\vec{r})-\vec{\alpha}_{\rm R}(\vec{r})\times\vec{k}\right]\hat{\tau}_z+ \nonumber \\
+V_{\rm Z}\hat{\sigma}_x\hat{\tau}_z -i\Delta(\vec{r})\hat{\sigma}_y\hat{\tau}_y,
\end{eqnarray}
where $\vec{r}=(x,y,z)$ and $\vec{k}=(k_x,k_y,k_z)$. Here $m^*$ is the effective mass of the conduction band of the InAs nanowire, $\phi(\vec{r})$ the electrostatic potential inside the wire, $E_{\rm F}$ the wire's Fermi energy, $\vec{\alpha}_{\rm R}(\vec{r})$ the vector of Rashba couplings in the three spatial directions, $V_{\rm Z}$ the Zeeman energy produced by an external magnetic field in the $x$-direction, $\Delta(\vec{r})$ the induced superconducting pair potential, and $\sigma$ and $\tau$ the Pauli matrices in spin and electron-hole space, respectively. The specific wire, electrostatic and geometrical parameters used in our simulations are summarized in Table \ref{Table_parameters}. We note that there are three quantities entering the Hamiltonian as inhomogeneous functions: the potential profile $\phi$ (that controls the local wire's band bottom), the spin-orbit coupling $\vec{\alpha}_{\rm R}$, and the induced pairing $\Delta$. On the other hand, we consider other quantities constant in space: the Zeeman splitting $V_{\rm Z}$, assuming that the applied magnetic field does not suffer from SC finger screening, and the effective mass $m^*$, which is taken as an effective renormalized parameter. In the remainder of this section we explain in detail how we model the spatial-dependent quantities. For a description of the precise numerical methods used to solve the Hamiltonian, see App. \ref{SI1}.

The electrostatic potential $\phi(\vec{r})$ is found by solving self-consistently the Poisson's equation
\begin{eqnarray}
\vec{\nabla}(\epsilon(\vec{r})\cdot\vec{\nabla}\phi(\vec{r}))=\rho_{\rm tot}[\phi(\vec{r})],
\label{Poisson}
\end{eqnarray}
where $\epsilon(\vec{r})$ is the dielectric permittivity in the entire system and $\rho_{\rm tot}[\phi(\vec{r})]$ is the total charge density of the wire, which itself depends on $\phi(\vec{r})$. The two superlattice geometries considered in this work, Figs. \ref{Fig1}(a,b), are taken into account through piecewise functions of $\epsilon(\vec{r})$, where each material is characterized by a different constant permittivity, as shown in Fig. \ref{Fig1}(a), leading to abrupt changes at the interfaces. Following Ref. \onlinecite{Winkler:arxiv18}, we model the total charge density of the wire as
\begin{equation}
\rho_{\rm tot}=\rho_{\rm surf}+\rho_{\rm mobile}.
\label{charge_density}
\end{equation}
Here $\rho_{\rm surf}$ represents the charge density of a thin layer of  donor states that typically forms at the surface of the InAs wire exposed to air \cite{Olsson:PRL96}. It depends on the details of the surface chemistry and its precise value is difficult to know \cite{Thelander:Nano10}. We model it as a $1$nm layer of \emph{positive} charge fixed at the wire's surface that is independent of the applied gate voltage. We consider two possible values compatible with existent literature, one larger, $\rho_{\rm surf}/e=2\times 10^{18}$cm$^{-3}$, and the other smaller, $\rho_{\rm surf}/e=2\times 10^{17}$cm$^{-3}$. The main effect of this charge is to produce an accumulation of electrons in the wire close to the surface and thus an {\it{intrinsic}} average doping in the absence of applied gate voltage. Hence, it conditions the values of $V_{\rm gate}$ necessary to deplete or charge the wire.

On the other hand, $\rho_{\rm mobile}$ represents the mobile charges inside the wire. For the range of $V_{\rm gate}$ values that we are going to explore in this work $\rho_{\rm mobile}=\rho_{\rm e}$, i.e., it is the charge density produced by the electrons in the InAs conduction band. Should we consider stronger (negative) gate voltages, we would need to also take into account mobile charges coming from the InAs heavy hole and light hole bands (separated from the conduction band by the semiconducting gap energy), but this is not the case here (see App. \ref{SI1}.1 for more details). The spatial distribution of $\rho_{\rm e}$ depends on $\phi(\vec{r})$, in contrast to the surface charge $\rho_{\rm surf}$ that is localized at the nanowire facets not covered by the Al, as explained before. In our calculations we use the Thomas-Fermi approximation for a 3D electron gas and take
\begin{equation}
\rho_{\rm e}(\vec{r})=-\frac{e}{3\pi^2}\left(\frac{2m^*|e\phi(\vec{r})+E_{\rm F}|f(-(e\phi(\vec{r})+E_{\rm F}))}{\hbar^2}\right)^\frac{3}{2},
\label{Thomas-Fermi}
\end{equation}
where $f$ is the Fermi-Dirac distribution (we assume $T=10$mK) and we set to zero the wire's Fermi energy ($E_{\rm F}=0$). We use the Thomas-Fermi approximation instead of performing a full Schr\"odinger-Poisson calculation because it is less demanding computationally. It has nevertheless been shown recently \cite{Mikkelsen:PRX18} that this approximation gives results in good agreement with the full treatment in similar simulations of InAs/Al heterostructures. To check this, we perform Schr\"odinger-Poisson self-consistent calculations for some specific cases in App. \ref{SI2} and quantify the deviations of the wire's charge distribution between the two. We find that Thomas-Fermi approximation slightly overestimates the electron charge density close to the SC fingers and at the wire's boundaries, but otherwise produce very similar results for the electrostatic potential.

In the bottom-right panel of Fig. \ref{Fig1}(b) we show schematically the boundary conditions used in our simulations. A voltage $V_{\rm gate}$ is applied to the back gate that is at a distance from the SC fingers/nanowire structure given by the width of the substrate (which we take as SiO$_2$). This back gate is used to tune the average chemical potential inside the wire. We assume that $\rho_{\rm surf}$ covers all the wire's facets except for those in direct contact to the SC fingers. The boundary condition between the nanowire and the SC superlattice depends on several microscopic details such as their material composition, their sizes, the type and quality of the interface, etc. Certainly, the proximity effect will also depend on these details. A detailed description of this problem is beyond the scope of this work.
Concerning its electrostatic effect, we shall assume that there is a perfect Ohmic contact between the SC and the semiconductor that imposes a constant potential at the interface that we call $V_{\rm SC}$. It represents the band bending with respect to the Fermi level of the InAs conduction band in the vicinity of the SC-semiconductor interface due to the work function difference between both materials. For an extended epitaxial InAs-Al interface, this quantity has been recently analysed in Refs. \onlinecite{Antipov:PRX18, Mikkelsen:PRX18}. Following those studies, here we will take $V_{\rm SC}=0.2$eV. However, the precise number is not important for the qualitative analysis that we present here. It will create an accumulation of electrons close to the SCs very similar to the one created by $\rho_{\rm surf}$, contributing to the {\it{intrinsic}} doping of the wire in the absence of $V_{\rm gate}$. It will thus have an influence on the values of back gate's voltages needed to deplete or charge the wire.

To visualize the effect of SC superlattice and $\rho_{\rm surf}$, we show for the top-superlattice setup an example of the potential energy profile (in red) and the electron charge density profile (in blue) along the wire ($x$-direction) in Fig. \ref{Fig1}(c), and across the wire's section ($z$-direction) in Fig. \ref{Fig1}(d). These curves are calculated with the self-consistent Thomas-Fermi approximation for some particular representative values of $V_{\rm gate}$, $V_{\rm SC}$ and $\rho_{\rm surf}$. As expected, the potential energy profile (that represents the local band bottom energy) oscillates along the wire with the periodicity set by the SC superlattice. It is minimum below the SC fingers and maximum between them. Conversely, the charge density profile is maximum below the fingers and minimum between them. In the transverse direction we can see that the charge density localizes close to the SC finger, right where the band-bottom energy is minimum, forming an electron accumulation layer.

The second inhomogeneous quantity that enters the Hamiltonian of Eq. (\ref{Hamiltonian}) is the spin-orbit coupling. We assume that it is locally proportional to the electric field $\vec{\alpha}(\vec{r})\propto\vec{E}(\vec{r})=-\vec{\nabla}\phi(\vec{r})$. According to Refs. \onlinecite{Wojcik:PRB18, Winkler:arxiv18} and using an 8-band $k\cdot p$ theory \cite{Winkler:03}, it can be modelled as
\begin{equation}
\label{alpha}
\vec{\alpha}(\vec{r})=\vec{\alpha}_{\rm int}+\frac{eP^2}{3}\left[\frac{1}{E_{\rm cv}^2}-\frac{1}{(E_{\rm cv}+E_{\rm vv})^2}\right] \vec{\nabla} \phi(\vec{r}),
\end{equation}
where $P$ is the coupling between the lowest-energy conduction band and the highest-energy valence band, $E_{\mathrm{cv}}$ is the semiconductor gap (energy difference between the conduction and valence bands), and  $E_{\mathrm{vv}}$ is the energy gap between the highest-energy and lowest-energy valence bands (split-off gap). For an InAs nanowire with wurzite crystal structure these values are \cite{Winkler:03}  $P=919.7$meV$\cdot$nm, $E_{\mathrm{cv}}=418$meV  and $E_{\mathrm{vv}}=380$meV.  Additionally, since we are considering a wurtzite InAs nanowire, there is an intrinsic Rashba constant contribution in the x-direction \cite{Voon:PRB96,Gmitra:PRB16} of the order of $\alpha_{\rm int}\simeq30$meV$\cdot$nm.

Finally, the last inhomogeneous quantity is the induced superconducting pairing $\Delta(\vec{r})$, which we model as a telegraph function with a constant value $\Delta_0$ (of the order of the bulk gap in the parent superconductor) at the wire's facets in contact to the SC fingers and zero otherwise. 

\begin{table}[htb]
\centering
\caption{Parameters used in this work.}

\renewcommand{\arraystretch}{1.5}
\newcolumntype{s}{>{\hsize=0.82\hsize}X}
\newcolumntype{l}{>{\hsize=1.09\hsize}X}

\begin{tabularx}{0.47\textwidth}{ |X|X|X| }
  \multicolumn{3}{c}{Wurzite InAs parameters \cite{Levinshtein:00, Gmitra:PRB16} }  \\
  \hline \hline
  $m^*=0.023m_{\rm e}$ & $E_{\rm F}=0$ & $\alpha_{\rm int}=30$meV$\cdot$nm \\
  \hline
\end{tabularx}

\vspace*{0.22 cm}

\begin{tabularx}{0.48\textwidth}{ |s|l|l| }
  \multicolumn{3}{c}{Electrostatic parameters \cite{Thelander:Nano10, Mikkelsen:PRX18, Perry:11, Winkler:arxiv18} } \\
  \hline \hline
  $\epsilon_{\mathrm{InAs}}=17.7\epsilon_0$ & $\epsilon_{\mathrm{SiO}}=5.5\epsilon_0$ & $\epsilon_{\mathrm{vacuum}}=\epsilon_0$ \\
  \hline
  $V_{\rm SC}=200$meV  & $\rho_{\rm surf}^{(1)}=2\cdot10^{-3}\frac{e}{nm^3}$  & $\rho_{\rm surf}^{(2)}=2\cdot10^{-4}\frac{e}{nm^3}$  \\
  \hline
\end{tabularx}

\vspace*{0.22 cm}

\begin{tabularx}{0.48\textwidth}{ |X|X|X| }
  \multicolumn{3}{c}{Geometrical parameters}  \\
  \hline \hline
  $W_{\mathrm{InAs}}=80$nm & $W_{\mathrm{Al}}=10$nm & $W_{\mathrm{SiO}}=20$nm \\
  \hline
\end{tabularx}

\vspace*{0.22 cm}

\begin{tabularx}{0.32\textwidth}{ |X|X| }
  \multicolumn{2}{c}{Other parameters}  \\
  \hline \hline
  $\Delta_0=0.2$meV  \cite{Chang:Nnano15} & $T=10$mK \\
  \hline
\end{tabularx}

\label{Table_parameters}
\end{table}

\begin{figure}
\includegraphics{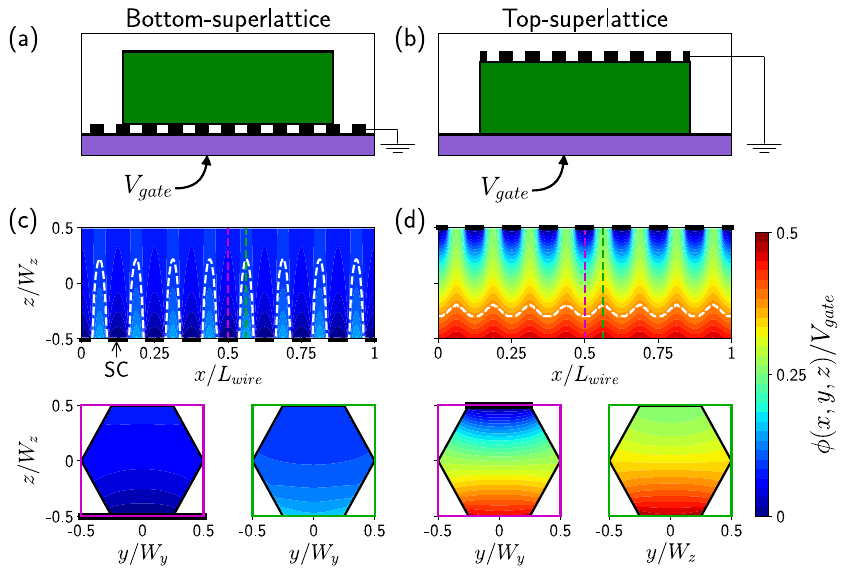}
\caption{(Colour online) Electrostatic potential profile created inside an InAs wire in contact to Al SC fingers due to the voltage applied to the back gate. Here $V_{\rm SC}=0$, $\rho_{\rm surf}=0$ and $\rho_{\rm e}$ is neglected. Two setups are considered, bottom-superlattice to the left and top-superlattice to the right, with $L_{\rm cell}=150$nm and $r_{\rm SC}=0.5$. (a,b) Sketches of both systems. (c,d) Electrostatic profile normalized to $V_{\rm gate}$ along the wire (top), and across the wire's section (bottom), both for sections with SC finger (enclosed by a purple square) and between SC fingers (enclosed by a green square). A white dotted line is used in (c,d) to highlight the shape of the potential oscillations in each setup for one particular isopotential.}
\label{Fig2}
\end{figure}

\begin{figure}
\includegraphics{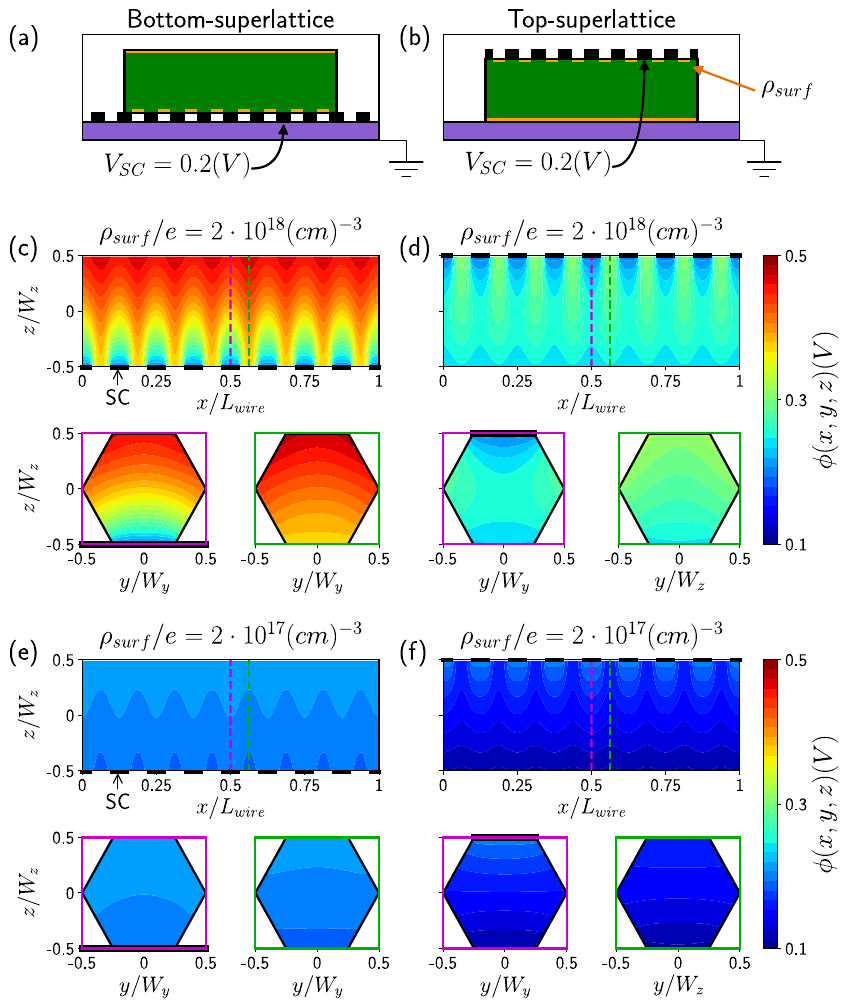}
\caption{(Colour online) Electrostatic potential profile created inside an InAs wire in contact to Al SC fingers due to the wire's band offset with respect to the Fermi level at the interface with the SC ($V_{\rm SC}=0.2$V) and the surface charge layer at the rest of the facets. Here $V_{\rm gate}=0$ and $\rho_{\rm e}$ is neglected. Two setups are considered, bottom-superlattice to the left and top-superlattice to the right, with $L_{\rm cell}=150$nm and $r_{\rm SC}=0.5$. (a,b) Sketches of both systems. (c,d) Electrostatic profile along the wire (top), and across the wire's section (bottom) for a surface charge density of $\rho_{\rm surf}=2\cdot 10^{18}(e/cm^{3})$. (e,f) Same for $\rho_{\rm surf}=2\cdot 10^{17}(e/cm^{3})$.}
\label{Fig3}
\end{figure}

\section{Electrostatic effects}
\label{Electrostatic}

\subsection{Electrostatic potential profile}
\label{Potential}
We want to study the impact of a realistic electrostatic potential profile along and across the 3D wire on the topological phase diagram and the formation of MBSs. Since we are interested in understanding the effect of the superlattice structure, we consider throughout this work periodic boundary conditions in the $x$ direction (and thus ignore border effects in the electrostatic problem). Moreover, in this section we ignore the screening effect of the mobile charges inside the wire, $\rho_{\rm e}$, because we want to isolate the impact of the electrostatic environment on the wire's potential profile (see App. \ref{SI1-1} and Fig. \ref{FigSI1}). Nevertheless, they are included self-consistently in Sec. \ref{3D_results}.

In Fig. \ref{Fig2} we plot the potential profile $\phi$ created by the bottom gate normalized to $V_{\rm gate}$, both for the bottom-superlattice device to the left and for the top-superlattice one to the right. In this case we ignore the presence of the Al-InAs band offset and the surface charge layer and take $V_{\rm SC}=0$ and $\rho_{\rm surf}=0$.  The potential oscillates along the wire with the periodicity of the superlattice, but the oscillations are very different for each setup, see white dotted guidelines in Figs. \ref{Fig2}(c,d) that highlight some isopotentials. In the bottom-superlattice device the potential maximum oscillates between the top and the bottom of the wire depending on whether the wire's section is between or on top of the SC fingers, while in the top-superlattice setup the maximum is always at the bottom of the wire, leading to smaller oscillations along the $x$ direction. This can be better appreciated in the bottom panels of Figs. \ref{Fig2}(c,d), where the potential profile across the wire's section is depicted both for sections with a SC finger (purple squared) and between SC fingers (green squared). The oscillations thus produce stronger potential wells in the first setup and subsequent bound states localized over the SCs. When present, these states are detrimental for the stability of the topological phase as we will analyse in Sec. \ref{Impact}.

Another difference between the two setups is the ability of the gate to control the potential inside the wire (and, therefore, to produce a certain doping) in the presence of the electrostatic environment. Gating is more difficult in the bottom-superlattice device because the metallic fingers are closer to the gate and thus they screen its potential more efficiently. This is why $\phi/V_{\rm gate}$ is closer to zero (blue color) in Fig. \ref{Fig2}(c) whereas in  Fig. \ref{Fig2}(d) the potential better approaches $V_{\rm gate}$ (red color) at the bottom of the wire, away from the SC fingers.

Now we explore the electrostatic potential profile created by the surface charge density $\rho_{\rm surf}$ and the potential boundary condition at the interface with the SC fingers ($V_{\rm SC}=0.2$V). As illustrated in Fig. \ref{Fig3}, we perform this study setting the back gate potential to zero. As before, the potential oscillates along the wire with the periodicity of the superlattice and across the wire's section it varies depending on whether that section is on or between the SC fingers. Since the potential profile times the electron charge $-e$ represents the wire's conduction band bottom, the wire's doping is proportional to the electrostatic potential. The main effect of the wire's band-offset with respect to the Fermi level at the SC interface and the surface charge at the other interfaces is to increase the wire's doping by a quantity that we call $\mu_{\rm int}$, which is the spatial average of the potential energy profile created by $V_{\rm SC}$ and $\rho_{\rm surf}$. This is more pronounced for the case with a larger $\rho_{\rm surf}$. We note that for realistic parameters $\mu_{\rm int}$ is always positive. On the other hand, the total doping of the wire $\mu$ coming both from the intrinsic charge and the gate voltage can be positive or negative depending on the sign and magnitude of $V_{\rm gate}$.

\begin{figure}
\includegraphics{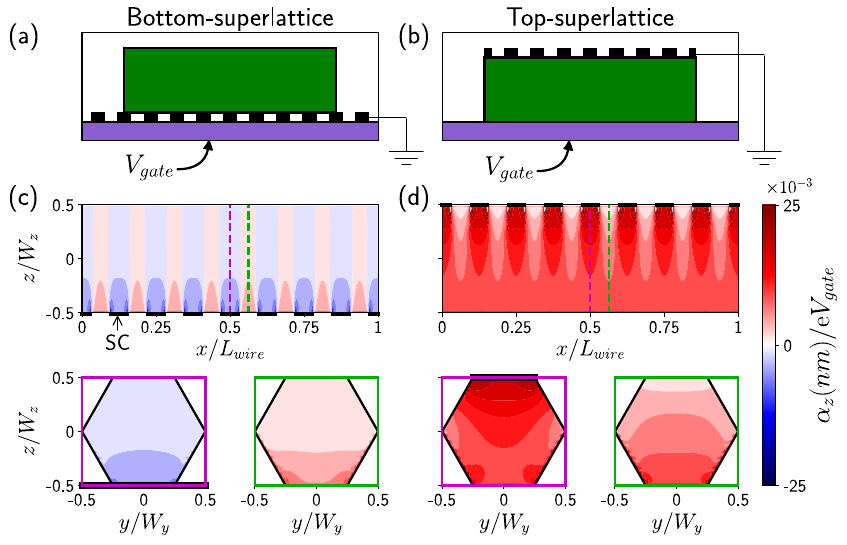}
\caption{(Colour online) Contribution of the back gate potential to the local longitudinal Rashba coupling inside the wire. $V_{\rm SC}$ and $\rho_{\rm surf}$ are fixed to zero, and $\rho_{\rm e}$ is neglected. Two setups are considered, bottom-superlattice to the left and top-superlattice to the right, with $L_{\rm cell}=150$nm and $r_{\rm SC}=0.5$. (a,b) Sketches of the two setups. (c,d) $\alpha_z$ along the wire (top) and across the wire's section (bottom), both for sections on with SC finger (inside a purple square) and between SC fingers (inside a green square).}
\label{Fig4}
\end{figure}

\begin{figure}
\includegraphics{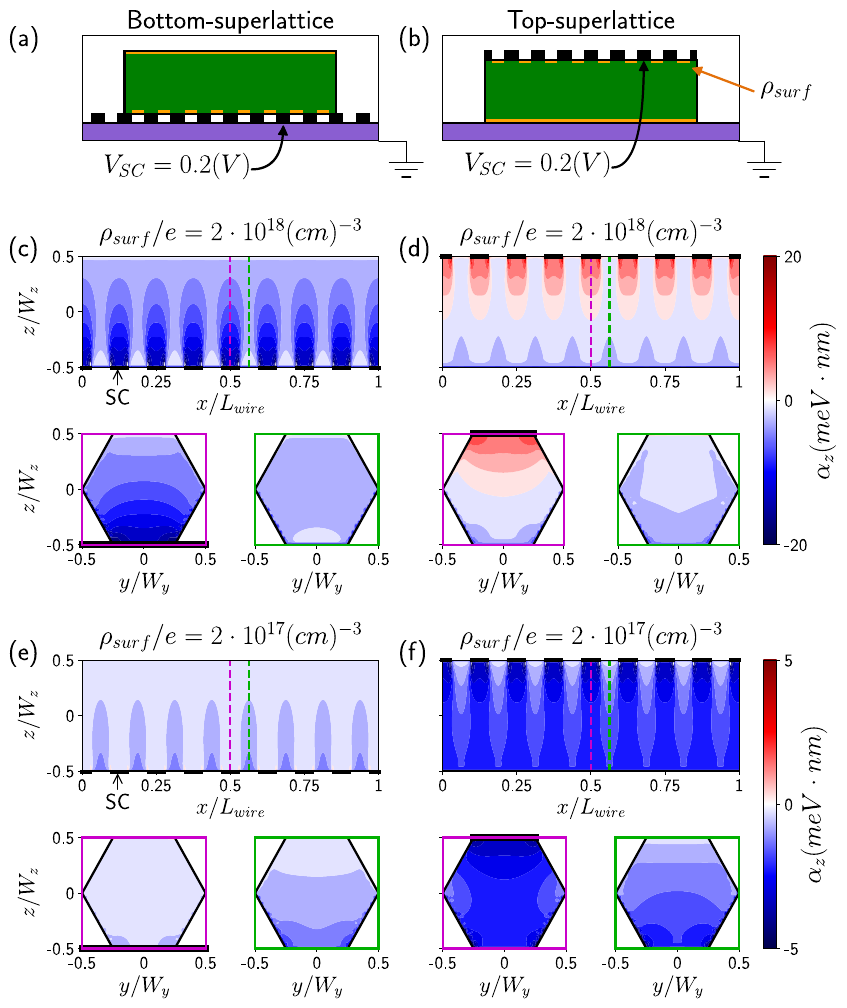}
\caption{(Colour online) Contribution of the Al-InAs band offset ($V_{\rm SC}=0.2$V) and surface charge layer to the to the local longitudinal Rashba coupling inside the wire. Here $V_{\rm gate}=0$V and $\rho_{\rm e}$ is neglected. Two setups are considered, bottom-superlattice to the left and top-superlattice to the right, with $L_{\rm cell}=150$nm and $r_{\rm SC}=0.5$. (a,b) Sketches of the two setups. (c,d) $\alpha_z$ along the wire (top), and across the wire's section (bottom) for a surface charge density of $\rho_{\rm surf}=2\cdot 10^{18}(e/cm^{3})$. (e,f) Same for $\rho_{\rm surf}=2\cdot 10^{17}(e/cm^{3})$.}
\label{Fig5}
\end{figure}

\subsection{Inhomogeneous Rashba coupling}
\label{Rashba}
The inhomogeneous electrostatic potential calculated in the previous section creates an inhomogeneous electric field that, in turn, generates an inhomogeneous Rashba coupling along and across the wire. We assume that the Rashba coupling is locally proportional to the electric field, as explained in Sec. \ref{Methods}. There are three Rashba couplings, $\alpha_{x,y,z}$, giving rise to six terms in the Hamiltonian of Eq. \ref{Hamiltonian}. Considering that the magnetic field in our model points in the $x$ direction, only two of those terms contribute to the opening of a topological minigap. These are proportional to $\alpha_z \sigma_y k_x$ and $\alpha_y \sigma_z k_x$. The effect of the other four Rashba terms is basically to produce hybridization of the transverse subbands and the subsequent even-odd effect for the appearance of Majoranas \cite{Lim:PRB12}. It turns out that $\alpha_y$ is negligible in these wire setups (due to the back gate-superlattice parallel disposition). Thus, we focus here on analysing the spatial behaviour of the transverse $\alpha_z$ coupling, shown in Figs. \ref{Fig4} and \ref{Fig5}. Following the rationale of the previous section, in the first figure we explore the Rashba coupling behaviour against the back gate potential (normalized to $V_{\rm gate}$) setting $V_{\rm SC}=0$V and $\rho_{\rm surf}=0$. Conversely, in the second one we study the contribution of the Al-InAs band offset and surface charge density setting $V_{\rm gate}=0$V.

For the top-superlattice setup we can see in Fig. \ref{Fig4}(d) that $\alpha_z$ exhibits some oscillations along the wire with the periodicity of the lattice, specially close to the SC fingers, but it is on average large and positive. This is beneficial for the formation of a robust topological minigap for $V_{\rm Z}>V_{\rm c}$. On the contrary, for the bottom-superlattice device $\alpha_z$ oscillates between positive and negative values along the $x$ direction, see Fig. \ref{Fig4}(d), averaging to a smaller number, which is detrimental for the protection of MBSs.

The Rashba coupling produced by the back gate electric field has to be supplemented with the one created by the Al-InAs band offset and surface charge layer, shown in Fig. \ref{Fig5}. On average this is proportional to the magnitude of $\rho_{\rm surf}$, see the different color bar ranges in (c,d) and (e,f). For the bottom-superlattice device shown in Figs. \ref{Fig5}(c,e), $\alpha_z$ oscillates along $x$ as before but with the same sign, giving a finite contribution to the topological gap (specially close to the SC fingers). This is also true for the top-superlattice device in the case of the smaller $\rho_{\rm surf}$, Fig. \ref{Fig5}(f), but it changes sign along and across the wire for the larger one, Fig. \ref{Fig5}(d).

According to these results and unless there are other sources of electric fields, the Rashba spin-orbit coupling relevant for Majoranas in the bottom-superlattice setup is going to be dominated by boundary conditions rather than by the voltage applied to the back gate. On the contrary, for the top-superlattice device $\alpha_z$ is going to be dominated by the back gate except for small values of $V_{\rm gate}$, in which case its qualitative behaviour is strongly dependent on the magnitude of $\rho_{\rm surf}$.


\section{Impact of the superlattice on the nanowire spectral properties}
\label{Impact}
We focus now on the impact of the superlattice, in particular the inhomogeneous electrochemical potential and the inhomogeneous induced superconductivity, on the spectral properties of a finite-length nanowire. In the calculations of this section we consider that all the charge density is at the wire's symmetry axis, so that we effectively solve a 1D problem. We do this for two reasons. One is that it is computationally less expensive and still useful to understand the impact of the superlattice on the formation of MBSs. It is also a way to isolate the effect of the longitudinal subbands created by the superlattice, which is what we seek here, from the transverse subbands, which introduce further phenomenology \cite{Stanescu:PRB11, Lutchyn:PRL11, Lim:PRB12} unrelated to the superlattice. Nevertheless, as explained in the Introduction, in the final section we will solve the complete 3D problem.

Since we aim to understand qualitatively the effect of each {\it{kind}} of inhomogeneity, in the following subsections we study their contribution separately, fixing other parameters to constant homogeneous values. For example, to find the spectrum in Subsecs. \ref{Impact-mu} and \ref{Impact-Eint} we diagonalize the Hamiltonian of Eq. (\ref{Hamiltonian}) for constant $\Delta$ and $\alpha_{\rm R}=\alpha_z$, but for the potential profile along $x$ calculated in Sec. \ref{Potential}, which is the result of a 3D Poisson calculation (but taken at  $y=0$ and $z=0$). In Subsec. \ref{Impact-SC} we consider an inhomogeneous induced pairing in $x$ and fix $\mu$ and again $\alpha_z$ to constant values. We have also analysed the case of an inhomogeneous superlattice Rashba coupling with other parameters constant (not shown), but the effect on the wire's spectrum is small, although it does influence the Majorana wave function shape.

\begin{figure}
\includegraphics{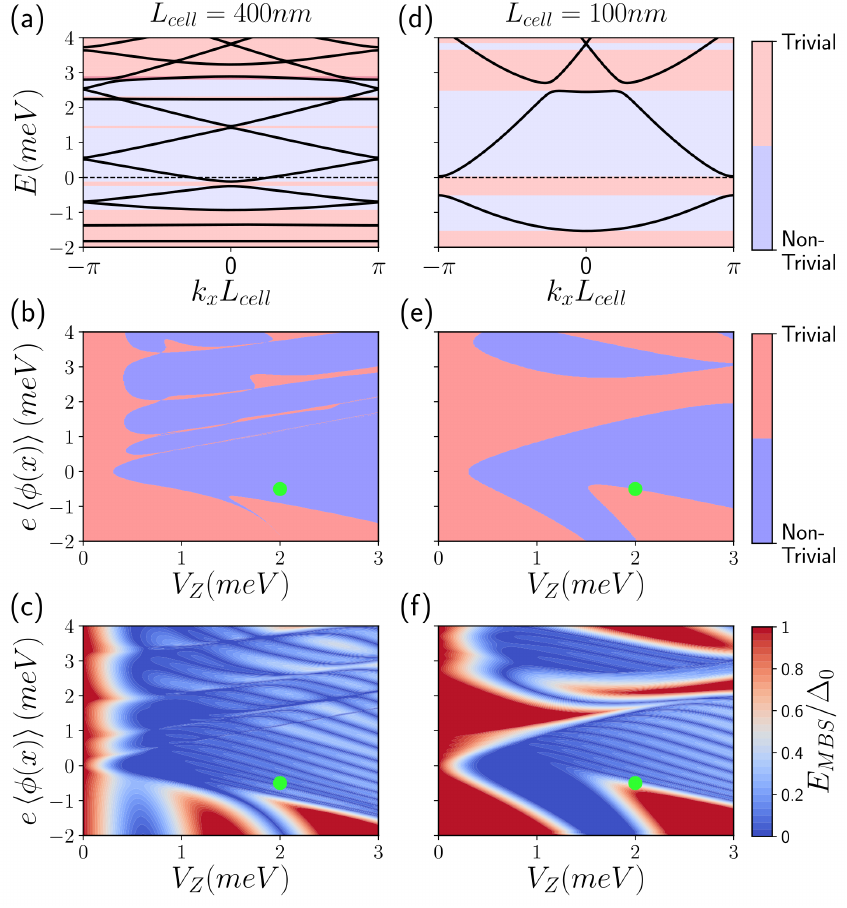}
\caption{(Colour online) (a) Dispersion relation for a 1D superlattice nanowire with superlattice parameters $L_{\rm cell}=400$nm and $r_{\rm SC}=0.5$. The electrostatic potential profile oscillates along the wire's axis following a similar profile as the one shown in Fig. \ref{Fig2} but evaluated at $(y,z)=(0,0)$. Here we take homogeneous in $x$ values for the induced SC pairing, Rashba coupling and intrinsic doping: $\Delta_0=0.2$meV, $\alpha_z=40$meV$\cdot$nm, and $\mu_{\rm int}=5$meV. (b) Corresponding topological phase diagram for the bulk system. (c) Lowest level energy for a finite-length nanowire of $L_{\rm wire}=1.2\mu$m. (d-f) The same but for $L_{\rm cell}=100$nm. The green dots mark the values of $V_{\rm Z}$ and $\mu=e\langle\phi(x)\rangle$ for which the top figures are plotted.}
\label{Fig6}
\end{figure}

\subsection{Impact of the inhomogeneous electrochemical potential}
\label{Impact-mu}
We start by analysing the effect on the wire's spectrum of the superlattice chemical potential. We take a similar potential profile as the ones of Figs. \ref{Fig2}(c,d) (but with different $L_{\rm cell}$ values), i.e. ignoring the inhomogeneous intrinsic doping of the wire, at (y,z)=(0,0). On the other hand, we take constant values for the induced pairing and Rashba coupling ($\Delta_0=0.2$meV and $\alpha_z=40$meV$\cdot$nm).

Due to the superlattice structure, the real space unit cell is larger than for a homogeneous potential wire, leading to the formation of longitudinal subbands in the dispersion relation, see Figs. \ref{Fig6}(a,d) for two values of $L_{\rm cell}$. The number of these longitudinal subbands per unit energy increases with $L_{\rm cell}$. As stated in Ref. \onlinecite{Levine:PRB17}, only when the Fermi energy crosses an odd number of Fermi pair points, the system is topologically non-trivial (light blue regions). Otherwise it is trivial (light pink regions). The electrostatic potential can open a gap between longitudinal subbands, whose size depends on the strength of the potential oscillations, leading to energy ranges where the Fermi energy crosses no band \cite{Malard:PRB16} (see Fig. \ref{Fig6}(d)). This causes the wire to exit the topological phase.

In Figs. \ref{Fig6}(b,e) we plot the wire's phase diagram versus Zeeman field $V_{\rm Z}$ and chemical potential, given by the space average of the electrostatic potential times the electric charge $e\langle\phi(x)\rangle$. The green dots mark the values of these parameters for which the dispersion relations in (a,d) are plotted. This phase diagram is certainly more complex than the one of an homogeneous 1D Majorana nanowire, characterized by a single solid hyperbolic topological zone corresponding to one topological band (whose boundary is given by the condition $\mu=\pm\sqrt{V_{\rm Z}^2-\Delta^2}$). Here, since we have several longitudinal subbands, we have several more or less contiguous topological zones (with shapes that only slightly resemble the single-band hyperbolic one) separated by trivial regions whenever the Fermi energy crosses an even number of Fermi pair points, see Fig. \ref{Fig6}(b). Moreover, whenever the Fermi energy lies at the gaps between longitudinal subbands, the phase diagram develops trivial {\it{holes}} within the topological phase, see for instance the light pink region at the bottom-right corner in Fig. \ref{Fig6}(e). At the boundaries of this trivial holes we have the condition $\lambda_{\rm F}=L_{\rm cell}$, as pointed out in Refs. \onlinecite{Lu:PRB16,Malard:PRB16}. Additionally, we note that a change in the back gate potential will not only move the subbands upwards or downwards in a rigid way, but it will also change the hybridization between the longitudinal subbands, leading to a change in the trivial hole sizes.

It is known that, for a finite-length nanowire, Majorana zero modes appear in the wire's spectrum in the topological phase. These states are localized at the edges of the wire and decay exponentially towards its center with the so called Majorana localization length, that is proportional to the SC coherence length \cite{Klinovaja:PRB12}. When the wire's length is not much greater than the Majorana localization length, left and right MBSs overlap and their energy lifts from zero producing characteristic Majorana oscillations as a function of $V_{\rm Z}$ and $\mu$. The lowest level energy of a finite-length nanowire ($L_{\rm wire}=1.2\mu$m) is shown in Figs. \ref{Fig6}(c,f), where we see the impact of the electrostatic potential superlattice on the Majorana oscillations. As it can be observed, the regions where the lowest-energy modes approach zero energy in Figs. \ref{Fig6}(c,f), coincide (roughly) with the non-trivial regions in the phase diagrams of Figs. \ref{Fig6} (b,e).

\begin{figure}
\includegraphics{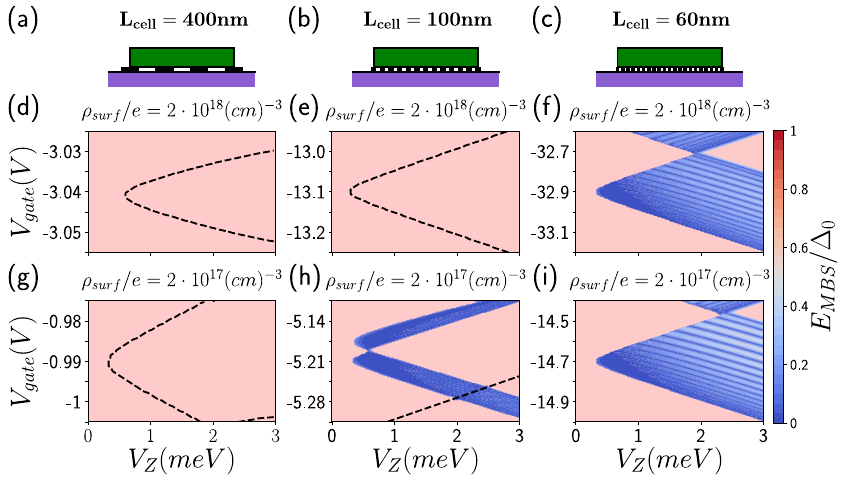}
\caption{(Colour online) Lowest level energy as a function of applied gate voltage and Zeeman field for a finite-length 1D bottom-superlattice nanowire in the presence of an inhomogeneous potential profile. This potential is taken from a 3D calculation with Al-InAs band offset $V_{\rm SC}=0.2$V and different surface charge density values, evaluated at $(y,z)=(0,0)$. Different superlattice cell sizes (with $r_{\rm SC}=0.5$) are considered. Topologically trivial regions are coloured in light pink, non-trivial regions are plotted in blue-red scale given by the color bar (where the MBS energy is normalized to $\Delta_0$), and the black dashed lines mark localized trivial zero energy modes. The Rashba coupling and induced pairing in the Hamiltonian are fixed to the homogeneous quantities $\alpha_{\rm R}=30$meV$\cdot$nm and $\Delta_0=0.2$meV. The length of the wire is $L_{\rm wire}=1.2\mu m$.}
\label{Fig7}
\end{figure}

\begin{figure}
\includegraphics{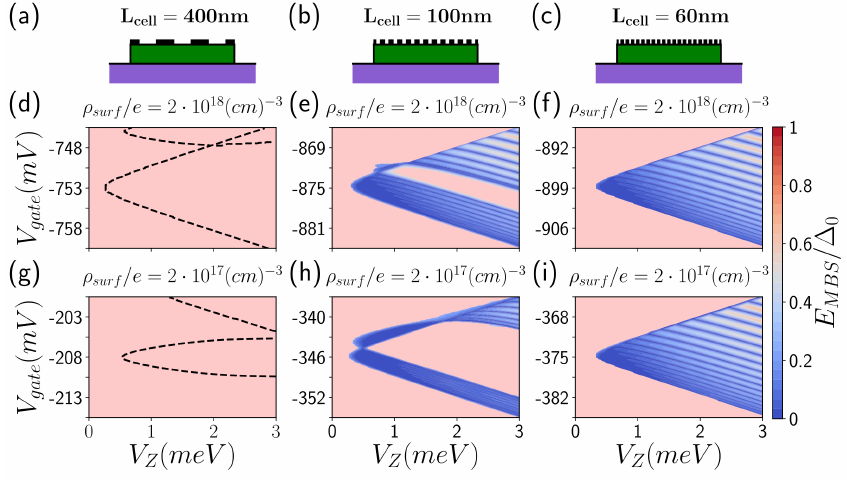}
\caption{(Colour online) Same as Fig. \ref{Fig7} but for a top-superlattice setup.}
\label{Fig8}
\end{figure}

\subsection{Role of the intrinsic doping}
\label{Impact-Eint}
In this subsection we solve the same problem as in the previous one, but we now include the effect of the inhomogeneous doping $\mu_{\rm int}$ created by the SC-semiconductor band offset and the surface charge density. Fig. \ref{Fig7} and Fig. \ref{Fig8} show, for the bottom- and top-superlattice devices, the lowest level energy as a function of the Zeeman field and the back gate voltage for different superlattice cell sizes (with $r_{\rm SC}=0.5$) and for different surface charge densities. Note that trivial regions are coloured in light pink, as in the phase diagrams of Fig. \ref{Fig6}.

The different columns in Fig. \ref{Fig7} and Fig. \ref{Fig8} correspond to different sizes of $L_{\rm cell}$. Notice that the size of the topological regions increases as the superlattice cell decreases. Actually, for a large enough $L_{\rm cell}$ the topological phase is inexistent, see Figs. \ref{Fig7}, \ref{Fig8} (d,g). For large superlattice cell sizes, topologically trivial localized states are present (black dashed lines), which may interfere with the MBSs. This effect is more pronounced in the bottom-superlattice setup because the back gate voltages needed to enter the topological phase are larger due to the screening of the SC fingers. This in turn produces stronger potential oscillations and subsequent localized states, as explained in Sec. \ref{Potential}. At smaller $L_{\rm cell}$ sizes, the localized states disappear.

For medium cell sizes $L_{\rm cell}$, which are probably more appropriate for experimental realization, we typically encounter the condition $\lambda_{\rm F}=L_{\rm cell}$ explained in the previous subsection and trivial holes appear in the topological phase, both in the bottom and top-superlattice setups. However, the top-superlattice setup develops larger topological regions and they are present for the two values of $\rho_{\rm surf}$ considered, see Figs. \ref{Fig8}(e,h). In the bottom-superlattice case no topological region is found for the larger $\rho_{\rm surf}$, see Fig. \ref{Fig7}(e).

For small $L_{\rm cell}$ sizes the topological phase is more stable, meaning that there are no trivial holes. This is so because for small and short potential oscillations the electrons in the wire feel an effective homogeneous potential \cite{Levine:PRB17}. Moreover, the performance of both setups (top and bottom) is comparable, although the back gate voltages needed for the bottom one are much larger.


\subsection{Impact of inhomogeneous induced pairing}
\label{Impact-SC}
Finally, we consider the impact of the inhomogeneous superconductivity. For this purpose, we solve a 1D wire where we fix the chemical potential and Rashba coupling to constant values. The superconducting pairing amplitude is taken as a telegraph function that oscillates between $\Delta_0=0.2$meV and zero with a period given by $L_{\rm cell}$ and $r_{\rm SC}$. As done in the previous sections, this is a simplified model to understand qualitatively the effect of inhomogeneous superconductivity. When we consider the realistic 3D model later on, the induced pairing will be only present at the surface of the wire in the regions where it is close to the SC fingers.

\begin{figure}
\includegraphics{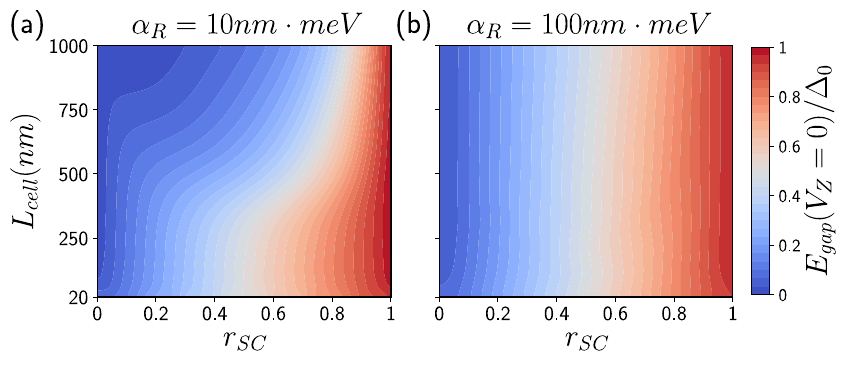}
\caption{(Colour online) Energy gap (at Zeeman energy $V_{\rm Z}=0$ and $k=0$) versus $L_{\rm cell}$ and $r_{\rm SC}=L_{\rm SC}/L_{\rm cell}$ for a 1D nanowire with a telegraph superconducting pairing that oscillates between $\Delta_0=0.2$meV and zero along $x$.  The chemical potential and Rashba coupling are fixed to homogeneous values $\mu=0$ and (a) $\alpha_{\rm R}=10$meV$\cdot$nm, (b) $\alpha_{\rm R}=100$meV$\cdot$nm.}
\label{Fig9}
\end{figure}

Fig. \ref{Fig9} shows the energy gap (energy of the lowest-energy state at $k=0$) normalized to $\Delta_0$ for an infinite 1D wire against the superlattice parameters $L_{\rm cell}$ and $r_{\rm SC}=L_{\rm SC}/L_{\rm cell}$. For small coverage $r_{\rm SC}<0.5$ the induced superconductivity is poor and it improves as $r_{\rm SC}$ increases. For $r_{\rm SC}\rightarrow 1$ we recover a perfect induced gap $\Delta_0$ corresponding to a wire covered by an uniform SC at $V_{\rm Z}=0$. Interestingly, for strong spin-orbit coupling the gap energy basically does not depend on $L_{\rm cell}$, see Fig. \ref{Fig9}(b). However, for small $\alpha_{\rm R}$ the induced gap worsens considerably with $L_{\rm cell}$, as shown in  Fig. \ref{Fig9}(a).


\begin{figure}
\includegraphics{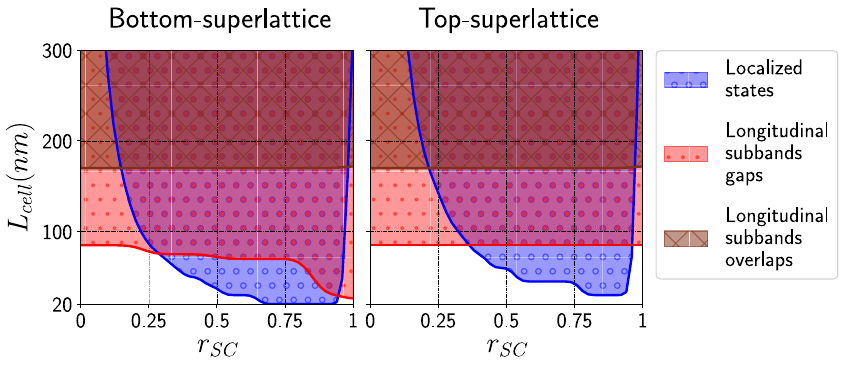}
\caption{(Colour online) Approximate regions in superlattice parameter space $L_{\rm cell}$ and $r_{\rm SC}$ where different mechanisms that spoil the topological phase appear, such as the formation of longitudinal subband overlaps, longitudinal subband gaps and localized states; marked in brown, red and blue, respectively. We have taken $V_{\rm Z}=0.6$meV, $\langle\mu_{\rm int}\rangle=200$meV, $\langle\mu_{\rm V_{\rm gate}}\rangle\in[0,3]$meV, and $\langle\alpha_{z}\rangle\in[5,50]$meV$\cdot$nm.}
\label{Fig10}
\end{figure}

\subsection{Superlattice features in parameter space}
\label{Optimal_parameters}
We can summarize our previous findings by plotting a diagram in parameter space that shows the different features caused by the superlattice and that interfere with the topological phase. This is done in Fig. \ref{Fig10} versus $L_{\rm cell}$ and $r_{\rm SC}$ for $V_{\rm Z}=0.6$meV and $\Delta_0=0.2$meV, and taking the following (realistic) spatial average values for other parameters: $\langle\mu_{int}\rangle=200$meV, $\langle\mu_{\rm V_{\rm gate}}\rangle\in[0,3]$meV and $\langle\alpha_{z}\rangle\in[5,50]$meV$\cdot$nm.

In the brown area we have values of $L_{\rm cell}$ and $r_{\rm SC}$ for which the Fermi energy crosses an even number of Fermi pair points in the nanowire dispersion relation. This happens when the level spacing between longitudinal subbands is smaller than the (energy) size of the topological phase ($\frac{\pi^2\hbar^2}{2mL_{\rm cell}^2}\le \sqrt{V_{\rm Z}^2-\Delta^2}$). In this case the topological regions of  contiguous longitudinal subbands interfere and the system exits the topological phase (there is an annihilation of an even number of Majoranas at each wire's edge). See, for instance, the upper subbands plotted in Figs. \ref{Fig6}(a,b). 

In the red area we have values of $L_{\rm cell}$ and $r_{\rm SC}$ for which there appear gaps between (the lowest) longitudinal subbands in the nanowire's dispersion relation. As we explained in Sec. \ref{Impact-mu}, when the Fermi energy is within these gaps, trivial holes emerge in the topological regions of the phase diagram. See for example the bottom-right corner of Fig. \ref{Fig6}(e). This happens when there is a resonance between the Fermi wavelength $\lambda_{\rm F}$ and the superlattice length $L_{\rm cell}$.
The red area is somewhat larger for the bottom-superlattice than for the top one. This is because the appearance and size of the longitudinal subbands gaps depends on the strength of the potential oscillations, which is larger for the bottom-superlattice due to the back gate's screening by the metallic fingers.

Finally, in the blue area localized states are formed. As we saw in Sec. \ref{Potential}, the superlattice of fingers creates potential oscillations along the wire. When the height of these oscillations is large enough ($\frac{\pi^2\hbar^2}{2mL_{\rm SC}^2}\le \frac{\sigma_{\rm \phi}}{<\phi>}\langle\mu_{\rm int}\rangle$), there appear potential wells for the electrons that create localized states (see App. \ref{SI3} for the $L_{\rm cell}$-$r_{\rm SC}$ dependence of $\langle\phi\rangle$ and $\sigma_{\rm \phi}$). These states interfere with the MBSs detaching them from zero energy. Moreover, when the potential oscillations are very strong, they divide effectively the nanowire into regions of smaller length, destroying the Majoranas. Again, the blue area is slightly larger for the bottom-superlattice than for the top-one.

This diagram gives us an idea of different detrimental mechanisms for a robust topological phase that may appear as a function of superlattice parameters. This does not mean that we cannot find topological regions for those $L_{\rm cell}$ and $r_{\rm SC}$ values, but that those regions will be interrupted at some points instead of extending more widely as a function of nanowire parameters. To complete this study  we should also consider the size of the topological minigap. As we saw in Sec. \ref{Impact-SC} (see Fig. \ref{Fig9}), it decreases when the superconducting partial coverage $r_{\rm SC}$ does, which is additionally Rashba coupling dependent (see App. \ref{SI3} for more details). Moreover, we have to bear in mind that the qualitative analysis of Fig. \ref{Fig10} is performed for a 1D model of the nanowire. When a 3D wire is considered, several transverse modes can be occupied. In this case there will be an interplay between longitudinal and transverse subbands that will introduce further complexity to the determination of the optimal superlattice parameters.

\section{3D results}
\label{3D_results}
In this section we consider together all the different ingredients that have been analysed separately in the previous sections and for a realistic 3D nanowire. In particular, we take representative superlattice parameters $L_{\rm cell}=100$nm and $r_{\rm SC}=0.5$. To calculate the electrostatic potential profile we perform self-consistent Poisson simulations in the Thomas-Fermi approximation for the bottom and top-superlattice setups. We find the wire states by diagonalizing the Bogoliubov-de Gennes Hamiltonian for a $2\mu$m long wire using the previous potential. As mentioned in Sec. \ref{Impact-SC}, we model the induced pairing as a telegraph function with $\Delta_0=0.2$meV in the regions of the wire close to the SC fingers and zero otherwise. In particular, for these 3D calculations we consider $\Delta_0\neq 0$ for a certain depth ($\sim 25\%$ of the wire's width) close to the SC fingers in the transverse direction.

\begin{figure*}
\includegraphics{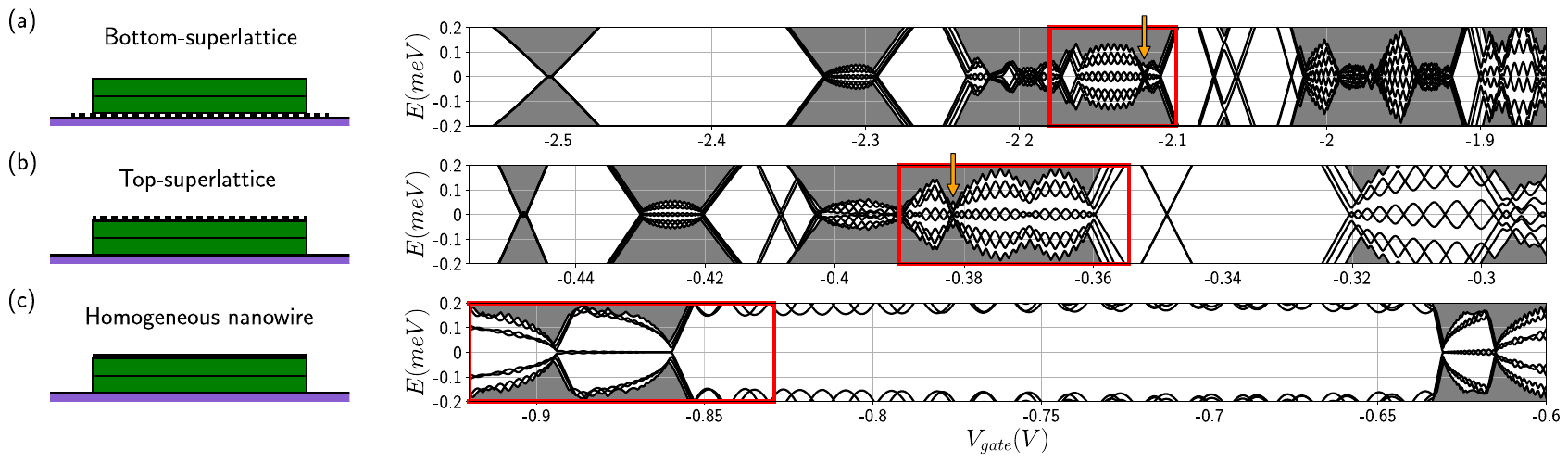}
\caption{(Colour online) Low energy spectrum versus back gate voltage for a $2\mu$m long 3D  top-superlattice nanowire (a), bottom-superlattice nanowire (b), and homogeneous nanowire (c). Superlattice parameters are $L_{\rm cell}=100$nm and $r_{\rm SC}=0.5$. Wire parameters are $V_{\rm Z}=0.6$meV, $\Delta_0=0.2$meV, $V_{\rm SC}=0.2$V and $\rho_{\rm surf}/e=2\cdot 10^{17}(cm)^{-3}$. The red rectangles represent the $V_{\rm gate}$ values for which Fig. \ref{Fig12} is plotted.}
\label{Fig11}
\end{figure*}

In Fig. \ref{Fig11} we show the low-energy spectrum as a function of back gate voltage for a particular value of Zeeman splitting, $V_{\rm Z}=0.6$meV, both for the bottom-superlattice in (a) and the top-superlattice setup in (b). We explore a wide range of $V_{\rm gate}$ values that corresponds to the first transverse occupied subband that develops topological states (seen as quasi-zero energy  states whose energies split from zero in an oscillating manner). As explained before, this subband appears for larger negative values of $V_{\rm gate}$ in the bottom-superlattice case due to the screening effects of the SC fingers. We note that, strictly speaking, in these systems one cannot really label subbands as purely transverse or longitudinal because the spin-orbit term in the Hamiltonian mixes the two momenta. However, and due to the small cross-section of the wires, groups of subbands have still a dominant weight on a particular quasi-transverse subband.

In these spectra we can observe all the phenomenology that we have been discussing in previous sections. For the most negative values of $V_{\rm gate}$, left part of Figs. \ref{Fig11}(a,b), the wire is almost empty except for very flat bands that appear at the quantum wells of the electrostatic potential oscillations. As a function of $V_{\rm gate}$ these create quick gap closings and reopenings and the topological phase cannot be developed. As $V_{\rm gate}$ is increased, middle part of Figs. \ref{Fig11}(a,b), different dispersing longitudinal subbands become populated. When the topological conditions are satisfied, we find extended $V_{\rm gate}$ regions with oscillating low energy modes separated by a minigap from the quasicontinuum of states (dark grey). These are the regions of interest because those oscillating states correspond to (more or less overlapping) MBSs localized at the left and right edges of the finite-length wire. The size of the oscillations and the minigap depends on the longitudinal subband. Sometimes, these topological regions are crossed by a localized state that closes the minigap at a certain $V_{\rm gate}$ point (see arrows in Figs. \ref{Fig11}(a,b)). The localized states disperse linearly with $V_{\rm gate}$ and cross zero energy displaying an $x$ shape. Other times we find trivial regions (without Majorana oscillations) between two topological ones due to topological subbands gaps or to topological subbands overlaps, as explained in Sec. \ref{Impact-mu}. Finally, at the right-most values of $V_{\rm gate}$ an additional transverse subband crosses below the Fermi level and the spectrum becomes more intricate, with the even-odd effect playing a role (not shown).

For comparison, we also show in Fig. \ref{Fig11}(c) the case of a nanowire homogeneously covered by a SC at the top of the wire. The range of $V_{\rm gate}$ values displayed in this case is chosen so that no hole states appear in the system. For more negative voltages the lower part of the nanowire becomes populated by hole quasiparticles from the valence band and a proper description of the system would require to consider an extended version of the model Hamiltonian of Eq. (\ref{Hamiltonian}) where electrons and holes coexist. To avoid this complication, we analyse higher voltages for which several transverse subbands are already populated. Note that here there are no longitudinal subbands. At the left and right parts of panel (c) we observe the well known even-odd effect between overlapping topological regions of different subbands. In the middle part, however, and for a pretty wide range of gate voltages, we have a region with no subgap states that corresponds to the trivial phase developed between two well separated transverse subbands.

\begin{figure*}
\includegraphics{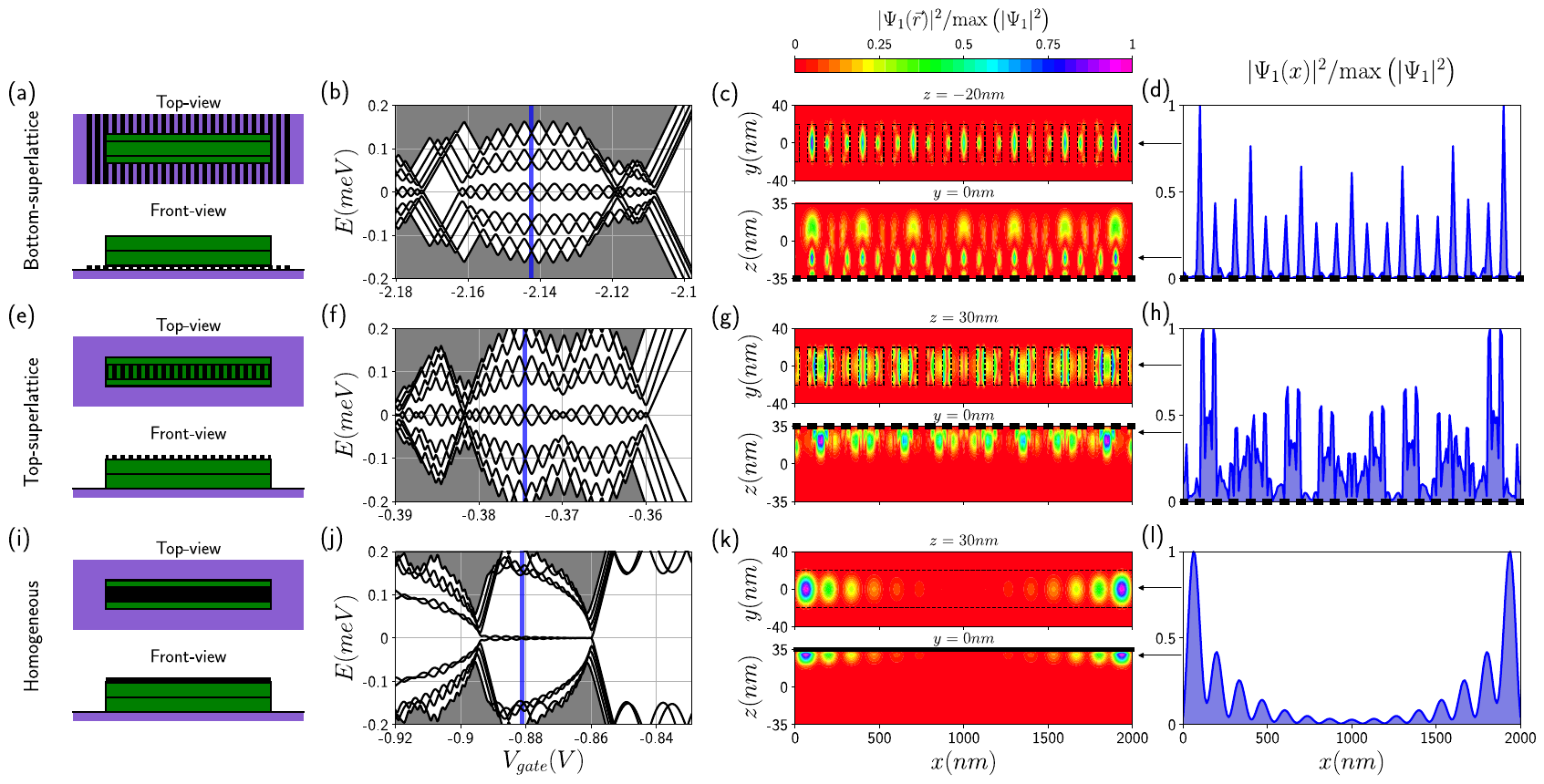}
\caption{(Colour online) Low energy spectrum and Majorana wave function for the same devices of Fig. \ref{Fig11} at the region of the red rectangles. (a,e) Top and front view of the two setups considered throughout this work: bottom- and top-superlattice of SC fingers. (b,f) Low-energy spectrum versus back gate voltage. (c,g) Probability density of the lowest-energy eigenstate at the voltage marked with a blue line in (b,f). (d,h) Longitudinal cut of the probability density of (c,g) at the (y,z) cross-section values marked by arrows. For comparison, we show equivalent results for an homogeneous nanowire in (i-l).}
\label{Fig12}
\end{figure*}

\begin{figure}
\includegraphics{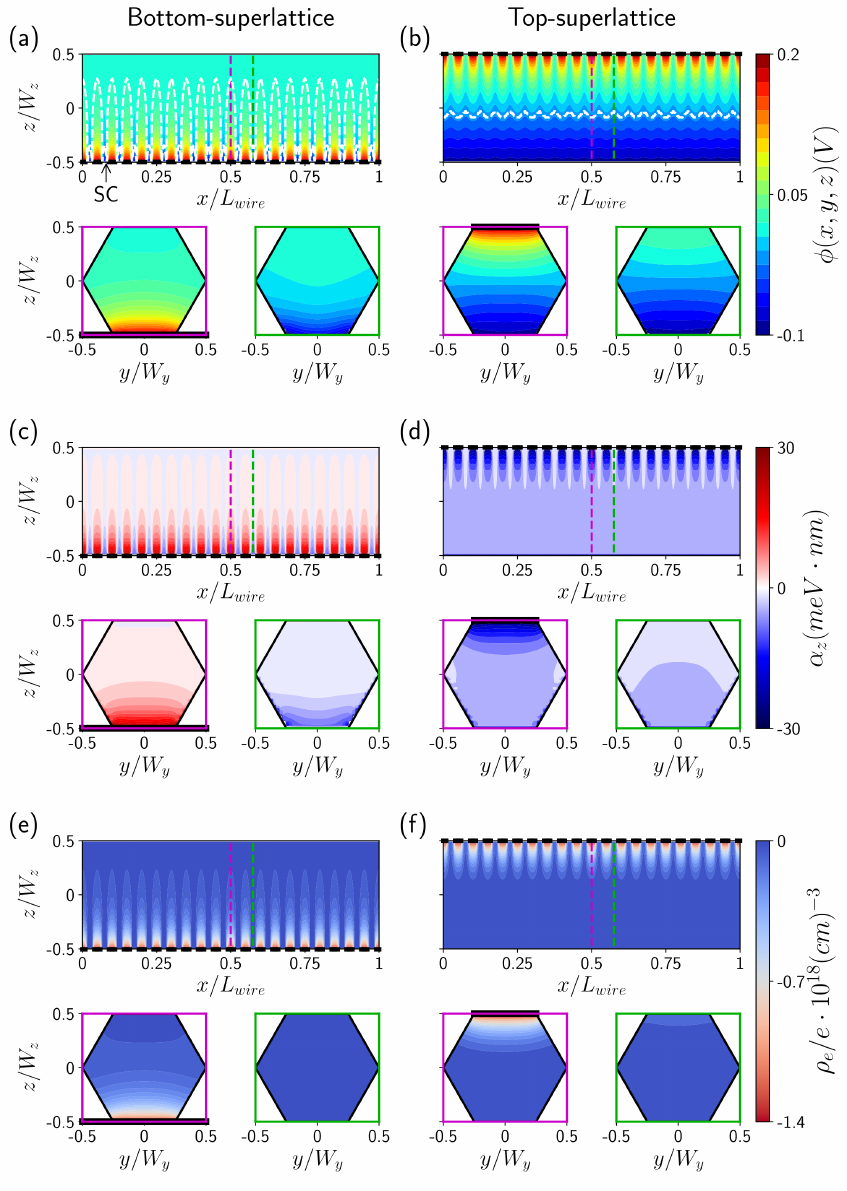}
\caption{(Colour online) Electrostatic potential (a,b), Rashba coupling $\alpha_z$ (c,d) and charge density profiles (e,f) for the same bottom and top-superlattice nanowires of Fig. \ref{Fig12}. Here, $V_{\rm gate}=-2.142$V for the bottom-superlattice and $V_{\rm gate}=-0.376$V for the top one, marked by blue lines in Figs. \ref{Fig12}(b,f). The total wire's charge is $Q_{\rm tot}/e=809$ for (e) and $Q_{\rm tot}/e=633$ for (f).}
\label{Fig13}
\end{figure}

Now we focus more specifically on one of the topological regions and analyse the location and shape of its MBSs. In Fig. \ref{Fig12} we show with more detail the low-energy spectrum as a function of back gate voltage for the regions marked by a red rectangle in Fig. \ref{Fig11}. To understand their behaviour, in Fig. \ref{Fig13} we plot the corresponding electrostatic potential, Rashba coupling $\alpha_z$ and charge density profiles for the $V_{\rm gate}$ voltage marked by a blue line in the corresponding spectrum.

The topological minigap is somewhat larger for the top-superlattice setup than for the bottom one. In the top-superlattice device it reaches approximately $\Delta_0/2$, which corresponds to the maximum possible induced gap for a superlattice with $r_{\rm SC}=0.5$, see analysis of Fig. \ref{Fig9}. This relatively large value can be understood by looking at the Rashba coupling profile in Fig. \ref{Fig13}(d). We see that $\alpha_z$ has a pretty homogeneous finite value all over the wire and it gets specially sizeable ($\sim -30$meV$\cdot$nm) below the SC fingers, precisely where most of the charge density is located according to Fig. \ref{Fig13}(f). However, the minigap in the bottom-superlattice is smaller than in the top's one. In this case $\alpha_z$ strongly oscillates between positive and negative values along the wire's axis, see Fig. \ref{Fig13}(c), resulting in a smaller average Rashba coupling. In the homogeneous case the minigap is the largest, close to $\Delta_0$ in the middle region of panel Fig. \ref{Fig12}(j). Here the induced effective gap has to be necessarily better since the SC covers the whole length of the wire. Moreover, there is a homogeneous and large Rashba coupling along the wire ($\sim -30$meV$\cdot$nm) close to the SC where the charge density concentrates (not shown here).
Concerning the Majorana oscillations, they are pretty comparable for the two types of superlattices and definitively bigger than for the homogeneous case.

In Figs. \ref{Fig12}(c,g,k) we plot the Majorana probability density of the different setups along and across the wire for the values of $V_{\rm gate}$ marked by the blue lines in (b,f,j), respectively. We find that in all cases the MBSs are localized at the edges of the wire, but with different longitudinal and transverse profiles. Across the wire's section the wave function tends to be close to the SC fingers in the top-superlattice setup. This is consistent with the charge density profile of Fig. \ref{Fig13}(f). On the other hand, the probability density oscillates from top to bottom in the bottom-superlattice one, see lower panel of Fig. \ref{Fig12}(c). As we noticed in Sec. \ref{Potential}, this is related to the shape of the potential profile due to the strong gate voltages needed to deplete the wire in this setup. The probability density accommodates to the isopotential curves, which for the bottom-superlattice device oscillate from top to bottom in the $z$-direction as highlighted with a white guideline in Fig. \ref{Fig13}(a) for a particular $\phi$ value.

Figures \ref{Fig12}(d,h,l) show longitudinal cuts of the probability density at the (y,z) cross-section values marked by arrows in Figs.  \ref{Fig12}(c,g,k). As expected, in the homogeneous case the wave function decays exponentially towards the wire's center with the Majorana localization length $\xi_{\rm M}$ \cite{Klinovaja:PRB12}. For the parameters of this case we obtain $\xi_{\rm M}=350$nm, which is consistent with panel (l). On the other hand, for the superlattice nanowires the decay length is characterized by the interplay between two scales, the Majorana length and the superlattice length $L_{\rm cell}$. The decay length in the homogeneous case is shorter and the probability density is pretty localized at the wire edges and almost zero at its center. This is not the case for the superlattices since their wave functions decay more slowly. To quantify this, we finally compute the absolute value of the Majorana charge $Q_{\rm M}$ that measures the wave function overlap between the right and the left Majoranas \cite{Ben-Sach:PRB15, Escribano:BJN18,Penaranda:PRB18}
\begin{equation}
|Q_{\rm M}|=e\left|\int u_{\mathrm{L}}(\vec{r})u_{\mathrm{R}}(\vec{r}) dr^3\right|,
\end{equation}
where $u_{\rm L,R}$ are the electron components of the left and right Majorana wavefunctions, given by $\gamma_{\rm L}=\psi_{+1}+\psi_{-1}$ and $\gamma_{\rm R}=-i(\psi_{+1}-\psi_{-1})$, being $\psi_{\pm1}$ the even/odd lowest-energy eigenstates. We get the values $\left|Q_{\rm M}^{\rm BS}\right|/e=0.93$, $\left|Q_{\rm M}^{\rm TS}\right|/e=0.88$ and $\left|Q_{\rm M}^{\rm h}\right|/e=0.63$ for the bottom-superlattice, top-superlattice and homogeneous cases, respectively. As expected, the Majorana charge is larger for both superlattice devices compared to the homogeneous case.

\subsection{Alternative superlattice configuration}
\label{Alternative}

\begin{figure}
\centering
\includegraphics{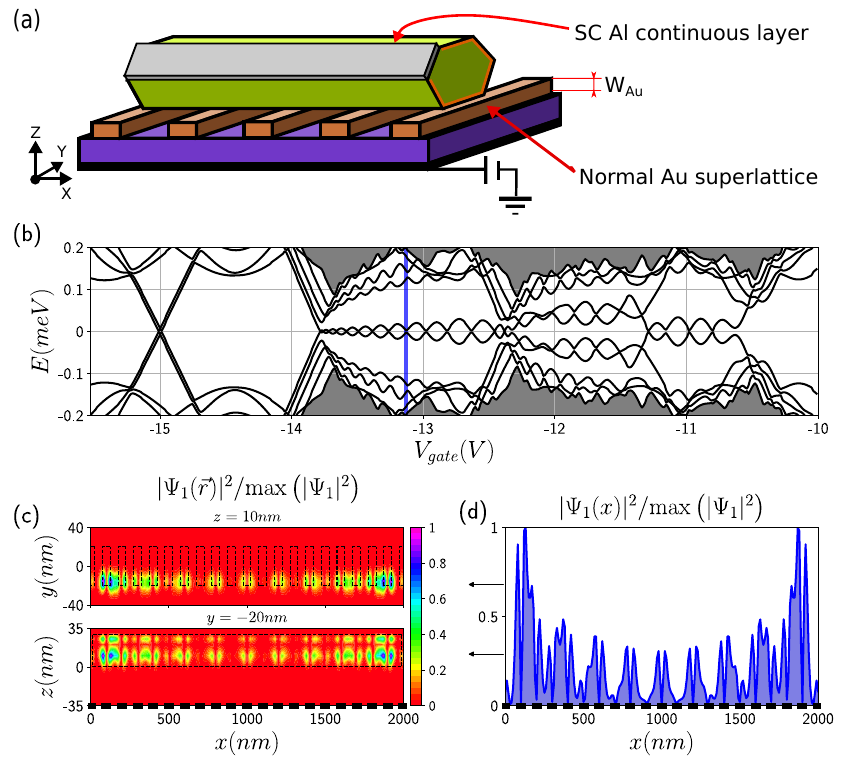}
\caption{\label{Fig14} (Colour online) (a) Alternative superlattice nanowire configuration designed to increase the MBSs topological protection. It combines a semiconducting nanowire (green) with one facet covered uniformly by SC layer (grey) and a superlattice of (non-superconducting) fingers (brown). (b) Low-energy spectrum versus back gate voltage. (c) Probability density of the lowest-energy eigenstate at the voltage marked with a blue line in (b). (d) Longitudinal cut of the probability density of (c) at the (y,z) cross-section values marked by arrows.  Parameters are the same as in Fig. \ref{Fig12}: $L_{\rm wire}=2\mu$m, $L_{\rm cell}=100$nm, $W_{\rm Au}=W_{\rm Al}=10$nm, $r_{\rm SC}=0.5$, $V_{\rm Z}=0.6$meV, $\Delta_0=0.2$meV, $V_{\rm SC}=0.2$V and $\rho_{\rm surf}/e=2\cdot 10^{17}(cm)^{-3}$. We take $V_{\rm N}=0$V as the boundary condition for the fingers.}
\end{figure}

We have seen that the main inconvenience of the Majorana superlattice nanowires analysed in this work comes from the partial superconducting coverage produced by the SC superlattice (specially as $r_{\rm SC}$ diminishes). This leads to a reduced induced SC gap that, in turn, produces a smaller topological minigap and a larger Majorana charge. We could improve this scenario by covering one of the wire's facets continuously with a thin SC layer, like in a conventional epitaxial Majorana nanowire, while still placing the hybrid structure on a superlattice. We analyse this alternative configuration in Fig.  \ref{Fig14} for the case of a bottom-superlattice setup. Now the superlattice can be either superconducting or normal (since the induced superconductivity is already provided by the SC layer). We choose here a set of normal metal fingers, such as gold, that could be used as tunneling local probes along the wire by driving a current between each finger and the SC homogeneous layer.
The tunneling coupling in this case is advantageous because it leads to a smaller wire's intrinsic doping and to a larger localization of the MBS wavefunctions close to the Al SC layer, where the electrostatic potential and induced pairing are larger.

In Fig. \ref{Fig14}(b) we show the low-energy spectrum of this setup for the same parameters of Fig. \ref{Fig12} except for the boundary condition between the (normal) bottom superlattice and the wire, which we take as $V_{\rm N}\simeq0$V. The values of $V_{\rm gate}$ for which we find the first topological subbands are pretty negative since the continuous Al layer induces a large intrinsic doping in the wire. The structure of this spectrum is a combination of the homogeneous and superlattice ones. From $V_{\rm gate}\simeq-13.7$V to $\simeq-12.3$V one transverse topological subband is occupied. At that point a different transverse subband populates the wire and the even-odd effect destroys the topological phase (as it occurs in the homogeneous wire). However, at $V_{\rm gate}\simeq-11.3$V a zero energy mode appears again but without a (prominent) gap closing. This is the signature of a gap between different longitudinal subbands, which allows one of the last two transverse subbands to re-enter into the topological phase. The interplay between longitudinal and transverse subbands gives rise to a wider $V_{\rm gate}-V_{\rm Z}$ space where topological states emerge, in comparison to a homogeneous nanowire, as it was previously stated in Ref. \onlinecite{Levine:PRB17}.

Now, as was our intention, in the topological regions we get a topological minigap that is comparable to the one of the homogeneous case, see Fig. \ref{Fig12}(j). The probability density of the lowest energy mode at the $V_{\rm gate}$ value marked with a blue line in (b) can be seen in (c). As expected, it is located close to the Al thin layer in the transverse direction. A longitudinal cut at the (z,y) values marked by arrows is shown in (d). The MBSs, that still display a doubling of the oscillating peaks characteristic of the superlattice, decay exponentially from the edges towards the wire's center faster than for the top- and bottom-superlattices analysed before. The Majorana charge is now $\left|Q_{\rm M}\right|/e=0.71$, considerably smaller than for the bottom superlattice alone (0.93) and closer to that of the homogeneous case (0.61). The sizeable minigap in this case protects the system from quasiparticle excitations, separating the Majorana modes from the quasi-continuum of states and preventing transitions into it due to, e.g., temperature or out of equilibrium perturbations.

To finish this section, we would like to mention that in this study and for simplicity we have ignored the orbital effects of the magnetic field. According to the literature (see for instance Ref. \onlinecite{Winkler:arxiv18}), the orbital effects are important when the electron's wave function is spread across the wire's section, especially when it has a ring-like shape. In this case, the electrons circulate around the magnetic field that points along the wire and interference \emph{orbital} effects appear. Furthermore, orbital effects are also enhanced for high electron densities, since most high transverse subbands have large angular momentum that couples strongly to the magnetic field. Conversely, the orbital effects diminish both for low transverse subbands and when the electron's wave function is pushed towards the SCs (by the action of the back gate), since it then occupies only a small region of the wire's section. We note that this is precisely the region in the spectrum that we focus on in Figs. \ref{Fig12} and \ref{Fig14}. We have explored the first occupied transverse subband (that displays MBSs) for the different superlattice structures, since it is the best behaved for Majorana purposes. For the back gate voltages involved, the wave function is indeed pushed towards the SC fingers (which is beneficial for the stability of the Majoranas since the induced pairing, and consequently the minigap, are larger there). Admittedly, this is not the case for the bottom superlattice setup, Fig. \ref{Fig12}(c), where the electron probability density oscillates from top to bottom in the transverse direction. Therefore, we expect that the orbital effects might be important in that case and beyond the current analysis performed in our work.

\section{Summary and Conclusions}
\label{Conclusions}
In this work we have analysed in detail the proposal of Levine \textit{et al.}\cite{Levine:PRB17} to look for topological superconductivity in Majorana nanowires in which the induced superconductivity is achieved by proximity to a superlattice of SC fingers (instead of having the SC cover continuously the length of the semiconducting wire). This configuration can have practical benefits to manipulate the Majorana wave function and to measure it. For instance, one could use an STM tip to drive a current between the tip and each of the SC fingers to measure the local density of states along the wire. The fingers could also work as local probes themselves.

Specifically, here we study the impact of the three-dimensionality of the system and the electrostatic environment on the spectral properties of the nanowire. To this end, we compute self-consistently the 3D Schr\"odinger-Poisson equations in the Thomas-Fermi approximation, where we include the Rashba coupling as a term locally proportional to the electric field. We consider two types of experimental setups, one in which the SC superlattice is on top of the nanowire and the other where it is below with respect to the back gate. We find that an accurate description of the nanowire boundary conditions and the surrounding media are crucial for a proper understanding of the system's properties. In particular, the interface of the nanowire with the SC, vacuum or substrate, creates an accumulation of electrons around the wire's cross-section. Its main effect is to contribute to the average intrinsic doping of the wire (that has to be compensated with an external gate when looking for the first populated subbands). On the other hand, the extrinsic doping produced by the applied gate voltage is dominated by the SC superlattice structure, giving rise to an inhomogeneous (oscillating) electrostatic  potential.

Depending on the location of the SC superlattice and the number and width of the SC fingers, we find a rich phenomenology that includes the emergence of trivial holes in the topological phase diagram and the formation of localized states near the SC fingers that may interfere with the topological states. Moreover, since the Rashba coupling is proportional to the electric field, the spin-orbit coupling also becomes an inhomogeneous quantity in this system. 
This results in a reduction of the topological minigap, specially in the bottom-superlattice device, owing to a lower spatial average Rashba value. In the same vein, the induced superconducting gap is  smaller than in a conventional homogeneous Majorana nanowire due to the smaller superconducting coverage of the nanowire. 

In contrast, the system develops a wider topological phase as a function of magnetic field and average chemical potential as a consequence of the emergence of additional (longitudinal) subbands. In the topological regions, MBSs do appear at the edges of the superlattice nanowire. Their probability density across the wire's section is concentrated close to the SC fingers in the top-superlattice setup. They extend further into the wire's bulk in the bottom-superlattice one due to the stronger potential oscillations created in this case by the back gate. Along the wire, the MBSs decay exponentially towards its center with a decay length characterized by the interplay between the superconducting coherence length and the superlattice length.

In general, we find that the performance of the two types of setups considered here is quite similar, although the bottom-superlattice nanowire is slightly worse because of the larger potential oscillations that appear in this case. In both cases, the main disadvantage is the poor topological protection of the MBSs (manifested in a small topological minigap and large left and right Majorana wave function overlap), arising essentially from the low superconducting coverage. This could be solved by covering one of the lateral wire's facets with a continuous SC layer while still placing it on a superlattice of fingers (that could be superconducting or not). This kind of device benefits from the superlattice structure (with a wider topological phase in $V_{\rm gate}-V_{\rm Z}$ space and the possibility to use the fingers as probes), and furthermore displays a sizeable topological minigap and small Majorana charge comparable to those of a conventional homogeneous Majorana nanowire. We thus believe that the use of mixed setups of this type is probably the best route towards creating Majorana states in the presence of superlattices.

The dataset and scripts required to plot the figures of this publication are available at the Zenodo repository \cite{Zenodo}.

\begin{acknowledgments}
We thank Eduardo J. H. Lee, Haim Beidenkopf, Enrique G. Michel, Nurit Avraham, Hadas Shtrikman and Jesper Nyg\aa rd for valuable discussions. Research supported by the Spanish MINECO through Grants FIS2016-80434-P, BES-2017-080374 and FIS2017-84860-R (AEI/FEDER, EU), the European Union's Horizon 2020 research and innovation programme under the FETOPEN grant agreement No 828948 and grant agreement LEGOTOP No 788715, the Ram\'on y Cajal programme RYC-2011-09345, the Mar\'ia de Maeztu Programme for Units of Excellence in R\&D (MDM-2014-0377), the DFG (CRC/Transregio 183, EI 519/7- 1), the Israel Science Foundation (ISF) and the Binational Science Foundation (BSF).
\end{acknowledgments}

\bibliography{superlattice}

\begin{thebibliography}{78}%
\makeatletter
\providecommand \@ifxundefined [1]{%
 \@ifx{#1\undefined}
}%
\providecommand \@ifnum [1]{%
 \ifnum #1\expandafter \@firstoftwo
 \else \expandafter \@secondoftwo
 \fi
}%
\providecommand \@ifx [1]{%
 \ifx #1\expandafter \@firstoftwo
 \else \expandafter \@secondoftwo
 \fi
}%
\providecommand \natexlab [1]{#1}%
\providecommand \enquote  [1]{``#1''}%
\providecommand \bibnamefont  [1]{#1}%
\providecommand \bibfnamefont [1]{#1}%
\providecommand \citenamefont [1]{#1}%
\providecommand \href@noop [0]{\@secondoftwo}%
\providecommand \href [0]{\begingroup \@sanitize@url \@href}%
\providecommand \@href[1]{\@@startlink{#1}\@@href}%
\providecommand \@@href[1]{\endgroup#1\@@endlink}%
\providecommand \@sanitize@url [0]{\catcode `\\12\catcode `\$12\catcode
  `\&12\catcode `\#12\catcode `\^12\catcode `\_12\catcode `\%12\relax}%
\providecommand \@@startlink[1]{}%
\providecommand \@@endlink[0]{}%
\providecommand \url  [0]{\begingroup\@sanitize@url \@url }%
\providecommand \@url [1]{\endgroup\@href {#1}{\urlprefix }}%
\providecommand \urlprefix  [0]{URL }%
\providecommand \Eprint [0]{\href }%
\providecommand \doibase [0]{http://dx.doi.org/}%
\providecommand \selectlanguage [0]{\@gobble}%
\providecommand \bibinfo  [0]{\@secondoftwo}%
\providecommand \bibfield  [0]{\@secondoftwo}%
\providecommand \translation [1]{[#1]}%
\providecommand \BibitemOpen [0]{}%
\providecommand \bibitemStop [0]{}%
\providecommand \bibitemNoStop [0]{.\EOS\space}%
\providecommand \EOS [0]{\spacefactor3000\relax}%
\providecommand \BibitemShut  [1]{\csname bibitem#1\endcsname}%
\let\auto@bib@innerbib\@empty
\bibitem [{\citenamefont {Hasan}\ and\ \citenamefont
  {Kane}(2010)}]{Hasan:RMP10}%
  \BibitemOpen
  \bibfield  {author} {\bibinfo {author} {\bibfnamefont {M.~Z.}\ \bibnamefont
  {Hasan}}\ and\ \bibinfo {author} {\bibfnamefont {C.~L.}\ \bibnamefont
  {Kane}},\ }\href {\doibase 10.1103/RevModPhys.82.3045} {\bibfield  {journal}
  {\bibinfo  {journal} {Rev. Mod. Phys.}\ }\textbf {\bibinfo {volume} {82}},\
  \bibinfo {pages} {3045} (\bibinfo {year} {2010})}\BibitemShut {NoStop}%
\bibitem [{\citenamefont {Alicea}(2012)}]{Alicea:RPP12}%
  \BibitemOpen
  \bibfield  {author} {\bibinfo {author} {\bibfnamefont {J.}~\bibnamefont
  {Alicea}},\ }\href {\doibase 10.1088/0034-4885/75/7/076501} {\bibfield
  {journal} {\bibinfo  {journal} {Rep. Prog. Phys.}\ }\textbf {\bibinfo
  {volume} {75}},\ \bibinfo {pages} {076501} (\bibinfo {year}
  {2012})}\BibitemShut {NoStop}%
\bibitem [{\citenamefont {Beenakker}(2013)}]{Beenakker:arxiv11}%
  \BibitemOpen
  \bibfield  {author} {\bibinfo {author} {\bibfnamefont {C.~W.~J.}\
  \bibnamefont {Beenakker}},\ }\href {\doibase
  10.1146/annurev-conmatphys-030212-184337} {\bibfield  {journal} {\bibinfo
  {journal} {Annual Review of Condensed Matter Physics}\ }\textbf {\bibinfo
  {volume} {4}},\ \bibinfo {pages} {113} (\bibinfo {year} {2013})}\BibitemShut
  {NoStop}%
\bibitem [{\citenamefont {Sato}\ and\ \citenamefont
  {Fujimoto}(2016)}]{Sato:JPSJ16}%
  \BibitemOpen
  \bibfield  {author} {\bibinfo {author} {\bibfnamefont {M.}~\bibnamefont
  {Sato}}\ and\ \bibinfo {author} {\bibfnamefont {S.}~\bibnamefont
  {Fujimoto}},\ }\href {\doibase 10.7566/JPSJ.85.072001} {\bibfield  {journal}
  {\bibinfo  {journal} {J. Phys. Soc. Jpn}\ }\textbf {\bibinfo {volume} {85}},\
  \bibinfo {pages} {072001} (\bibinfo {year} {2016})}\BibitemShut {NoStop}%
\bibitem [{\citenamefont {Aguado}(2017)}]{Aguado:rnc17}%
  \BibitemOpen
  \bibfield  {author} {\bibinfo {author} {\bibfnamefont {R.}~\bibnamefont
  {Aguado}},\ }\href {\doibase 10.1393/ncr/i2017-10141-9} {\bibfield  {journal}
  {\bibinfo  {journal} {La Rivista del Nuovo Cimento}\ }\textbf {\bibinfo
  {volume} {40}},\ \bibinfo {pages} {523} (\bibinfo {year} {2017})}\BibitemShut
  {NoStop}%
\bibitem [{\citenamefont {Lutchyn}\ \emph {et~al.}(2018)\citenamefont
  {Lutchyn}, \citenamefont {Bakkers}, \citenamefont {Kouwenhoven},
  \citenamefont {Krogstrup}, \citenamefont {Marcus},\ and\ \citenamefont
  {Oreg}}]{Lutchyn:NRM18}%
  \BibitemOpen
  \bibfield  {author} {\bibinfo {author} {\bibfnamefont {R.~M.}\ \bibnamefont
  {Lutchyn}}, \bibinfo {author} {\bibfnamefont {E.~P. A.~M.}\ \bibnamefont
  {Bakkers}}, \bibinfo {author} {\bibfnamefont {L.~P.}\ \bibnamefont
  {Kouwenhoven}}, \bibinfo {author} {\bibfnamefont {P.}~\bibnamefont
  {Krogstrup}}, \bibinfo {author} {\bibfnamefont {C.~M.}\ \bibnamefont
  {Marcus}}, \ and\ \bibinfo {author} {\bibfnamefont {Y.}~\bibnamefont
  {Oreg}},\ }\href {\doibase 10.1038/s41578-018-0003-1} {\bibfield  {journal}
  {\bibinfo  {journal} {Nature Reviews Materials}\ }\textbf {\bibinfo {volume}
  {3}},\ \bibinfo {pages} {52} (\bibinfo {year} {2018})}\BibitemShut {NoStop}%
\bibitem [{\citenamefont {Nayak}\ \emph {et~al.}(2008)\citenamefont {Nayak},
  \citenamefont {Simon}, \citenamefont {Stern}, \citenamefont {Freedman},\ and\
  \citenamefont {Das~Sarma}}]{Nayak:RMP08}%
  \BibitemOpen
  \bibfield  {author} {\bibinfo {author} {\bibfnamefont {C.}~\bibnamefont
  {Nayak}}, \bibinfo {author} {\bibfnamefont {S.~H.}\ \bibnamefont {Simon}},
  \bibinfo {author} {\bibfnamefont {A.}~\bibnamefont {Stern}}, \bibinfo
  {author} {\bibfnamefont {M.}~\bibnamefont {Freedman}}, \ and\ \bibinfo
  {author} {\bibfnamefont {S.}~\bibnamefont {Das~Sarma}},\ }\href {\doibase
  10.1103/RevModPhys.80.1083} {\bibfield  {journal} {\bibinfo  {journal} {Rev.
  Mod. Phys.}\ }\textbf {\bibinfo {volume} {80}},\ \bibinfo {pages} {1083}
  (\bibinfo {year} {2008})}\BibitemShut {NoStop}%
\bibitem [{\citenamefont {Aasen}\ \emph {et~al.}(2016)\citenamefont {Aasen},
  \citenamefont {Hell}, \citenamefont {Mishmash}, \citenamefont {Higginbotham},
  \citenamefont {Danon}, \citenamefont {Leijnse}, \citenamefont {Jespersen},
  \citenamefont {Folk}, \citenamefont {Marcus}, \citenamefont {Flensberg},\
  and\ \citenamefont {Alicea}}]{Aasen:PRX16}%
  \BibitemOpen
  \bibfield  {author} {\bibinfo {author} {\bibfnamefont {D.}~\bibnamefont
  {Aasen}}, \bibinfo {author} {\bibfnamefont {M.}~\bibnamefont {Hell}},
  \bibinfo {author} {\bibfnamefont {R.~V.}\ \bibnamefont {Mishmash}}, \bibinfo
  {author} {\bibfnamefont {A.}~\bibnamefont {Higginbotham}}, \bibinfo {author}
  {\bibfnamefont {J.}~\bibnamefont {Danon}}, \bibinfo {author} {\bibfnamefont
  {M.}~\bibnamefont {Leijnse}}, \bibinfo {author} {\bibfnamefont {T.~S.}\
  \bibnamefont {Jespersen}}, \bibinfo {author} {\bibfnamefont {J.~A.}\
  \bibnamefont {Folk}}, \bibinfo {author} {\bibfnamefont {C.~M.}\ \bibnamefont
  {Marcus}}, \bibinfo {author} {\bibfnamefont {K.}~\bibnamefont {Flensberg}}, \
  and\ \bibinfo {author} {\bibfnamefont {J.}~\bibnamefont {Alicea}},\ }\href
  {\doibase 10.1103/PhysRevX.6.031016} {\bibfield  {journal} {\bibinfo
  {journal} {Phys. Rev. X}\ }\textbf {\bibinfo {volume} {6}},\ \bibinfo {pages}
  {031016} (\bibinfo {year} {2016})}\BibitemShut {NoStop}%
\bibitem [{\citenamefont {Sarma}\ \emph {et~al.}(2015)\citenamefont {Sarma},
  \citenamefont {Freedman},\ and\ \citenamefont {Nayak}}]{Das:NPJ15}%
  \BibitemOpen
  \bibfield  {author} {\bibinfo {author} {\bibfnamefont {S.~D.}\ \bibnamefont
  {Sarma}}, \bibinfo {author} {\bibfnamefont {M.}~\bibnamefont {Freedman}}, \
  and\ \bibinfo {author} {\bibfnamefont {C.}~\bibnamefont {Nayak}},\ }\href
  {\doibase 10.1038/npjqi.2015.1} {\bibfield  {journal} {\bibinfo  {journal}
  {Npj Quantum Information}\ }\textbf {\bibinfo {volume} {1}},\ \bibinfo
  {pages} {15001} (\bibinfo {year} {2015})}\BibitemShut {NoStop}%
\bibitem [{\citenamefont {Nadj-Perge}\ \emph {et~al.}(2014)\citenamefont
  {Nadj-Perge}, \citenamefont {Drozdov}, \citenamefont {Li}, \citenamefont
  {Chen}, \citenamefont {Jeon}, \citenamefont {Seo}, \citenamefont {MacDonald},
  \citenamefont {Bernevig},\ and\ \citenamefont
  {Yazdani}}]{Nadj-Perge:Science14}%
  \BibitemOpen
  \bibfield  {author} {\bibinfo {author} {\bibfnamefont {S.}~\bibnamefont
  {Nadj-Perge}}, \bibinfo {author} {\bibfnamefont {I.~K.}\ \bibnamefont
  {Drozdov}}, \bibinfo {author} {\bibfnamefont {J.}~\bibnamefont {Li}},
  \bibinfo {author} {\bibfnamefont {H.}~\bibnamefont {Chen}}, \bibinfo {author}
  {\bibfnamefont {S.}~\bibnamefont {Jeon}}, \bibinfo {author} {\bibfnamefont
  {J.}~\bibnamefont {Seo}}, \bibinfo {author} {\bibfnamefont {A.~H.}\
  \bibnamefont {MacDonald}}, \bibinfo {author} {\bibfnamefont {B.~A.}\
  \bibnamefont {Bernevig}}, \ and\ \bibinfo {author} {\bibfnamefont
  {A.}~\bibnamefont {Yazdani}},\ }\href {\doibase 10.1126/science.1259327}
  {\bibfield  {journal} {\bibinfo  {journal} {Science}\ }\textbf {\bibinfo
  {volume} {346}},\ \bibinfo {pages} {602} (\bibinfo {year}
  {2014})}\BibitemShut {NoStop}%
\bibitem [{\citenamefont {Ruby}\ \emph {et~al.}(2015)\citenamefont {Ruby},
  \citenamefont {Pientka}, \citenamefont {Peng}, \citenamefont {von Oppen},
  \citenamefont {Heinrich},\ and\ \citenamefont {Franke}}]{Ruby:PRL15}%
  \BibitemOpen
  \bibfield  {author} {\bibinfo {author} {\bibfnamefont {M.}~\bibnamefont
  {Ruby}}, \bibinfo {author} {\bibfnamefont {F.}~\bibnamefont {Pientka}},
  \bibinfo {author} {\bibfnamefont {Y.}~\bibnamefont {Peng}}, \bibinfo {author}
  {\bibfnamefont {F.}~\bibnamefont {von Oppen}}, \bibinfo {author}
  {\bibfnamefont {B.~W.}\ \bibnamefont {Heinrich}}, \ and\ \bibinfo {author}
  {\bibfnamefont {K.~J.}\ \bibnamefont {Franke}},\ }\href {\doibase
  10.1103/PhysRevLett.115.197204} {\bibfield  {journal} {\bibinfo  {journal}
  {Phys. Rev. Lett.}\ }\textbf {\bibinfo {volume} {115}},\ \bibinfo {pages}
  {197204} (\bibinfo {year} {2015})}\BibitemShut {NoStop}%
\bibitem [{\citenamefont {Feldman}\ \emph {et~al.}(2017)\citenamefont
  {Feldman}, \citenamefont {Randeria}, \citenamefont {Li}, \citenamefont
  {Jeon}, \citenamefont {Xie}, \citenamefont {Wang}, \citenamefont {Drozdov},
  \citenamefont {Andrei~Bernevig},\ and\ \citenamefont {Yazdani}}]{Feldman:17}%
  \BibitemOpen
  \bibfield  {author} {\bibinfo {author} {\bibfnamefont {B.~E.}\ \bibnamefont
  {Feldman}}, \bibinfo {author} {\bibfnamefont {M.~T.}\ \bibnamefont
  {Randeria}}, \bibinfo {author} {\bibfnamefont {J.}~\bibnamefont {Li}},
  \bibinfo {author} {\bibfnamefont {S.}~\bibnamefont {Jeon}}, \bibinfo {author}
  {\bibfnamefont {Y.}~\bibnamefont {Xie}}, \bibinfo {author} {\bibfnamefont
  {Z.}~\bibnamefont {Wang}}, \bibinfo {author} {\bibfnamefont {I.~K.}\
  \bibnamefont {Drozdov}}, \bibinfo {author} {\bibfnamefont {B.}~\bibnamefont
  {Andrei~Bernevig}}, \ and\ \bibinfo {author} {\bibfnamefont {A.}~\bibnamefont
  {Yazdani}},\ }\href {\doibase doi:10.1038/nphys3947} {\bibfield  {journal}
  {\bibinfo  {journal} {Nature Physics}\ }\textbf {\bibinfo {volume} {13}},\
  \bibinfo {pages} {286} (\bibinfo {year} {2017})}\BibitemShut {NoStop}%
\bibitem [{\citenamefont {Pawlak}\ \emph {et~al.}(2016)\citenamefont {Pawlak},
  \citenamefont {Kisiel}, \citenamefont {Klinovaja}, \citenamefont {Meier},
  \citenamefont {Kawai}, \citenamefont {Glatzel}, \citenamefont {Loss},\ and\
  \citenamefont {Meyer}}]{Pawlak:17}%
  \BibitemOpen
  \bibfield  {author} {\bibinfo {author} {\bibfnamefont {R.}~\bibnamefont
  {Pawlak}}, \bibinfo {author} {\bibfnamefont {M.}~\bibnamefont {Kisiel}},
  \bibinfo {author} {\bibfnamefont {J.}~\bibnamefont {Klinovaja}}, \bibinfo
  {author} {\bibfnamefont {T.}~\bibnamefont {Meier}}, \bibinfo {author}
  {\bibfnamefont {S.}~\bibnamefont {Kawai}}, \bibinfo {author} {\bibfnamefont
  {T.}~\bibnamefont {Glatzel}}, \bibinfo {author} {\bibfnamefont
  {D.}~\bibnamefont {Loss}}, \ and\ \bibinfo {author} {\bibfnamefont
  {E.}~\bibnamefont {Meyer}},\ }\href {\doibase doi:10.1038/npjqi.2016.35}
  {\bibfield  {journal} {\bibinfo  {journal} {Npj Quantum Information}\
  }\textbf {\bibinfo {volume} {2}},\ \bibinfo {pages} {16035} (\bibinfo {year}
  {2016})}\BibitemShut {NoStop}%
\bibitem [{\citenamefont {Mourik}\ \emph {et~al.}(2012)\citenamefont {Mourik},
  \citenamefont {Zuo}, \citenamefont {Frolov}, \citenamefont {Plissard},
  \citenamefont {Bakkers},\ and\ \citenamefont
  {Kouwenhoven}}]{Mourik:Science12}%
  \BibitemOpen
  \bibfield  {author} {\bibinfo {author} {\bibfnamefont {V.}~\bibnamefont
  {Mourik}}, \bibinfo {author} {\bibfnamefont {K.}~\bibnamefont {Zuo}},
  \bibinfo {author} {\bibfnamefont {S.~M.}\ \bibnamefont {Frolov}}, \bibinfo
  {author} {\bibfnamefont {S.~R.}\ \bibnamefont {Plissard}}, \bibinfo {author}
  {\bibfnamefont {E.~P. A.~M.}\ \bibnamefont {Bakkers}}, \ and\ \bibinfo
  {author} {\bibfnamefont {L.~P.}\ \bibnamefont {Kouwenhoven}},\ }\href
  {\doibase 10.1126/science.1222360} {\bibfield  {journal} {\bibinfo  {journal}
  {Science}\ }\textbf {\bibinfo {volume} {336}},\ \bibinfo {pages} {1003}
  (\bibinfo {year} {2012})}\BibitemShut {NoStop}%
\bibitem [{\citenamefont {Deng}\ \emph {et~al.}(2016)\citenamefont {Deng},
  \citenamefont {Vaitiekenas}, \citenamefont {Hansen}, \citenamefont {Danon},
  \citenamefont {Leijnse}, \citenamefont {Flensberg}, \citenamefont
  {Nyg\r{a}rd}, \citenamefont {Krogstrup},\ and\ \citenamefont
  {Marcus}}]{Deng:Science16}%
  \BibitemOpen
  \bibfield  {author} {\bibinfo {author} {\bibfnamefont {M.~T.}\ \bibnamefont
  {Deng}}, \bibinfo {author} {\bibfnamefont {S.}~\bibnamefont {Vaitiekenas}},
  \bibinfo {author} {\bibfnamefont {E.~B.}\ \bibnamefont {Hansen}}, \bibinfo
  {author} {\bibfnamefont {J.}~\bibnamefont {Danon}}, \bibinfo {author}
  {\bibfnamefont {M.}~\bibnamefont {Leijnse}}, \bibinfo {author} {\bibfnamefont
  {K.}~\bibnamefont {Flensberg}}, \bibinfo {author} {\bibfnamefont
  {J.}~\bibnamefont {Nyg\r{a}rd}}, \bibinfo {author} {\bibfnamefont
  {P.}~\bibnamefont {Krogstrup}}, \ and\ \bibinfo {author} {\bibfnamefont
  {C.~M.}\ \bibnamefont {Marcus}},\ }\href {\doibase 10.1126/science.aaf3961}
  {\bibfield  {journal} {\bibinfo  {journal} {Science}\ }\textbf {\bibinfo
  {volume} {354}},\ \bibinfo {pages} {1557} (\bibinfo {year}
  {2016})}\BibitemShut {NoStop}%
\bibitem [{\citenamefont {Zhang}\ \emph {et~al.}(2018)\citenamefont {Zhang},
  \citenamefont {Liu}, \citenamefont {Gazibegovic}, \citenamefont {Xu},
  \citenamefont {Logan}, \citenamefont {Wang}, \citenamefont {van Loo},
  \citenamefont {Bommer}, \citenamefont {de~Moor}, \citenamefont {Car},
  \citenamefont {Veld}, \citenamefont {van Veldhoven}, \citenamefont
  {Koelling}, \citenamefont {Verheijen}, \citenamefont {Pendharkar},
  \citenamefont {Pennachio}, \citenamefont {Shojaei}, \citenamefont {Lee},
  \citenamefont {Palmstr{\o}m}, \citenamefont {Bakkers}, \citenamefont
  {Das~Sarma},\ and\ \citenamefont {Kouwenhoven}}]{Zhang:Nat17a}%
  \BibitemOpen
  \bibfield  {author} {\bibinfo {author} {\bibfnamefont {H.}~\bibnamefont
  {Zhang}}, \bibinfo {author} {\bibfnamefont {C.-X.}\ \bibnamefont {Liu}},
  \bibinfo {author} {\bibfnamefont {S.}~\bibnamefont {Gazibegovic}}, \bibinfo
  {author} {\bibfnamefont {D.}~\bibnamefont {Xu}}, \bibinfo {author}
  {\bibfnamefont {J.~A.}\ \bibnamefont {Logan}}, \bibinfo {author}
  {\bibfnamefont {G.}~\bibnamefont {Wang}}, \bibinfo {author} {\bibfnamefont
  {N.}~\bibnamefont {van Loo}}, \bibinfo {author} {\bibfnamefont {J.~D.}\
  \bibnamefont {Bommer}}, \bibinfo {author} {\bibfnamefont {M.~W.}\
  \bibnamefont {de~Moor}}, \bibinfo {author} {\bibfnamefont {D.}~\bibnamefont
  {Car}}, \bibinfo {author} {\bibfnamefont {R.~L. M. O.~h.}\ \bibnamefont
  {Veld}}, \bibinfo {author} {\bibfnamefont {P.~J.}\ \bibnamefont {van
  Veldhoven}}, \bibinfo {author} {\bibfnamefont {S.}~\bibnamefont {Koelling}},
  \bibinfo {author} {\bibfnamefont {M.~A.}\ \bibnamefont {Verheijen}}, \bibinfo
  {author} {\bibfnamefont {M.}~\bibnamefont {Pendharkar}}, \bibinfo {author}
  {\bibfnamefont {D.~J.}\ \bibnamefont {Pennachio}}, \bibinfo {author}
  {\bibfnamefont {B.}~\bibnamefont {Shojaei}}, \bibinfo {author} {\bibfnamefont
  {J.~S.}\ \bibnamefont {Lee}}, \bibinfo {author} {\bibfnamefont {C.~J.}\
  \bibnamefont {Palmstr{\o}m}}, \bibinfo {author} {\bibfnamefont {E.~P.}\
  \bibnamefont {Bakkers}}, \bibinfo {author} {\bibfnamefont {S.}~\bibnamefont
  {Das~Sarma}}, \ and\ \bibinfo {author} {\bibfnamefont {L.~P.}\ \bibnamefont
  {Kouwenhoven}},\ }\href {https://doi.org/10.1038/nature26142} {\bibfield
  {journal} {\bibinfo  {journal} {Nature}\ }\textbf {\bibinfo {volume} {556}},\
  \bibinfo {pages} {74} (\bibinfo {year} {2018})}\BibitemShut {NoStop}%
\bibitem [{\citenamefont {Chen}\ \emph {et~al.}(2017)\citenamefont {Chen},
  \citenamefont {Yu}, \citenamefont {Stenger}, \citenamefont {Hocevar},
  \citenamefont {Car}, \citenamefont {Plissard}, \citenamefont {Bakkers},
  \citenamefont {Stanescu},\ and\ \citenamefont {Frolov}}]{Chen:Science17}%
  \BibitemOpen
  \bibfield  {author} {\bibinfo {author} {\bibfnamefont {J.}~\bibnamefont
  {Chen}}, \bibinfo {author} {\bibfnamefont {P.}~\bibnamefont {Yu}}, \bibinfo
  {author} {\bibfnamefont {J.}~\bibnamefont {Stenger}}, \bibinfo {author}
  {\bibfnamefont {M.}~\bibnamefont {Hocevar}}, \bibinfo {author} {\bibfnamefont
  {D.}~\bibnamefont {Car}}, \bibinfo {author} {\bibfnamefont {S.~R.}\
  \bibnamefont {Plissard}}, \bibinfo {author} {\bibfnamefont {E.~P. A.~M.}\
  \bibnamefont {Bakkers}}, \bibinfo {author} {\bibfnamefont {T.~D.}\
  \bibnamefont {Stanescu}}, \ and\ \bibinfo {author} {\bibfnamefont {S.~M.}\
  \bibnamefont {Frolov}},\ }\href {\doibase 10.1126/sciadv.1701476} {\bibfield
  {journal} {\bibinfo  {journal} {Science Advances}\ }\textbf {\bibinfo
  {volume} {3}},\ \bibinfo {pages} {e1701476} (\bibinfo {year}
  {2017})}\BibitemShut {NoStop}%
\bibitem [{\citenamefont {Deng}\ \emph {et~al.}(2018)\citenamefont {Deng},
  \citenamefont {Vaitiek\.{e}nas}, \citenamefont {Prada}, \citenamefont
  {San-Jose}, \citenamefont {Nyg\r{a}rd}, \citenamefont {Krogstrup},
  \citenamefont {Aguado},\ and\ \citenamefont {Marcus}}]{Deng:PRB18}%
  \BibitemOpen
  \bibfield  {author} {\bibinfo {author} {\bibfnamefont {M.-T.}\ \bibnamefont
  {Deng}}, \bibinfo {author} {\bibfnamefont {S.}~\bibnamefont
  {Vaitiek\.{e}nas}}, \bibinfo {author} {\bibfnamefont {E.}~\bibnamefont
  {Prada}}, \bibinfo {author} {\bibfnamefont {P.}~\bibnamefont {San-Jose}},
  \bibinfo {author} {\bibfnamefont {J.}~\bibnamefont {Nyg\r{a}rd}}, \bibinfo
  {author} {\bibfnamefont {P.}~\bibnamefont {Krogstrup}}, \bibinfo {author}
  {\bibfnamefont {R.}~\bibnamefont {Aguado}}, \ and\ \bibinfo {author}
  {\bibfnamefont {C.~M.}\ \bibnamefont {Marcus}},\ }\href {\doibase
  10.1103/PhysRevB.98.085125} {\bibfield  {journal} {\bibinfo  {journal} {Phys.
  Rev. B}\ }\textbf {\bibinfo {volume} {98}},\ \bibinfo {pages} {085125}
  (\bibinfo {year} {2018})}\BibitemShut {NoStop}%
\bibitem [{\citenamefont {Vaitiek\.{e}nas}\ \emph
  {et~al.}(2018{\natexlab{a}})\citenamefont {Vaitiek\.{e}nas}, \citenamefont
  {Deng}, \citenamefont {Nyg\r{a}rd}, \citenamefont {Krogstrup},\ and\
  \citenamefont {Marcus}}]{Vaitiekenas:PRL18}%
  \BibitemOpen
  \bibfield  {author} {\bibinfo {author} {\bibfnamefont {S.}~\bibnamefont
  {Vaitiek\.{e}nas}}, \bibinfo {author} {\bibfnamefont {M.-T.}\ \bibnamefont
  {Deng}}, \bibinfo {author} {\bibfnamefont {J.}~\bibnamefont {Nyg\r{a}rd}},
  \bibinfo {author} {\bibfnamefont {P.}~\bibnamefont {Krogstrup}}, \ and\
  \bibinfo {author} {\bibfnamefont {C.~M.}\ \bibnamefont {Marcus}},\ }\href
  {\doibase 10.1103/PhysRevLett.121.037703} {\bibfield  {journal} {\bibinfo
  {journal} {Phys. Rev. Lett.}\ }\textbf {\bibinfo {volume} {121}},\ \bibinfo
  {pages} {037703} (\bibinfo {year} {2018}{\natexlab{a}})}\BibitemShut
  {NoStop}%
\bibitem [{\citenamefont {G\"{u}l}\ \emph {et~al.}(2018)\citenamefont
  {G\"{u}l}, \citenamefont {Zhang}, \citenamefont {Bommer}, \citenamefont
  {Moor}, \citenamefont {Car}, \citenamefont {Plissard}, \citenamefont
  {Bakkers}, \citenamefont {Geresdi}, \citenamefont {Watanabe}, \citenamefont
  {Taniguchi},\ and\ \citenamefont {Kouwenhoven}}]{Gul:NNano18}%
  \BibitemOpen
  \bibfield  {author} {\bibinfo {author} {\bibfnamefont {O.}~\bibnamefont
  {G\"{u}l}}, \bibinfo {author} {\bibfnamefont {H.}~\bibnamefont {Zhang}},
  \bibinfo {author} {\bibfnamefont {J.~D.~S.}\ \bibnamefont {Bommer}}, \bibinfo
  {author} {\bibfnamefont {M.~W. A.~d.}\ \bibnamefont {Moor}}, \bibinfo
  {author} {\bibfnamefont {D.}~\bibnamefont {Car}}, \bibinfo {author}
  {\bibfnamefont {S.~R.}\ \bibnamefont {Plissard}}, \bibinfo {author}
  {\bibfnamefont {E.~P. A.~M.}\ \bibnamefont {Bakkers}}, \bibinfo {author}
  {\bibfnamefont {A.}~\bibnamefont {Geresdi}}, \bibinfo {author} {\bibfnamefont
  {K.}~\bibnamefont {Watanabe}}, \bibinfo {author} {\bibfnamefont
  {T.}~\bibnamefont {Taniguchi}}, \ and\ \bibinfo {author} {\bibfnamefont
  {L.~P.}\ \bibnamefont {Kouwenhoven}},\ }\href
  {https://www.nature.com/articles/s41565-017-0032-8} {\bibfield  {journal}
  {\bibinfo  {journal} {Nature News}\ }\textbf {\bibinfo {volume} {13}},\
  \bibinfo {pages} {192} (\bibinfo {year} {2018})}\BibitemShut {NoStop}%
\bibitem [{\citenamefont {Grivnin}\ \emph {et~al.}(2018)\citenamefont
  {Grivnin}, \citenamefont {Bor}, \citenamefont {Heiblum}, \citenamefont
  {Oreg},\ and\ \citenamefont {Shtrikman}}]{Grivnin:arxiv18}%
  \BibitemOpen
  \bibfield  {author} {\bibinfo {author} {\bibfnamefont {A.}~\bibnamefont
  {Grivnin}}, \bibinfo {author} {\bibfnamefont {E.}~\bibnamefont {Bor}},
  \bibinfo {author} {\bibfnamefont {M.}~\bibnamefont {Heiblum}}, \bibinfo
  {author} {\bibfnamefont {Y.}~\bibnamefont {Oreg}}, \ and\ \bibinfo {author}
  {\bibfnamefont {H.}~\bibnamefont {Shtrikman}},\ }\href
  {https://arxiv.org/abs/1807.06632} {\bibfield  {journal} {\bibinfo  {journal}
  {arXiv:1807.06632}\ } (\bibinfo {year} {2018})}\BibitemShut {NoStop}%
\bibitem [{\citenamefont {Vaitiek\.{e}nas}\ \emph
  {et~al.}(2018{\natexlab{b}})\citenamefont {Vaitiek\.{e}nas}, \citenamefont
  {Deng}, \citenamefont {Krogstrup},\ and\ \citenamefont
  {Marcus}}]{Vaitiekenas:arxiv18}%
  \BibitemOpen
  \bibfield  {author} {\bibinfo {author} {\bibfnamefont {S.}~\bibnamefont
  {Vaitiek\.{e}nas}}, \bibinfo {author} {\bibfnamefont {J.}~\bibnamefont
  {Deng}}, \bibinfo {author} {\bibfnamefont {P.}~\bibnamefont {Krogstrup}}, \
  and\ \bibinfo {author} {\bibfnamefont {C.~M.}\ \bibnamefont {Marcus}},\
  }\href {https://arxiv.org/abs/1809.05513} {\bibfield  {journal} {\bibinfo
  {journal} {arXiv:1809.05513}\ } (\bibinfo {year}
  {2018}{\natexlab{b}})}\BibitemShut {NoStop}%
\bibitem [{\citenamefont {Setiawan}\ \emph {et~al.}(2017)\citenamefont
  {Setiawan}, \citenamefont {Liu}, \citenamefont {Sau},\ and\ \citenamefont
  {Das~Sarma}}]{Setiawan:PRB17}%
  \BibitemOpen
  \bibfield  {author} {\bibinfo {author} {\bibfnamefont {F.}~\bibnamefont
  {Setiawan}}, \bibinfo {author} {\bibfnamefont {C.-X.}\ \bibnamefont {Liu}},
  \bibinfo {author} {\bibfnamefont {J.~D.}\ \bibnamefont {Sau}}, \ and\
  \bibinfo {author} {\bibfnamefont {S.}~\bibnamefont {Das~Sarma}},\ }\href
  {\doibase 10.1103/PhysRevB.96.184520} {\bibfield  {journal} {\bibinfo
  {journal} {Phys. Rev. B}\ }\textbf {\bibinfo {volume} {96}},\ \bibinfo
  {pages} {184520} (\bibinfo {year} {2017})}\BibitemShut {NoStop}%
\bibitem [{\citenamefont {Liu}\ \emph {et~al.}(2017)\citenamefont {Liu},
  \citenamefont {Sau}, \citenamefont {Stanescu},\ and\ \citenamefont
  {Das~Sarma}}]{Liu:PRB17}%
  \BibitemOpen
  \bibfield  {author} {\bibinfo {author} {\bibfnamefont {C.-X.}\ \bibnamefont
  {Liu}}, \bibinfo {author} {\bibfnamefont {J.~D.}\ \bibnamefont {Sau}},
  \bibinfo {author} {\bibfnamefont {T.~D.}\ \bibnamefont {Stanescu}}, \ and\
  \bibinfo {author} {\bibfnamefont {S.}~\bibnamefont {Das~Sarma}},\ }\href
  {\doibase 10.1103/PhysRevB.96.075161} {\bibfield  {journal} {\bibinfo
  {journal} {Phys. Rev. B}\ }\textbf {\bibinfo {volume} {96}},\ \bibinfo
  {pages} {075161} (\bibinfo {year} {2017})}\BibitemShut {NoStop}%
\bibitem [{\citenamefont {Reeg}\ \emph
  {et~al.}(2018{\natexlab{a}})\citenamefont {Reeg}, \citenamefont {Dmytruk},
  \citenamefont {Chevallier}, \citenamefont {Loss},\ and\ \citenamefont
  {Klinovaja}}]{Reeg:PRB18b}%
  \BibitemOpen
  \bibfield  {author} {\bibinfo {author} {\bibfnamefont {C.}~\bibnamefont
  {Reeg}}, \bibinfo {author} {\bibfnamefont {O.}~\bibnamefont {Dmytruk}},
  \bibinfo {author} {\bibfnamefont {D.}~\bibnamefont {Chevallier}}, \bibinfo
  {author} {\bibfnamefont {D.}~\bibnamefont {Loss}}, \ and\ \bibinfo {author}
  {\bibfnamefont {J.}~\bibnamefont {Klinovaja}},\ }\href {\doibase
  10.1103/PhysRevB.98.245407} {\bibfield  {journal} {\bibinfo  {journal} {Phys.
  Rev. B}\ }\textbf {\bibinfo {volume} {98}},\ \bibinfo {pages} {245407}
  (\bibinfo {year} {2018}{\natexlab{a}})}\BibitemShut {NoStop}%
\bibitem [{\citenamefont {Moore}\ \emph {et~al.}(2018)\citenamefont {Moore},
  \citenamefont {Zeng}, \citenamefont {Stanescu},\ and\ \citenamefont
  {Tewari}}]{Moore:PRB18}%
  \BibitemOpen
  \bibfield  {author} {\bibinfo {author} {\bibfnamefont {C.}~\bibnamefont
  {Moore}}, \bibinfo {author} {\bibfnamefont {C.}~\bibnamefont {Zeng}},
  \bibinfo {author} {\bibfnamefont {T.~D.}\ \bibnamefont {Stanescu}}, \ and\
  \bibinfo {author} {\bibfnamefont {S.}~\bibnamefont {Tewari}},\ }\href
  {\doibase 10.1103/PhysRevB.98.155314} {\bibfield  {journal} {\bibinfo
  {journal} {Phys. Rev. B}\ }\textbf {\bibinfo {volume} {98}},\ \bibinfo
  {pages} {155314} (\bibinfo {year} {2018})}\BibitemShut {NoStop}%
\bibitem [{\citenamefont {Avila}\ \emph {et~al.}(2018)\citenamefont {Avila},
  \citenamefont {Pe\~{n}aranda}, \citenamefont {Prada}, \citenamefont
  {San-Jose},\ and\ \citenamefont {Aguado}}]{Avila:arxiv18}%
  \BibitemOpen
  \bibfield  {author} {\bibinfo {author} {\bibfnamefont {J.}~\bibnamefont
  {Avila}}, \bibinfo {author} {\bibfnamefont {F.}~\bibnamefont
  {Pe\~{n}aranda}}, \bibinfo {author} {\bibfnamefont {E.}~\bibnamefont
  {Prada}}, \bibinfo {author} {\bibfnamefont {P.}~\bibnamefont {San-Jose}}, \
  and\ \bibinfo {author} {\bibfnamefont {R.}~\bibnamefont {Aguado}},\ }\href
  {https://arxiv.org/abs/1807.04677} {\bibfield  {journal} {\bibinfo  {journal}
  {arXiv:1807.04677}\ } (\bibinfo {year} {2018})}\BibitemShut {NoStop}%
\bibitem [{\citenamefont {Vuik}\ \emph {et~al.}(2018)\citenamefont {Vuik},
  \citenamefont {Nijholt}, \citenamefont {Akhmerov},\ and\ \citenamefont
  {Wimmer}}]{Vuik:arxiv18}%
  \BibitemOpen
  \bibfield  {author} {\bibinfo {author} {\bibfnamefont {A.}~\bibnamefont
  {Vuik}}, \bibinfo {author} {\bibfnamefont {B.}~\bibnamefont {Nijholt}},
  \bibinfo {author} {\bibfnamefont {A.~R.}\ \bibnamefont {Akhmerov}}, \ and\
  \bibinfo {author} {\bibfnamefont {M.}~\bibnamefont {Wimmer}},\ }\href
  {https://arxiv.org/abs/1806.02801} {\bibfield  {journal} {\bibinfo  {journal}
  {arXiv:1806.02801}\ } (\bibinfo {year} {2018})}\BibitemShut {NoStop}%
\bibitem [{\citenamefont {Klinovaja}\ and\ \citenamefont
  {Loss}(2012)}]{Klinovaja:PRB12}%
  \BibitemOpen
  \bibfield  {author} {\bibinfo {author} {\bibfnamefont {J.}~\bibnamefont
  {Klinovaja}}\ and\ \bibinfo {author} {\bibfnamefont {D.}~\bibnamefont
  {Loss}},\ }\href {\doibase 10.1103/PhysRevB.86.085408} {\bibfield  {journal}
  {\bibinfo  {journal} {Phys. Rev. B}\ }\textbf {\bibinfo {volume} {86}},\
  \bibinfo {pages} {085408} (\bibinfo {year} {2012})}\BibitemShut {NoStop}%
\bibitem [{\citenamefont {Ben-Shach}\ \emph {et~al.}(2015)\citenamefont
  {Ben-Shach}, \citenamefont {Haim}, \citenamefont {Appelbaum}, \citenamefont
  {Oreg}, \citenamefont {Yacoby},\ and\ \citenamefont
  {Halperin}}]{Ben-Sach:PRB15}%
  \BibitemOpen
  \bibfield  {author} {\bibinfo {author} {\bibfnamefont {G.}~\bibnamefont
  {Ben-Shach}}, \bibinfo {author} {\bibfnamefont {A.}~\bibnamefont {Haim}},
  \bibinfo {author} {\bibfnamefont {I.}~\bibnamefont {Appelbaum}}, \bibinfo
  {author} {\bibfnamefont {Y.}~\bibnamefont {Oreg}}, \bibinfo {author}
  {\bibfnamefont {A.}~\bibnamefont {Yacoby}}, \ and\ \bibinfo {author}
  {\bibfnamefont {B.~I.}\ \bibnamefont {Halperin}},\ }\href {\doibase
  10.1103/PhysRevB.91.045403} {\bibfield  {journal} {\bibinfo  {journal} {Phys.
  Rev. B}\ }\textbf {\bibinfo {volume} {91}},\ \bibinfo {pages} {045403}
  (\bibinfo {year} {2015})}\BibitemShut {NoStop}%
\bibitem [{\citenamefont {Chang}\ \emph {et~al.}(2018)\citenamefont {Chang},
  \citenamefont {Albrecht}, \citenamefont {Jespersen}, \citenamefont
  {Kuemmeth}, \citenamefont {Krogstrup}, \citenamefont {Nyg\r{a}rd},\ and\
  \citenamefont {Marcus}}]{Chang:Nnano15}%
  \BibitemOpen
  \bibfield  {author} {\bibinfo {author} {\bibfnamefont {W.}~\bibnamefont
  {Chang}}, \bibinfo {author} {\bibfnamefont {S.~M.}\ \bibnamefont {Albrecht}},
  \bibinfo {author} {\bibfnamefont {T.~S.}\ \bibnamefont {Jespersen}}, \bibinfo
  {author} {\bibfnamefont {F.}~\bibnamefont {Kuemmeth}}, \bibinfo {author}
  {\bibfnamefont {P.}~\bibnamefont {Krogstrup}}, \bibinfo {author}
  {\bibfnamefont {J.}~\bibnamefont {Nyg\r{a}rd}}, \ and\ \bibinfo {author}
  {\bibfnamefont {C.~M.}\ \bibnamefont {Marcus}},\ }\href {\doibase
  10.1038/nnano.2014.306} {\bibfield  {journal} {\bibinfo  {journal} {Nature
  Nanotechnology}\ }\textbf {\bibinfo {volume} {10}},\ \bibinfo {pages} {232}
  (\bibinfo {year} {2018})}\BibitemShut {NoStop}%
\bibitem [{\citenamefont {G\"{u}l}\ \emph {et~al.}(2017)\citenamefont
  {G\"{u}l}, \citenamefont {Zhang}, \citenamefont {de~Vries}, \citenamefont
  {van Veen}, \citenamefont {Zuo}, \citenamefont {Mourik}, \citenamefont
  {Conesa-Boj}, \citenamefont {Nowak}, \citenamefont {van Woerkom},
  \citenamefont {Quintero-P\'{e}rez}, \citenamefont {Cassidy}, \citenamefont
  {Geresdi}, \citenamefont {Koelling}, \citenamefont {Car}, \citenamefont
  {Plissard}, \citenamefont {Bakkers},\ and\ \citenamefont
  {Kouwenhoven}}]{Gul:NNano17}%
  \BibitemOpen
  \bibfield  {author} {\bibinfo {author} {\bibfnamefont {O.}~\bibnamefont
  {G\"{u}l}}, \bibinfo {author} {\bibfnamefont {H.}~\bibnamefont {Zhang}},
  \bibinfo {author} {\bibfnamefont {F.~K.}\ \bibnamefont {de~Vries}}, \bibinfo
  {author} {\bibfnamefont {J.}~\bibnamefont {van Veen}}, \bibinfo {author}
  {\bibfnamefont {K.}~\bibnamefont {Zuo}}, \bibinfo {author} {\bibfnamefont
  {V.}~\bibnamefont {Mourik}}, \bibinfo {author} {\bibfnamefont
  {S.}~\bibnamefont {Conesa-Boj}}, \bibinfo {author} {\bibfnamefont {M.~P.}\
  \bibnamefont {Nowak}}, \bibinfo {author} {\bibfnamefont {D.~J.}\ \bibnamefont
  {van Woerkom}}, \bibinfo {author} {\bibfnamefont {M.}~\bibnamefont
  {Quintero-P\'{e}rez}}, \bibinfo {author} {\bibfnamefont {M.~C.}\ \bibnamefont
  {Cassidy}}, \bibinfo {author} {\bibfnamefont {A.}~\bibnamefont {Geresdi}},
  \bibinfo {author} {\bibfnamefont {S.}~\bibnamefont {Koelling}}, \bibinfo
  {author} {\bibfnamefont {D.}~\bibnamefont {Car}}, \bibinfo {author}
  {\bibfnamefont {S.~R.}\ \bibnamefont {Plissard}}, \bibinfo {author}
  {\bibfnamefont {E.~P. A.~M.}\ \bibnamefont {Bakkers}}, \ and\ \bibinfo
  {author} {\bibfnamefont {L.~P.}\ \bibnamefont {Kouwenhoven}},\ }\href
  {\doibase 10.1021/acs.nanolett.7b00540} {\bibfield  {journal} {\bibinfo
  {journal} {Nano Letters}\ }\textbf {\bibinfo {volume} {17}},\ \bibinfo
  {pages} {2690} (\bibinfo {year} {2017})}\BibitemShut {NoStop}%
\bibitem [{\citenamefont {Lutchyn}\ \emph {et~al.}(2010)\citenamefont
  {Lutchyn}, \citenamefont {Sau},\ and\ \citenamefont
  {Das~Sarma}}]{Lutchyn:PRL10}%
  \BibitemOpen
  \bibfield  {author} {\bibinfo {author} {\bibfnamefont {R.~M.}\ \bibnamefont
  {Lutchyn}}, \bibinfo {author} {\bibfnamefont {J.~D.}\ \bibnamefont {Sau}}, \
  and\ \bibinfo {author} {\bibfnamefont {S.}~\bibnamefont {Das~Sarma}},\ }\href
  {\doibase 10.1103/PhysRevLett.105.077001} {\bibfield  {journal} {\bibinfo
  {journal} {Phys. Rev. Lett.}\ }\textbf {\bibinfo {volume} {105}},\ \bibinfo
  {pages} {077001} (\bibinfo {year} {2010})}\BibitemShut {NoStop}%
\bibitem [{\citenamefont {Oreg}\ \emph {et~al.}(2010)\citenamefont {Oreg},
  \citenamefont {Refael},\ and\ \citenamefont {von Oppen}}]{Oreg:PRL10}%
  \BibitemOpen
  \bibfield  {author} {\bibinfo {author} {\bibfnamefont {Y.}~\bibnamefont
  {Oreg}}, \bibinfo {author} {\bibfnamefont {G.}~\bibnamefont {Refael}}, \ and\
  \bibinfo {author} {\bibfnamefont {F.}~\bibnamefont {von Oppen}},\ }\href
  {\doibase 10.1103/PhysRevLett.105.177002} {\bibfield  {journal} {\bibinfo
  {journal} {Phys. Rev. Lett.}\ }\textbf {\bibinfo {volume} {105}},\ \bibinfo
  {pages} {177002} (\bibinfo {year} {2010})}\BibitemShut {NoStop}%
\bibitem [{\citenamefont {Beidenkopf}()}]{Beidenkopf:private}%
  \BibitemOpen
  \bibfield  {author} {\bibinfo {author} {\bibfnamefont {H.}~\bibnamefont
  {Beidenkopf}},\ }\href@noop {} {}\bibinfo {note} {Private
  communication}\BibitemShut {NoStop}%
\bibitem [{\citenamefont {Levine}\ \emph {et~al.}(2017)\citenamefont {Levine},
  \citenamefont {Haim},\ and\ \citenamefont {Oreg}}]{Levine:PRB17}%
  \BibitemOpen
  \bibfield  {author} {\bibinfo {author} {\bibfnamefont {Y.}~\bibnamefont
  {Levine}}, \bibinfo {author} {\bibfnamefont {A.}~\bibnamefont {Haim}}, \ and\
  \bibinfo {author} {\bibfnamefont {Y.}~\bibnamefont {Oreg}},\ }\href {\doibase
  10.1103/PhysRevB.96.165147} {\bibfield  {journal} {\bibinfo  {journal} {Phys.
  Rev. B}\ }\textbf {\bibinfo {volume} {96}},\ \bibinfo {pages} {165147}
  (\bibinfo {year} {2017})}\BibitemShut {NoStop}%
\bibitem [{\citenamefont {Sau}\ \emph {et~al.}(2012)\citenamefont {Sau},
  \citenamefont {Lin}, \citenamefont {Hui},\ and\ \citenamefont
  {Das~Sarma}}]{Sau:PRL12}%
  \BibitemOpen
  \bibfield  {author} {\bibinfo {author} {\bibfnamefont {J.~D.}\ \bibnamefont
  {Sau}}, \bibinfo {author} {\bibfnamefont {C.~H.}\ \bibnamefont {Lin}},
  \bibinfo {author} {\bibfnamefont {H.-Y.}\ \bibnamefont {Hui}}, \ and\
  \bibinfo {author} {\bibfnamefont {S.}~\bibnamefont {Das~Sarma}},\ }\href
  {\doibase 10.1103/PhysRevLett.108.067001} {\bibfield  {journal} {\bibinfo
  {journal} {Phys. Rev. Lett.}\ }\textbf {\bibinfo {volume} {108}},\ \bibinfo
  {pages} {067001} (\bibinfo {year} {2012})}\BibitemShut {NoStop}%
\bibitem [{\citenamefont {Malard}\ \emph {et~al.}(2016)\citenamefont {Malard},
  \citenamefont {Japaridze},\ and\ \citenamefont {Johannesson}}]{Malard:PRB16}%
  \BibitemOpen
  \bibfield  {author} {\bibinfo {author} {\bibfnamefont {M.}~\bibnamefont
  {Malard}}, \bibinfo {author} {\bibfnamefont {G.~I.}\ \bibnamefont
  {Japaridze}}, \ and\ \bibinfo {author} {\bibfnamefont {H.}~\bibnamefont
  {Johannesson}},\ }\href {\doibase 10.1103/PhysRevB.94.115128} {\bibfield
  {journal} {\bibinfo  {journal} {Phys. Rev. B}\ }\textbf {\bibinfo {volume}
  {94}},\ \bibinfo {pages} {115128} (\bibinfo {year} {2016})}\BibitemShut
  {NoStop}%
\bibitem [{\citenamefont {Hoffman}\ \emph {et~al.}(2016)\citenamefont
  {Hoffman}, \citenamefont {Klinovaja},\ and\ \citenamefont
  {Loss}}]{Hoffman:PRB16}%
  \BibitemOpen
  \bibfield  {author} {\bibinfo {author} {\bibfnamefont {S.}~\bibnamefont
  {Hoffman}}, \bibinfo {author} {\bibfnamefont {J.}~\bibnamefont {Klinovaja}},
  \ and\ \bibinfo {author} {\bibfnamefont {D.}~\bibnamefont {Loss}},\ }\href
  {\doibase 10.1103/PhysRevB.93.165418} {\bibfield  {journal} {\bibinfo
  {journal} {Phys. Rev. B}\ }\textbf {\bibinfo {volume} {93}},\ \bibinfo
  {pages} {165418} (\bibinfo {year} {2016})}\BibitemShut {NoStop}%
\bibitem [{\citenamefont {Lu}\ \emph {et~al.}(2016)\citenamefont {Lu},
  \citenamefont {He}, \citenamefont {Xu}, \citenamefont {Lin},\ and\
  \citenamefont {Law}}]{Lu:PRB16}%
  \BibitemOpen
  \bibfield  {author} {\bibinfo {author} {\bibfnamefont {Y.}~\bibnamefont
  {Lu}}, \bibinfo {author} {\bibfnamefont {W.-Y.}\ \bibnamefont {He}}, \bibinfo
  {author} {\bibfnamefont {D.-H.}\ \bibnamefont {Xu}}, \bibinfo {author}
  {\bibfnamefont {N.}~\bibnamefont {Lin}}, \ and\ \bibinfo {author}
  {\bibfnamefont {K.~T.}\ \bibnamefont {Law}},\ }\href {\doibase
  10.1103/PhysRevB.94.024507} {\bibfield  {journal} {\bibinfo  {journal} {Phys.
  Rev. B}\ }\textbf {\bibinfo {volume} {94}},\ \bibinfo {pages} {024507}
  (\bibinfo {year} {2016})}\BibitemShut {NoStop}%
\bibitem [{\citenamefont {Prada}\ \emph {et~al.}(2012)\citenamefont {Prada},
  \citenamefont {San-Jose},\ and\ \citenamefont {Aguado}}]{Prada:PRB12}%
  \BibitemOpen
  \bibfield  {author} {\bibinfo {author} {\bibfnamefont {E.}~\bibnamefont
  {Prada}}, \bibinfo {author} {\bibfnamefont {P.}~\bibnamefont {San-Jose}}, \
  and\ \bibinfo {author} {\bibfnamefont {R.}~\bibnamefont {Aguado}},\ }\href
  {\doibase 10.1103/PhysRevB.86.180503} {\bibfield  {journal} {\bibinfo
  {journal} {Phys. Rev. B}\ }\textbf {\bibinfo {volume} {86}},\ \bibinfo
  {pages} {180503(R)} (\bibinfo {year} {2012})}\BibitemShut {NoStop}%
\bibitem [{\citenamefont {Kells}\ \emph {et~al.}(2012)\citenamefont {Kells},
  \citenamefont {Meidan},\ and\ \citenamefont {Brouwer}}]{Kells:PRB12}%
  \BibitemOpen
  \bibfield  {author} {\bibinfo {author} {\bibfnamefont {G.}~\bibnamefont
  {Kells}}, \bibinfo {author} {\bibfnamefont {D.}~\bibnamefont {Meidan}}, \
  and\ \bibinfo {author} {\bibfnamefont {P.~W.}\ \bibnamefont {Brouwer}},\
  }\href {\doibase 10.1103/PhysRevB.86.100503} {\bibfield  {journal} {\bibinfo
  {journal} {Phys. Rev. B}\ }\textbf {\bibinfo {volume} {86}},\ \bibinfo
  {pages} {100503(R)} (\bibinfo {year} {2012})}\BibitemShut {NoStop}%
\bibitem [{\citenamefont {Liu}\ \emph {et~al.}(2018)\citenamefont {Liu},
  \citenamefont {Sau},\ and\ \citenamefont {Das~Sarma}}]{Liu:PRB18}%
  \BibitemOpen
  \bibfield  {author} {\bibinfo {author} {\bibfnamefont {C.-X.}\ \bibnamefont
  {Liu}}, \bibinfo {author} {\bibfnamefont {J.~D.}\ \bibnamefont {Sau}}, \ and\
  \bibinfo {author} {\bibfnamefont {S.}~\bibnamefont {Das~Sarma}},\ }\href
  {\doibase 10.1103/PhysRevB.97.214502} {\bibfield  {journal} {\bibinfo
  {journal} {Phys. Rev. B}\ }\textbf {\bibinfo {volume} {97}},\ \bibinfo
  {pages} {214502} (\bibinfo {year} {2018})}\BibitemShut {NoStop}%
\bibitem [{\citenamefont {Dom{\'i}nguez}\ \emph {et~al.}(2017)\citenamefont
  {Dom{\'i}nguez}, \citenamefont {Cayao}, \citenamefont {San-Jose},
  \citenamefont {Aguado}, \citenamefont {Yeyati},\ and\ \citenamefont
  {Prada}}]{Dominguez:NPJ17}%
  \BibitemOpen
  \bibfield  {author} {\bibinfo {author} {\bibfnamefont {F.}~\bibnamefont
  {Dom{\'i}nguez}}, \bibinfo {author} {\bibfnamefont {J.}~\bibnamefont
  {Cayao}}, \bibinfo {author} {\bibfnamefont {P.}~\bibnamefont {San-Jose}},
  \bibinfo {author} {\bibfnamefont {R.}~\bibnamefont {Aguado}}, \bibinfo
  {author} {\bibfnamefont {A.~L.}\ \bibnamefont {Yeyati}}, \ and\ \bibinfo
  {author} {\bibfnamefont {E.}~\bibnamefont {Prada}},\ }\href
  {https://doi.org/10.1038/s41535-017-0012-0} {\bibfield  {journal} {\bibinfo
  {journal} {npj Quantum Materials}\ }\textbf {\bibinfo {volume} {2}},\
  \bibinfo {pages} {13} (\bibinfo {year} {2017})}\BibitemShut {NoStop}%
\bibitem [{\citenamefont {Escribano}\ \emph {et~al.}(2018)\citenamefont
  {Escribano}, \citenamefont {Yeyati},\ and\ \citenamefont
  {Prada}}]{Escribano:BJN18}%
  \BibitemOpen
  \bibfield  {author} {\bibinfo {author} {\bibfnamefont {S.~D.}\ \bibnamefont
  {Escribano}}, \bibinfo {author} {\bibfnamefont {A.~L.}\ \bibnamefont
  {Yeyati}}, \ and\ \bibinfo {author} {\bibfnamefont {E.}~\bibnamefont
  {Prada}},\ }\href {\doibase doi:10.3762/bjnano.9.203} {\bibfield  {journal}
  {\bibinfo  {journal} {Beilstein Journal of Nanotechnology}\ }\textbf
  {\bibinfo {volume} {9}},\ \bibinfo {pages} {2171} (\bibinfo {year}
  {2018})}\BibitemShut {NoStop}%
\bibitem [{\citenamefont {Pe\~naranda}\ \emph {et~al.}(2018)\citenamefont
  {Pe\~naranda}, \citenamefont {Aguado}, \citenamefont {San-Jose},\ and\
  \citenamefont {Prada}}]{Penaranda:PRB18}%
  \BibitemOpen
  \bibfield  {author} {\bibinfo {author} {\bibfnamefont {F.}~\bibnamefont
  {Pe\~naranda}}, \bibinfo {author} {\bibfnamefont {R.}~\bibnamefont {Aguado}},
  \bibinfo {author} {\bibfnamefont {P.}~\bibnamefont {San-Jose}}, \ and\
  \bibinfo {author} {\bibfnamefont {E.}~\bibnamefont {Prada}},\ }\href
  {\doibase 10.1103/PhysRevB.98.235406} {\bibfield  {journal} {\bibinfo
  {journal} {Phys. Rev. B}\ }\textbf {\bibinfo {volume} {98}},\ \bibinfo
  {pages} {235406} (\bibinfo {year} {2018})}\BibitemShut {NoStop}%
\bibitem [{\citenamefont {Fleckenstein}\ \emph {et~al.}(2018)\citenamefont
  {Fleckenstein}, \citenamefont {Dom\'{\i}nguez}, \citenamefont
  {Traverso~Ziani},\ and\ \citenamefont {Trauzettel}}]{Fleckenstein:PRB18}%
  \BibitemOpen
  \bibfield  {author} {\bibinfo {author} {\bibfnamefont {C.}~\bibnamefont
  {Fleckenstein}}, \bibinfo {author} {\bibfnamefont {F.}~\bibnamefont
  {Dom\'{\i}nguez}}, \bibinfo {author} {\bibfnamefont {N.}~\bibnamefont
  {Traverso~Ziani}}, \ and\ \bibinfo {author} {\bibfnamefont {B.}~\bibnamefont
  {Trauzettel}},\ }\href {\doibase 10.1103/PhysRevB.97.155425} {\bibfield
  {journal} {\bibinfo  {journal} {Phys. Rev. B}\ }\textbf {\bibinfo {volume}
  {97}},\ \bibinfo {pages} {155425} (\bibinfo {year} {2018})}\BibitemShut
  {NoStop}%
\bibitem [{\citenamefont {W\'ojcik}\ \emph {et~al.}(2018)\citenamefont
  {W\'ojcik}, \citenamefont {Bertoni},\ and\ \citenamefont
  {Goldoni}}]{Wojcik:PRB18}%
  \BibitemOpen
  \bibfield  {author} {\bibinfo {author} {\bibfnamefont {P.}~\bibnamefont
  {W\'ojcik}}, \bibinfo {author} {\bibfnamefont {A.}~\bibnamefont {Bertoni}}, \
  and\ \bibinfo {author} {\bibfnamefont {G.}~\bibnamefont {Goldoni}},\ }\href
  {\doibase 10.1103/PhysRevB.97.165401} {\bibfield  {journal} {\bibinfo
  {journal} {Phys. Rev. B}\ }\textbf {\bibinfo {volume} {97}},\ \bibinfo
  {pages} {165401} (\bibinfo {year} {2018})}\BibitemShut {NoStop}%
\bibitem [{\citenamefont {de~Moor}\ \emph {et~al.}(2018)\citenamefont
  {de~Moor}, \citenamefont {Bommer}, \citenamefont {Xu}, \citenamefont
  {Winkler}, \citenamefont {Antipov}, \citenamefont {Bargerbos}, \citenamefont
  {Wang}, \citenamefont {van Loo}, \citenamefont {Veld}, \citenamefont
  {Gazibegovic}, \citenamefont {Car}, \citenamefont {Logan}, \citenamefont
  {Pendharkar}, \citenamefont {Lee}, \citenamefont {Bakkers}, \citenamefont
  {Palmstr{\o}m}, \citenamefont {Lutchyn}, \citenamefont {Kouwenhoven},\ and\
  \citenamefont {Zhang}}]{Moor:NJP18}%
  \BibitemOpen
  \bibfield  {author} {\bibinfo {author} {\bibfnamefont {M.~W.~A.}\
  \bibnamefont {de~Moor}}, \bibinfo {author} {\bibfnamefont {J.~D.~S.}\
  \bibnamefont {Bommer}}, \bibinfo {author} {\bibfnamefont {D.}~\bibnamefont
  {Xu}}, \bibinfo {author} {\bibfnamefont {G.~W.}\ \bibnamefont {Winkler}},
  \bibinfo {author} {\bibfnamefont {A.~E.}\ \bibnamefont {Antipov}}, \bibinfo
  {author} {\bibfnamefont {A.}~\bibnamefont {Bargerbos}}, \bibinfo {author}
  {\bibfnamefont {G.}~\bibnamefont {Wang}}, \bibinfo {author} {\bibfnamefont
  {N.}~\bibnamefont {van Loo}}, \bibinfo {author} {\bibfnamefont {R.~L. M.
  O.~h.}\ \bibnamefont {Veld}}, \bibinfo {author} {\bibfnamefont
  {S.}~\bibnamefont {Gazibegovic}}, \bibinfo {author} {\bibfnamefont
  {D.}~\bibnamefont {Car}}, \bibinfo {author} {\bibfnamefont {J.~A.}\
  \bibnamefont {Logan}}, \bibinfo {author} {\bibfnamefont {M.}~\bibnamefont
  {Pendharkar}}, \bibinfo {author} {\bibfnamefont {J.~S.}\ \bibnamefont {Lee}},
  \bibinfo {author} {\bibfnamefont {E.~P. A.~M.}\ \bibnamefont {Bakkers}},
  \bibinfo {author} {\bibfnamefont {C.~J.}\ \bibnamefont {Palmstr{\o}m}},
  \bibinfo {author} {\bibfnamefont {R.~M.}\ \bibnamefont {Lutchyn}}, \bibinfo
  {author} {\bibfnamefont {L.~P.}\ \bibnamefont {Kouwenhoven}}, \ and\ \bibinfo
  {author} {\bibfnamefont {H.}~\bibnamefont {Zhang}},\ }\href
  {http://stacks.iop.org/1367-2630/20/i=10/a=103049} {\bibfield  {journal}
  {\bibinfo  {journal} {New Journal of Physics}\ }\textbf {\bibinfo {volume}
  {20}},\ \bibinfo {pages} {103049} (\bibinfo {year} {2018})}\BibitemShut
  {NoStop}%
\bibitem [{\citenamefont {Bommer}\ \emph {et~al.}(2019)\citenamefont {Bommer},
  \citenamefont {Zhang}, \citenamefont {G\"ul}, \citenamefont {Nijholt},
  \citenamefont {Wimmer}, \citenamefont {Rybakov}, \citenamefont {Garaud},
  \citenamefont {Rodic}, \citenamefont {Babaev}, \citenamefont {Troyer},
  \citenamefont {Car}, \citenamefont {Plissard}, \citenamefont {Bakkers},
  \citenamefont {Watanabe}, \citenamefont {Taniguchi},\ and\ \citenamefont
  {Kouwenhoven}}]{Bommer:arxiv18}%
  \BibitemOpen
  \bibfield  {author} {\bibinfo {author} {\bibfnamefont {J.~D.~S.}\
  \bibnamefont {Bommer}}, \bibinfo {author} {\bibfnamefont {H.}~\bibnamefont
  {Zhang}}, \bibinfo {author} {\bibfnamefont {O.}~\bibnamefont {G\"ul}},
  \bibinfo {author} {\bibfnamefont {B.}~\bibnamefont {Nijholt}}, \bibinfo
  {author} {\bibfnamefont {M.}~\bibnamefont {Wimmer}}, \bibinfo {author}
  {\bibfnamefont {F.~N.}\ \bibnamefont {Rybakov}}, \bibinfo {author}
  {\bibfnamefont {J.}~\bibnamefont {Garaud}}, \bibinfo {author} {\bibfnamefont
  {D.}~\bibnamefont {Rodic}}, \bibinfo {author} {\bibfnamefont
  {E.}~\bibnamefont {Babaev}}, \bibinfo {author} {\bibfnamefont
  {M.}~\bibnamefont {Troyer}}, \bibinfo {author} {\bibfnamefont
  {D.}~\bibnamefont {Car}}, \bibinfo {author} {\bibfnamefont {S.~R.}\
  \bibnamefont {Plissard}}, \bibinfo {author} {\bibfnamefont {E.~P. A.~M.}\
  \bibnamefont {Bakkers}}, \bibinfo {author} {\bibfnamefont {K.}~\bibnamefont
  {Watanabe}}, \bibinfo {author} {\bibfnamefont {T.}~\bibnamefont {Taniguchi}},
  \ and\ \bibinfo {author} {\bibfnamefont {L.~P.}\ \bibnamefont
  {Kouwenhoven}},\ }\href {\doibase 10.1103/PhysRevLett.122.187702} {\bibfield
  {journal} {\bibinfo  {journal} {Phys. Rev. Lett.}\ }\textbf {\bibinfo
  {volume} {122}},\ \bibinfo {pages} {187702} (\bibinfo {year}
  {2019})}\BibitemShut {NoStop}%
\bibitem [{\citenamefont {Vuik}\ \emph {et~al.}(2016)\citenamefont {Vuik},
  \citenamefont {Eeltink}, \citenamefont {Akhmerov},\ and\ \citenamefont
  {Wimmer}}]{Vuik:NJP16}%
  \BibitemOpen
  \bibfield  {author} {\bibinfo {author} {\bibfnamefont {A.}~\bibnamefont
  {Vuik}}, \bibinfo {author} {\bibfnamefont {D.}~\bibnamefont {Eeltink}},
  \bibinfo {author} {\bibfnamefont {A.~R.}\ \bibnamefont {Akhmerov}}, \ and\
  \bibinfo {author} {\bibfnamefont {M.}~\bibnamefont {Wimmer}},\ }\href
  {https://doi.org/10.1088/1367-2630/18/3/033013} {\bibfield  {journal}
  {\bibinfo  {journal} {New J. Phys.}\ }\textbf {\bibinfo {volume} {18}},\
  \bibinfo {pages} {033013} (\bibinfo {year} {2016})}\BibitemShut {NoStop}%
\bibitem [{\citenamefont {Antipov}\ \emph {et~al.}(2018)\citenamefont
  {Antipov}, \citenamefont {Bargerbos}, \citenamefont {Winkler}, \citenamefont
  {Bauer}, \citenamefont {Rossi},\ and\ \citenamefont
  {Lutchyn}}]{Antipov:PRX18}%
  \BibitemOpen
  \bibfield  {author} {\bibinfo {author} {\bibfnamefont {A.~E.}\ \bibnamefont
  {Antipov}}, \bibinfo {author} {\bibfnamefont {A.}~\bibnamefont {Bargerbos}},
  \bibinfo {author} {\bibfnamefont {G.~W.}\ \bibnamefont {Winkler}}, \bibinfo
  {author} {\bibfnamefont {B.}~\bibnamefont {Bauer}}, \bibinfo {author}
  {\bibfnamefont {E.}~\bibnamefont {Rossi}}, \ and\ \bibinfo {author}
  {\bibfnamefont {R.~M.}\ \bibnamefont {Lutchyn}},\ }\href {\doibase
  10.1103/PhysRevX.8.031041} {\bibfield  {journal} {\bibinfo  {journal} {Phys.
  Rev. X}\ }\textbf {\bibinfo {volume} {8}},\ \bibinfo {pages} {031041}
  (\bibinfo {year} {2018})}\BibitemShut {NoStop}%
\bibitem [{\citenamefont {Mikkelsen}\ \emph {et~al.}(2018)\citenamefont
  {Mikkelsen}, \citenamefont {Kotetes}, \citenamefont {Krogstrup},\ and\
  \citenamefont {Flensberg}}]{Mikkelsen:PRX18}%
  \BibitemOpen
  \bibfield  {author} {\bibinfo {author} {\bibfnamefont {A.~E.~G.}\
  \bibnamefont {Mikkelsen}}, \bibinfo {author} {\bibfnamefont {P.}~\bibnamefont
  {Kotetes}}, \bibinfo {author} {\bibfnamefont {P.}~\bibnamefont {Krogstrup}},
  \ and\ \bibinfo {author} {\bibfnamefont {K.}~\bibnamefont {Flensberg}},\
  }\href {\doibase 10.1103/PhysRevX.8.031040} {\bibfield  {journal} {\bibinfo
  {journal} {Phys. Rev. X}\ }\textbf {\bibinfo {volume} {8}},\ \bibinfo {pages}
  {031040} (\bibinfo {year} {2018})}\BibitemShut {NoStop}%
\bibitem [{\citenamefont {Nijholt}\ and\ \citenamefont
  {Akhmerov}(2016)}]{Nijholt:PRB16}%
  \BibitemOpen
  \bibfield  {author} {\bibinfo {author} {\bibfnamefont {B.}~\bibnamefont
  {Nijholt}}\ and\ \bibinfo {author} {\bibfnamefont {A.~R.}\ \bibnamefont
  {Akhmerov}},\ }\href {\doibase 10.1103/PhysRevB.93.235434} {\bibfield
  {journal} {\bibinfo  {journal} {Phys. Rev. B}\ }\textbf {\bibinfo {volume}
  {93}},\ \bibinfo {pages} {235434} (\bibinfo {year} {2016})}\BibitemShut
  {NoStop}%
\bibitem [{\citenamefont {Kiczek}\ and\ \citenamefont
  {Ptok}(2017)}]{Kiczek:JoP17}%
  \BibitemOpen
  \bibfield  {author} {\bibinfo {author} {\bibfnamefont {B.}~\bibnamefont
  {Kiczek}}\ and\ \bibinfo {author} {\bibfnamefont {A.}~\bibnamefont {Ptok}},\
  }\href {\doibase 10.1088/1361-648x/aa93ab} {\bibfield  {journal} {\bibinfo
  {journal} {J. Phys.: Condens. Matter}\ }\textbf {\bibinfo {volume} {29}},\
  \bibinfo {pages} {495301} (\bibinfo {year} {2017})}\BibitemShut {NoStop}%
\bibitem [{\citenamefont {Winkler}\ \emph {et~al.}(2019)\citenamefont
  {Winkler}, \citenamefont {Antipov}, \citenamefont {van Heck}, \citenamefont
  {Soluyanov}, \citenamefont {Glazman}, \citenamefont {Wimmer},\ and\
  \citenamefont {Lutchyn}}]{Winkler:arxiv18}%
  \BibitemOpen
  \bibfield  {author} {\bibinfo {author} {\bibfnamefont {G.~W.}\ \bibnamefont
  {Winkler}}, \bibinfo {author} {\bibfnamefont {A.~E.}\ \bibnamefont
  {Antipov}}, \bibinfo {author} {\bibfnamefont {B.}~\bibnamefont {van Heck}},
  \bibinfo {author} {\bibfnamefont {A.~A.}\ \bibnamefont {Soluyanov}}, \bibinfo
  {author} {\bibfnamefont {L.~I.}\ \bibnamefont {Glazman}}, \bibinfo {author}
  {\bibfnamefont {M.}~\bibnamefont {Wimmer}}, \ and\ \bibinfo {author}
  {\bibfnamefont {R.~M.}\ \bibnamefont {Lutchyn}},\ }\href {\doibase
  10.1103/PhysRevB.99.245408} {\bibfield  {journal} {\bibinfo  {journal} {Phys.
  Rev. B}\ }\textbf {\bibinfo {volume} {99}},\ \bibinfo {pages} {245408}
  (\bibinfo {year} {2019})}\BibitemShut {NoStop}%
\bibitem [{\citenamefont {Huang}\ \emph {et~al.}(2018)\citenamefont {Huang},
  \citenamefont {Sau}, \citenamefont {Stanescu},\ and\ \citenamefont
  {Das~Sarma}}]{Huang:PRB18}%
  \BibitemOpen
  \bibfield  {author} {\bibinfo {author} {\bibfnamefont {Y.}~\bibnamefont
  {Huang}}, \bibinfo {author} {\bibfnamefont {J.~D.}\ \bibnamefont {Sau}},
  \bibinfo {author} {\bibfnamefont {T.~D.}\ \bibnamefont {Stanescu}}, \ and\
  \bibinfo {author} {\bibfnamefont {S.}~\bibnamefont {Das~Sarma}},\ }\href
  {\doibase 10.1103/PhysRevB.98.224512} {\bibfield  {journal} {\bibinfo
  {journal} {Phys. Rev. B}\ }\textbf {\bibinfo {volume} {98}},\ \bibinfo
  {pages} {224512} (\bibinfo {year} {2018})}\BibitemShut {NoStop}%
\bibitem [{\citenamefont {Sau}\ and\ \citenamefont
  {Das~Sarma}(2012)}]{Sau:NatCom12}%
  \BibitemOpen
  \bibfield  {author} {\bibinfo {author} {\bibfnamefont {J.~D.}\ \bibnamefont
  {Sau}}\ and\ \bibinfo {author} {\bibfnamefont {S.}~\bibnamefont
  {Das~Sarma}},\ }\href {\doibase 10.1038/ncomms1966} {\bibfield  {journal}
  {\bibinfo  {journal} {Nature Communications}\ }\textbf {\bibinfo {volume}
  {3}},\ \bibinfo {pages} {964} (\bibinfo {year} {2012})}\BibitemShut {NoStop}%
\bibitem [{\citenamefont {Fulga}\ \emph {et~al.}(2013)\citenamefont {Fulga},
  \citenamefont {Haim}, \citenamefont {Akhmerov},\ and\ \citenamefont
  {Oreg}}]{Fulga:NJP13}%
  \BibitemOpen
  \bibfield  {author} {\bibinfo {author} {\bibfnamefont {I.~C.}\ \bibnamefont
  {Fulga}}, \bibinfo {author} {\bibfnamefont {A.}~\bibnamefont {Haim}},
  \bibinfo {author} {\bibfnamefont {A.~R.}\ \bibnamefont {Akhmerov}}, \ and\
  \bibinfo {author} {\bibfnamefont {Y.}~\bibnamefont {Oreg}},\ }\href
  {http://iopscience.iop.org/article/10.1088/1367-2630/15/4/045020/meta}
  {\bibfield  {journal} {\bibinfo  {journal} {New Journal of Physics}\ }\textbf
  {\bibinfo {volume} {15}},\ \bibinfo {pages} {045020} (\bibinfo {year}
  {2013})}\BibitemShut {NoStop}%
\bibitem [{\citenamefont {Stenger}\ \emph {et~al.}(2018)\citenamefont
  {Stenger}, \citenamefont {Woods}, \citenamefont {Frolov},\ and\ \citenamefont
  {Stanescu}}]{Stenger:PRB18}%
  \BibitemOpen
  \bibfield  {author} {\bibinfo {author} {\bibfnamefont {J.~P.~T.}\
  \bibnamefont {Stenger}}, \bibinfo {author} {\bibfnamefont {B.~D.}\
  \bibnamefont {Woods}}, \bibinfo {author} {\bibfnamefont {S.~M.}\ \bibnamefont
  {Frolov}}, \ and\ \bibinfo {author} {\bibfnamefont {T.~D.}\ \bibnamefont
  {Stanescu}},\ }\href {\doibase 10.1103/PhysRevB.98.085407} {\bibfield
  {journal} {\bibinfo  {journal} {Phys. Rev. B}\ }\textbf {\bibinfo {volume}
  {98}},\ \bibinfo {pages} {085407} (\bibinfo {year} {2018})}\BibitemShut
  {NoStop}%
\bibitem [{\citenamefont {Woods}\ \emph {et~al.}(2018)\citenamefont {Woods},
  \citenamefont {Stanescu},\ and\ \citenamefont {Das~Sarma}}]{Woods:PRB18}%
  \BibitemOpen
  \bibfield  {author} {\bibinfo {author} {\bibfnamefont {B.~D.}\ \bibnamefont
  {Woods}}, \bibinfo {author} {\bibfnamefont {T.~D.}\ \bibnamefont {Stanescu}},
  \ and\ \bibinfo {author} {\bibfnamefont {S.}~\bibnamefont {Das~Sarma}},\
  }\href {\doibase 10.1103/PhysRevB.98.035428} {\bibfield  {journal} {\bibinfo
  {journal} {Phys. Rev. B}\ }\textbf {\bibinfo {volume} {98}},\ \bibinfo
  {pages} {035428} (\bibinfo {year} {2018})}\BibitemShut {NoStop}%
\bibitem [{\citenamefont {Reeg}\ \emph {et~al.}(2017)\citenamefont {Reeg},
  \citenamefont {Loss},\ and\ \citenamefont {Klinovaja}}]{Reeg:PRB17}%
  \BibitemOpen
  \bibfield  {author} {\bibinfo {author} {\bibfnamefont {C.}~\bibnamefont
  {Reeg}}, \bibinfo {author} {\bibfnamefont {D.}~\bibnamefont {Loss}}, \ and\
  \bibinfo {author} {\bibfnamefont {J.}~\bibnamefont {Klinovaja}},\ }\href
  {\doibase 10.1103/PhysRevB.96.125426} {\bibfield  {journal} {\bibinfo
  {journal} {Phys. Rev. B}\ }\textbf {\bibinfo {volume} {96}},\ \bibinfo
  {pages} {125426} (\bibinfo {year} {2017})}\BibitemShut {NoStop}%
\bibitem [{\citenamefont {Reeg}\ \emph
  {et~al.}(2018{\natexlab{b}})\citenamefont {Reeg}, \citenamefont {Loss},\ and\
  \citenamefont {Klinovaja}}]{Reeg:PRB18}%
  \BibitemOpen
  \bibfield  {author} {\bibinfo {author} {\bibfnamefont {C.}~\bibnamefont
  {Reeg}}, \bibinfo {author} {\bibfnamefont {D.}~\bibnamefont {Loss}}, \ and\
  \bibinfo {author} {\bibfnamefont {J.}~\bibnamefont {Klinovaja}},\ }\href
  {\doibase 10.1103/PhysRevB.97.165425} {\bibfield  {journal} {\bibinfo
  {journal} {Phys. Rev. B}\ }\textbf {\bibinfo {volume} {97}},\ \bibinfo
  {pages} {165425} (\bibinfo {year} {2018}{\natexlab{b}})}\BibitemShut
  {NoStop}%
\bibitem [{\citenamefont {Nyg\r{a}rd}()}]{Nygard:private}%
  \BibitemOpen
  \bibfield  {author} {\bibinfo {author} {\bibfnamefont {J.}~\bibnamefont
  {Nyg\r{a}rd}},\ }\href@noop {} {}\bibinfo {note} {Private
  communication}\BibitemShut {NoStop}%
\bibitem [{\citenamefont {Olsson}\ \emph {et~al.}(1996)\citenamefont {Olsson},
  \citenamefont {Andersson}, \citenamefont {H\aa{}kansson}, \citenamefont
  {Kanski}, \citenamefont {Ilver},\ and\ \citenamefont
  {Karlsson}}]{Olsson:PRL96}%
  \BibitemOpen
  \bibfield  {author} {\bibinfo {author} {\bibfnamefont {L.~O.}\ \bibnamefont
  {Olsson}}, \bibinfo {author} {\bibfnamefont {C.~B.~M.}\ \bibnamefont
  {Andersson}}, \bibinfo {author} {\bibfnamefont {M.~C.}\ \bibnamefont
  {H\aa{}kansson}}, \bibinfo {author} {\bibfnamefont {J.}~\bibnamefont
  {Kanski}}, \bibinfo {author} {\bibfnamefont {L.}~\bibnamefont {Ilver}}, \
  and\ \bibinfo {author} {\bibfnamefont {U.~O.}\ \bibnamefont {Karlsson}},\
  }\href {\doibase 10.1103/PhysRevLett.76.3626} {\bibfield  {journal} {\bibinfo
   {journal} {Phys. Rev. Lett.}\ }\textbf {\bibinfo {volume} {76}},\ \bibinfo
  {pages} {3626} (\bibinfo {year} {1996})}\BibitemShut {NoStop}%
\bibitem [{\citenamefont {Thelander}\ \emph {et~al.}(2010)\citenamefont
  {Thelander}, \citenamefont {Dick}, \citenamefont {Borgstr\"{o}m},
  \citenamefont {Fr\"{o}berg}, \citenamefont {Caroff}, \citenamefont
  {Nilsson},\ and\ \citenamefont {Samuelson}}]{Thelander:Nano10}%
  \BibitemOpen
  \bibfield  {author} {\bibinfo {author} {\bibfnamefont {C.}~\bibnamefont
  {Thelander}}, \bibinfo {author} {\bibfnamefont {K.~A.}\ \bibnamefont {Dick}},
  \bibinfo {author} {\bibfnamefont {M.~T.}\ \bibnamefont {Borgstr\"{o}m}},
  \bibinfo {author} {\bibfnamefont {L.~E.}\ \bibnamefont {Fr\"{o}berg}},
  \bibinfo {author} {\bibfnamefont {P.}~\bibnamefont {Caroff}}, \bibinfo
  {author} {\bibfnamefont {H.~A.}\ \bibnamefont {Nilsson}}, \ and\ \bibinfo
  {author} {\bibfnamefont {L.}~\bibnamefont {Samuelson}},\ }\href {\doibase
  10.1088/0957-4484/21/20/205703} {\bibfield  {journal} {\bibinfo  {journal}
  {Nanotechnology}\ }\textbf {\bibinfo {volume} {21}},\ \bibinfo {pages}
  {205703} (\bibinfo {year} {2010})}\BibitemShut {NoStop}%
\bibitem [{\citenamefont {Winkler}\ \emph {et~al.}(2003)\citenamefont
  {Winkler}, \citenamefont {Papadakis}, \citenamefont {De~Poortere},\ and\
  \citenamefont {Shayegan}}]{Winkler:03}%
  \BibitemOpen
  \bibfield  {author} {\bibinfo {author} {\bibfnamefont {R.}~\bibnamefont
  {Winkler}}, \bibinfo {author} {\bibfnamefont {S.}~\bibnamefont {Papadakis}},
  \bibinfo {author} {\bibfnamefont {E.}~\bibnamefont {De~Poortere}}, \ and\
  \bibinfo {author} {\bibfnamefont {M.}~\bibnamefont {Shayegan}},\ }\href@noop
  {} {\emph {\bibinfo {title} {Spin-Orbit Coupling in Two-Dimensional Electron
  and Hole Systems}}},\ Vol.~\bibinfo {volume} {41}\ (\bibinfo  {publisher}
  {Springer},\ \bibinfo {year} {2003})\BibitemShut {NoStop}%
\bibitem [{\citenamefont {Lew Yan~Voon}\ \emph {et~al.}(1996)\citenamefont {Lew
  Yan~Voon}, \citenamefont {Willatzen}, \citenamefont {Cardona},\ and\
  \citenamefont {Christensen}}]{Voon:PRB96}%
  \BibitemOpen
  \bibfield  {author} {\bibinfo {author} {\bibfnamefont {L.~C.}\ \bibnamefont
  {Lew Yan~Voon}}, \bibinfo {author} {\bibfnamefont {M.}~\bibnamefont
  {Willatzen}}, \bibinfo {author} {\bibfnamefont {M.}~\bibnamefont {Cardona}},
  \ and\ \bibinfo {author} {\bibfnamefont {N.~E.}\ \bibnamefont
  {Christensen}},\ }\href {\doibase 10.1103/PhysRevB.53.10703} {\bibfield
  {journal} {\bibinfo  {journal} {Phys. Rev. B}\ }\textbf {\bibinfo {volume}
  {53}},\ \bibinfo {pages} {10703} (\bibinfo {year} {1996})}\BibitemShut
  {NoStop}%
\bibitem [{\citenamefont {Gmitra}\ and\ \citenamefont
  {Fabian}(2016)}]{Gmitra:PRB16}%
  \BibitemOpen
  \bibfield  {author} {\bibinfo {author} {\bibfnamefont {M.}~\bibnamefont
  {Gmitra}}\ and\ \bibinfo {author} {\bibfnamefont {J.}~\bibnamefont
  {Fabian}},\ }\href {\doibase 10.1103/PhysRevB.94.165202} {\bibfield
  {journal} {\bibinfo  {journal} {Phys. Rev. B}\ }\textbf {\bibinfo {volume}
  {94}},\ \bibinfo {pages} {165202} (\bibinfo {year} {2016})}\BibitemShut
  {NoStop}%
\bibitem [{\citenamefont {Levinshtein}\ \emph {et~al.}(2000)\citenamefont
  {Levinshtein}, \citenamefont {Rumyantsev},\ and\ \citenamefont
  {Shur}}]{Levinshtein:00}%
  \BibitemOpen
  \bibfield  {author} {\bibinfo {author} {\bibfnamefont {M.}~\bibnamefont
  {Levinshtein}}, \bibinfo {author} {\bibfnamefont {S.}~\bibnamefont
  {Rumyantsev}}, \ and\ \bibinfo {author} {\bibfnamefont {M.}~\bibnamefont
  {Shur}},\ }\href {\doibase 10.1142/2046} {\emph {\bibinfo {title} {Handbook
  series on semiconductor parameters}}},\ Vol.~\bibinfo {volume} {1}\ (\bibinfo
   {publisher} {World Scientific Publishing},\ \bibinfo {year}
  {2000})\BibitemShut {NoStop}%
\bibitem [{\citenamefont {Perry}(2011)}]{Perry:11}%
  \BibitemOpen
  \bibfield  {author} {\bibinfo {author} {\bibfnamefont {D.~L.}\ \bibnamefont
  {Perry}},\ }\href@noop {} {\emph {\bibinfo {title} {Handbook of Inorganic
  Compounds}}},\ \bibinfo {edition} {2nd}\ ed.\ (\bibinfo  {publisher} {CRC
  Press},\ \bibinfo {year} {2011})\BibitemShut {NoStop}%
\bibitem [{\citenamefont {Lim}\ \emph {et~al.}(2012)\citenamefont {Lim},
  \citenamefont {Serra}, \citenamefont {L\'opez},\ and\ \citenamefont
  {Aguado}}]{Lim:PRB12}%
  \BibitemOpen
  \bibfield  {author} {\bibinfo {author} {\bibfnamefont {J.~S.}\ \bibnamefont
  {Lim}}, \bibinfo {author} {\bibfnamefont {L.}~\bibnamefont {Serra}}, \bibinfo
  {author} {\bibfnamefont {R.}~\bibnamefont {L\'opez}}, \ and\ \bibinfo
  {author} {\bibfnamefont {R.}~\bibnamefont {Aguado}},\ }\href {\doibase
  10.1103/PhysRevB.86.121103} {\bibfield  {journal} {\bibinfo  {journal} {Phys.
  Rev. B}\ }\textbf {\bibinfo {volume} {86}},\ \bibinfo {pages} {121103(R)}
  (\bibinfo {year} {2012})}\BibitemShut {NoStop}%
\bibitem [{\citenamefont {Stanescu}\ \emph {et~al.}(2011)\citenamefont
  {Stanescu}, \citenamefont {Lutchyn},\ and\ \citenamefont
  {Das~Sarma}}]{Stanescu:PRB11}%
  \BibitemOpen
  \bibfield  {author} {\bibinfo {author} {\bibfnamefont {T.~D.}\ \bibnamefont
  {Stanescu}}, \bibinfo {author} {\bibfnamefont {R.~M.}\ \bibnamefont
  {Lutchyn}}, \ and\ \bibinfo {author} {\bibfnamefont {S.}~\bibnamefont
  {Das~Sarma}},\ }\href {\doibase 10.1103/PhysRevB.84.144522} {\bibfield
  {journal} {\bibinfo  {journal} {Phys. Rev. B}\ }\textbf {\bibinfo {volume}
  {84}},\ \bibinfo {pages} {144522} (\bibinfo {year} {2011})}\BibitemShut
  {NoStop}%
\bibitem [{\citenamefont {Lutchyn}\ \emph {et~al.}(2011)\citenamefont
  {Lutchyn}, \citenamefont {Stanescu},\ and\ \citenamefont
  {Das~Sarma}}]{Lutchyn:PRL11}%
  \BibitemOpen
  \bibfield  {author} {\bibinfo {author} {\bibfnamefont {R.~M.}\ \bibnamefont
  {Lutchyn}}, \bibinfo {author} {\bibfnamefont {T.~D.}\ \bibnamefont
  {Stanescu}}, \ and\ \bibinfo {author} {\bibfnamefont {S.}~\bibnamefont
  {Das~Sarma}},\ }\href {\doibase 10.1103/PhysRevLett.106.127001} {\bibfield
  {journal} {\bibinfo  {journal} {Phys. Rev. Lett.}\ }\textbf {\bibinfo
  {volume} {106}},\ \bibinfo {pages} {127001} (\bibinfo {year}
  {2011})}\BibitemShut {NoStop}%
\bibitem [{\citenamefont {D.~Escribano}\ \emph {et~al.}(2019)\citenamefont
  {D.~Escribano}, \citenamefont {Levy~Yeyati}, \citenamefont {Oreg},\ and\
  \citenamefont {Prada}}]{Zenodo}%
  \BibitemOpen
  \bibfield  {author} {\bibinfo {author} {\bibfnamefont {S.}~\bibnamefont
  {D.~Escribano}}, \bibinfo {author} {\bibfnamefont {A.}~\bibnamefont
  {Levy~Yeyati}}, \bibinfo {author} {\bibfnamefont {Y.}~\bibnamefont {Oreg}}, \
  and\ \bibinfo {author} {\bibfnamefont {E.}~\bibnamefont {Prada}},\ }\href
  {\doibase 10.5281/zenodo.3250709} {\enquote {\bibinfo {title} {Effects of the
  electrostatic environment on superlattice majorana nanowires (dataset)},}\
  }\bibinfo {howpublished} {Zenodo repository
  \url{https://doi.org/10.5281/zenodo.3250709}} (\bibinfo {year}
  {2019})\BibitemShut {NoStop}%
\bibitem [{\citenamefont {Logg}\ and\ \citenamefont {Wells}(2010)}]{Logg:10}%
  \BibitemOpen
  \bibfield  {author} {\bibinfo {author} {\bibfnamefont {A.}~\bibnamefont
  {Logg}}\ and\ \bibinfo {author} {\bibfnamefont {G.~N.}\ \bibnamefont
  {Wells}},\ }\href {\doibase 10.1145/1731022.1731030} {\bibfield  {journal}
  {\bibinfo  {journal} {ACM Transactions on Mathematical Software}\ }\textbf
  {\bibinfo {volume} {37}},\ \bibinfo {pages} {1} (\bibinfo {year}
  {2010})}\BibitemShut {NoStop}%
\bibitem [{\citenamefont {Logg}\ \emph {et~al.}(2012)\citenamefont {Logg},
  \citenamefont {Mardal}, \citenamefont {Wells} \emph {et~al.}}]{Logg:12}%
  \BibitemOpen
  \bibfield  {author} {\bibinfo {author} {\bibfnamefont {A.}~\bibnamefont
  {Logg}}, \bibinfo {author} {\bibfnamefont {K.-A.}\ \bibnamefont {Mardal}},
  \bibinfo {author} {\bibfnamefont {G.~N.}\ \bibnamefont {Wells}},  \emph
  {et~al.},\ }\href {\doibase 10.1007/978-3-642-23099-8} {\emph {\bibinfo
  {title} {Automated Solution of Differential Equations by the Finite Element
  Method}}}\ (\bibinfo  {publisher} {Springer},\ \bibinfo {year}
  {2012})\BibitemShut {NoStop}%
\bibitem [{\citenamefont {Wimmer}(2012)}]{Wimmer:ACM12}%
  \BibitemOpen
  \bibfield  {author} {\bibinfo {author} {\bibfnamefont {M.}~\bibnamefont
  {Wimmer}},\ }\href {\doibase 10.1145/2331130.2331138} {\bibfield  {journal}
  {\bibinfo  {journal} {ACM Transactions on Mathematical Software (TOMS)}\
  }\textbf {\bibinfo {volume} {38}},\ \bibinfo {pages} {1} (\bibinfo {year}
  {2012})}\BibitemShut {NoStop}%
\end{thebibliography}%




\appendix

\section{Numerical details}
\label{SI1}
In this appendix we detail the numerical methods used to solve the Schr\"odinger-Poisson equation given by Eq. (\ref{Hamiltonian}) and Eq. (\ref{Poisson}) in the main text. As explained in Sec. \ref{Methods}, instead of solving the coupled equations, our general procedure consists of, first, computing self-consistently the electrostatic potential within the Thomas-Fermi approximation, and then, building and diagonalizing the Hamiltonian in order to obtain the eigenspectrum. The reliability of this procedure compared to a full Schr\"odinger-Poisson approach is discussed in App. \ref{SI2}.

\subsection{Electrostatic potential}
\label{SI1-1}
To obtain the electrostatic potential, we solve the Poisson equation (given by Eq. (\ref{Poisson}) in the main text) using a Partial Differential Equation solver for Python called \textit{FEniCS} \cite{Logg:10, Logg:12}, which uses finite element techniques. We use a mesh with Lagrange elements with a discretization of 1nm. Regarding the boundary conditions of the semiconducting nanowire, we impose $V_{\rm gate}$ at the back gate, $V_{\rm SC}$ at the boundaries with the SC fingers, $V_{\rm N}$ at the normal metal boundaries (if applicable), and periodic boundary conditions at the nanowire ends. This last condition eliminates border effects, which are well-known \cite{Escribano:BJN18, Penaranda:PRB18, Fleckenstein:PRB18} and do not change the qualitative physics introduced by the superlattice structure. The different geometries studied in this work (i.e. the bottom- and top-superlattices, the continuously covered nanowire and their combinations) are taken into account through an inhomogeneous dielectric permittivity $\epsilon(\vec{r})$. We model it as a piecewise function: constant inside each material and with abrupt changes at the interfaces. The specific values used in our simulations for the dielectric permittivity can be found in Table~\ref{Table_parameters} in the main text.

The source term $\rho_{\rm tot}=\rho_{\rm surf}+\rho_{\rm mobile}$ of the Poisson equation (shown in Eq. (\ref{charge_density}) in the main text) has two independent terms. The first one is the surface charge layer, that we model as a fix superficial positive charge density $\rho_{\rm surf}$ placed in the points of the mesh localized at the InAs-vacuum and InAs-SiO interfaces. The second source term is the 3D mobile charge density inside the wire, $\rho_{\rm mobile}=\rho_{\rm e}+\rho_{\rm lh}+\rho_{\rm hh}$, which in principle includes the contributions of the conduction band $\rho_{\rm e}$, and the light hole $\rho_{\rm lh}$ and  heavy hole $\rho_{\rm hh}$ bands. However, in this work we ignore the hole terms since they are not present for the gate potentials that we consider in our simulations. Specifically, they are relevant when $e\phi(x,y,z)\lesssim E_{\rm vv}$, that, for the specific geometries of this work, only occurs when $V_{\rm gate}<-3.5$V in the bottom-superlattice, $V_{\rm gate}<-0.8$V in the top one, $V_{\rm gate}<-1.8$V in the homogeneous nanowire, and  $V_{\rm gate}<-15.7$V in the alternative configuration that combines a bottom (normal) superlattice and a continuous SC layer. Therefore, we only compute the electron charge density corresponding to the wire's conduction band using to this end the Thomas-Fermi approximation for a 3D electron gas, as explained in Sec. \ref{Methods} of the main text. As the charge density depends in turn on the potential, the Poisson equation must be solved self-consistently. For this purpose, we use an iterative method to obtain the charge density using the Anderson mixing
\begin{equation}
\rho_{\rm mobile}^{(n)}=\beta\tilde{\rho}_{\rm mobile}^{(n)}+(1-\beta)\rho_{\rm mobile}^{(n-1)},
\label{Anderson_mixing}
\end{equation}
where $n$ is the step of the procedure and $\beta$ is the Anderson coefficient. In the first step of the process (i.e. $n=0$) we take $\rho_{\rm mobile}^{(0)}=0$ and we compute the electrostatic potential of the system. At any other arbitrary step $n$, we compute the charge density $\tilde{\rho}_{\rm mobile}^{(n)}$ using the electrostatic potential found in the previous iteration $n-1$. Then, we compute the electrostatic potential at the $n$ step using $\rho_{\rm mobile}^{(n)}$, given by the Anderson mixing of Eq. \ref{Anderson_mixing}. This charge density mixing between the $n$ and $n-1$ steps ensures the convergence to the solution. We repeat the procedure until the cumulative error is below the 1\%. Once the electrostatic potential is known, the Rashba coupling $\alpha_{\rm R}(\vec{r})$ is computed using Eq. (\ref{alpha}) of the main text.

\begin{figure}
\centering
\includegraphics{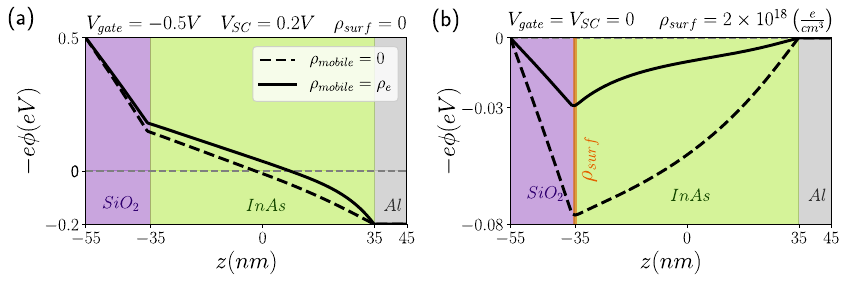}
\caption{\label{FigSI1}(Colour online) Representative examples of the electrostatic potential profile in a top-superlattice nanowire along the $z$-direction for $V_{\rm gate}=-0.5$V, $V_{\rm SC}=0.2$V and $\rho_{\rm acc}=0$ in (a); and for $\rho_{\rm acc}=2\times 10^{18}$e/cm$^3$ and $V_{\rm gate}=V_{\rm SC}=0$ in (b). The screening effect of the mobile charges inside the wire, $\rho_{\rm mobile}$, is ignored in the dashed line solution, whereas it is taken into account in the solid one. Geometric parameters are $W_{\rm Al}=10$nm, $W_{\rm SiO}=20$nm, $W_{\rm wire}=80$nm.}
\end{figure}

To provide more insight into the electrostatic potential created by the gate, the superficial charge density and the mobile charge density, we show in Fig. \ref{FigSI1}(a) the potential profile produced by some particular values of $V_{\rm gate}$ and $V_{\rm SC}$ (in the absence of $\rho_{\rm surf}$) along the wire's cross-section ($z$-direction) for a top-superlattice device. Separately, in Fig. \ref{FigSI1}(b) we show the effect of the surface charge layer for zero $V_{\rm gate}$ and $V_{\rm SC}$ for the same device. The solid curve corresponds to the self-consistent solution (in the Thomas-Fermi approximation), while for the dashed curve the presence of $\rho_{\rm mobile}$ in the Poisson equation has been ignored. The effect of $\rho_{\rm mobile}$ is small when the effective chemical potential is close to the bottom of the conduction band, as is the case of Fig. \ref{Fig1}(a). However, when this is not the case, the non self-consistent solution overestimates the band bottom displacement with respect to the Fermi level, see Fig. \ref{Fig1}(b). This happens because the screening effect of the mobile charges pushes the band bottom upwards reducing the wire's average doping.

\subsection{3D Hamiltonian}
The 3D Hamiltonian of Eq. (\ref{Hamiltonian}) in the main text is discretized using the finite difference method within the Bogoliubov-de Gennes formalism, using an inter-site distance (discretization) of $5$nm in the three directions. We find the eigenstates of the Hamiltonian using the ARPACK diagonalization tools implemented in the standard package \textit{Scipy} of Python. In order to reduce the computational cost, we only compute the 10 lowest-energy eigenstates, which are the relevant ones for Majorana physics.

\subsection{1D Hamiltonian and topological invariant}
We build the finite 1D Hamiltonian following the same method as for the 3D case, but taking the electrostatic potential at the center of the wire (i.e. $\phi(x,y=0,z=0)$). We exploit the periodic nature of this Hamiltonian to build the infinite 1D Hamiltonian in k-space $H(k)$, as explained in Ref. \onlinecite{Levine:PRB17}. From there, we can compute the class D topological invariant \cite{Levine:PRB17} $\mathcal{Q}$ as
\begin{equation}
\mathcal{Q}=\operatorname{sign}\left(\operatorname{Pf}\left\lbrace\Lambda H(k=0)\right\rbrace\right)\cdot \operatorname{sign}\left(\operatorname{Pf}\left\lbrace\Lambda H(k=\frac{\pi}{L})\right\rbrace\right),
\end{equation}
where $\operatorname{Pf}\left\lbrace M\right\rbrace$ is the Pfaffian of a matrix $M$, which we compute numerically using the Python package \textit{Pfaffian} provided by Ref. \onlinecite{Wimmer:ACM12}, and $\Lambda$ is the electron-hole symmetry matrix, that in our basis obeys
\begin{equation}
\Lambda H^*(-k) \Lambda^{-1}=-H(k) \leftarrow \Lambda=\mathcal{I}_{\mathrm{site}}\otimes\sigma_0\otimes\tau_x.
\end{equation}

\section{Reliability of Thomas-Fermi approximation}
\label{SI2}
The calculations shown in the main text have been performed using the Thomas-Fermi approximation for the charge density inside the wire. However, a more realistic and complete description consists of solving the coupled Schr\"odinger-Poisson equations, which requires to compute the charge density from the eigenspectrum of the Hamiltonian
\begin{equation}
\rho_{\rm e}^{\rm (SP)}(\vec{r})=e\sum_{i>0} \left|u_i(\vec{r})\right|^2 f(E_i)+ \left|v_i(\vec{r})\right|^2 f(-E_i),
\end{equation}
where $u_i(\vec{r})$ and $v_i(\vec{r})$ are the electron and hole components of the $i$-th eigenstate, $E_i$ its corresponding energy, $f(E)$ the Fermi-Dirac distribution, and the sum is done for every positive energy ($i>0$). Since the eigenspectrum is found by diagonalizing the Hamiltonian, which depends in turn on the charge density through the Poisson equation, the coupled Schr\"odinger-Poisson equations have to be solved self-consistently as well, following the same iterative procedure as described in App. \ref{SI1}. Nevertheless, this process is computationally more expensive because the Hamiltonian is diagonalized in each self-consistent step. Hence, when both methods provide similar results, it is justified to use the Thomas-Fermi approximation to reduce the computational cost.

It is a well-known fact that the Thomas-Fermi approximation ignores the kinetic terms, as well as the magnetic field dependence. Remarkably, some previous works \cite{Mikkelsen:PRX18} have shown that both approaches provide similar results, although for simplistic models of Majorana nanowires. However, this could not be true for superlattice ones since the superlattice leads to a stronger charge localization. In this appendix we compare the results obtained using both methods. First, we show that for $V_{\rm Z}=0$ both methods predict similar results for the lowest energy modes, in spite of ignoring the kinetic terms. And second, we show that the magnetic field dependence of the wire's charge density can be neglected.

\subsection{Comparison between Thomas-Fermi approximation and full Schr\"odinger-Poisson calculation}
\begin{figure}
\centering
\includegraphics{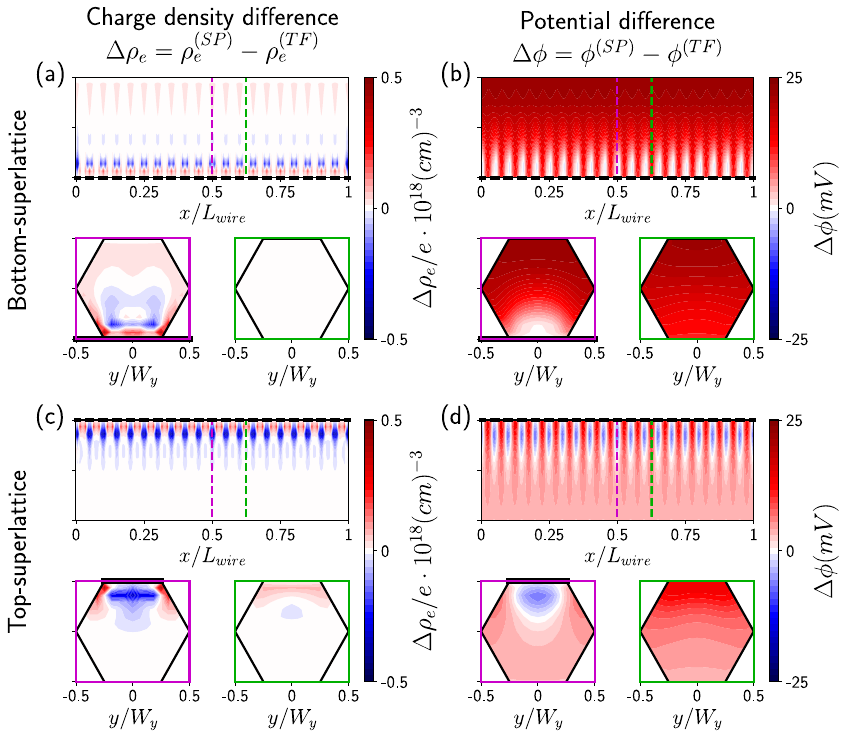}
\caption{\label{FigSI2-1}(Colour online) Difference between the electron charge densities inside the nanowire computed using the Schr\"odinger-Poisson approach and the Thomas-Fermi approximation, $\Delta\rho_{\rm e}$, for the bottom-superlattice setup (a) and the top-superlattice one (c). (b) and (d) show their corresponding electrostatic potential difference $\Delta\phi$. Parameters are the same as in Fig. \ref{Fig13}, except for $V_{\rm Z}=0$.}
\end{figure}

Fig. \ref{FigSI2-1} shows a comparison between both methods for the bottom (a-b) and top (c-d) setups of Sec. \ref{3D_results}, for the same parameters of Fig. \ref{Fig13}, except for $V_{\rm Z}=0$. The difference $\Delta\rho_{\rm e}$ between the charge densities computed using the Schr\"odinger-Poisson approach $\rho_{\rm e}^{\rm (SP)}$ and the Thomas-Fermi approximation $\rho_{\rm e}^{\rm (TF)}$ is shown in Figs. \ref{FigSI2-1}(a,c) for both devices. In both cases, the difference is a small positive quantity very close to the SC-InAs interface (more clearly seen in Fig.  \ref{FigSI2-1}(a)), which means that the Thomas-Fermi approximation slightly overestimates the electron density close to this interface, where the majority of the charge is located. Conversely, it is slightly negative further away from the interface. Everywhere else $\Delta\rho_{\rm e}\approx 0$.  The total charge obtained with both methods are $Q_{\rm tot}^{\rm (TF)}\simeq809e$ and $Q_{\rm tot}^{\rm (SP)}\simeq709e$ in the bottom-superlattice nanowire, and $Q_{\rm tot}^{\rm (TF)}\simeq633e$ and $Q_{\rm tot}^{\rm (SP)}\simeq624e$ in the top-superlattice one, which are pretty similar.

To obtain a quantitative estimation of the error made using the Thomas-Fermi approximation, we now analyse the electrostatic potential created by the charge density using both methods, which is the quantity that indeed enters into the Hamiltonian of Eq. (\ref{Hamiltonian}). The electrostatic potential difference $\Delta\phi=\phi^{\rm (SP)}-\phi^{\rm (TF)}$ between both methods is plotted in Figs. \ref{FigSI2-1}(b,d) for each device. Since the bare electrostatic interaction given by the Poisson equation is long-ranged, $\Delta\phi$ is very small (or zero) close to the SCs, despite the finite charge density difference there. By contrast, the maximum $\Delta\phi$ in both cases is found far apart from the back gate. It is roughly $20$mV and homogeneous for the bottom-superlattice nanowire, and around $10$mV between SC fingers for the top one.
Comparing with the electrostatic potential of Fig. \ref{Fig13}, which is computed using the Thomas-Fermi approximation for the same devices and for the same back gate voltages, we conclude that the average error is below 10\%, justifying the use of the Thomas-Fermi approximation for the range of gate voltages used in our simulations.

\subsection{Accuracy of the zero magnetic field Thomas-Fermi approximation}
\begin{figure}
\centering
\includegraphics{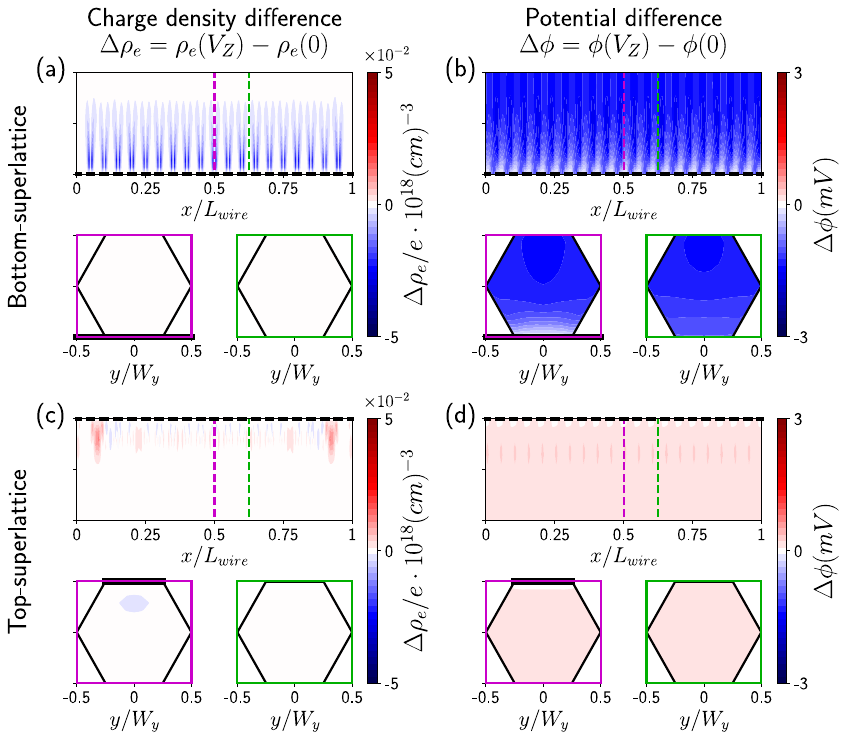}
\caption{\label{FigSI2-2}(Colour online) Difference between the charge densities computed using the Schr\"odinger-Poisson equation with magnetic field ($V_{\rm Z}=0.6meV$) and without it, for the bottom-superlattice nanowire (a) and for the top-superlattice one (c). (b) and (d) show their corresponding electrostatic potential difference. Parameters are the same as in Fig. \ref{Fig13}.}
\end{figure}

The previous analysis has been carried out for Zeeman splitting $V_{\rm Z}=0$, since the charge density computed using the Thomas-Fermi approximation (Eq. (\ref{Thomas-Fermi})) in the main text ignores the magnetic field dependence. To obtain a quantitative estimation of the error made due to this approximation, we show in Figs. \ref{FigSI2-2}(a,c) the difference between the charge densities computed using the Schr\"odinger-Poisson approach with and without an applied magnetic field (for both geometries). In addition, Figs. \ref{FigSI2-2}(b,d) show their corresponding electrostatic potential difference. Comparing with Fig. \ref{Fig13}, one can see that the error is below 1\%. This small difference is due to the fact that typical Zeeman splittings in these systems ($V_{\rm Z}\sim 1meV$) are much smaller than the electrochemical potentials ($e\phi-E_{\rm F}\sim 100meV$), so that last quantity dominates. Consequently, we conclude that neglecting the magnetic field dependency in the charge density is an adequate approximation for these calculations.

\section{Electrostatic potential and Rashba coupling in superlattice parameter space}
\label{SI3}
In this last appendix we show how the induced electrostatic potential and Rashba coupling behave versus the superlattice parameters $L_{\rm cell}-r_{\rm SC}$. This is relevant since, as we show below, we find that for some $L_{\rm cell}$ and $r_{\rm SC}$ values it is difficult to gate the wire due to screening effects, or the spin-orbit coupling induced by the back gate is negligible. We have (partially) used this information to plot Fig. \ref{Fig10} in the main text.

\begin{figure}
\centering
\includegraphics{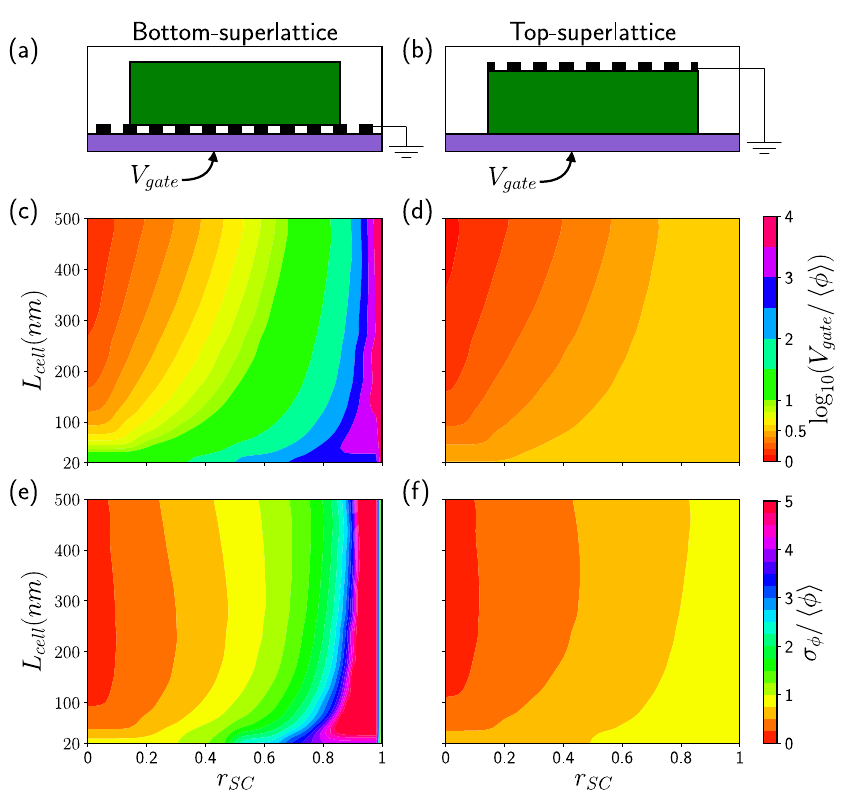}
\caption{\label{FigSI3-1} (Colour online) Variation of the spatially averaged electrostatic potential inside the wire due to the voltage applied to the back gate. Here $V_{\rm SC}=0$, $\rho_{\rm surf}=0$ and $\rho_{\rm e}$ is neglected. Two setups are considered, bottom-superlattice to the left and top-superlattice to the right. (a,b) Sketches of both systems. (c,d) Lever arm, defined as the back gate potential needed to change the spatially averaged electrostatic potential $\langle \phi\rangle$, versus superlattice parameters $L_{\rm cell}$ and $r_{\rm SC} = L_{\rm SC}/L_{\rm cell}$ in logarithmic scale. (e,f) Dispersion of the electrostatic potential variations along and across the wire, $\sigma_\phi=\sqrt{\left< \phi^2\right>-(\left< \phi\right>)^2}$.}
\end{figure}

Figures \ref{FigSI3-1}(c,d) show the lever arm (in logarithmic scale) versus the superlattice parameters for the bottom- (c) and top-superlattice (d) devices. This quantity is defined as the back gate potential needed to change the spatially averaged electrostatic potential $\langle \phi\rangle$. Here, this variation is independent of $V_{\rm gate}$ because for simplicity we ignore the screening produced by $\rho_{\rm e}$. In both setups, when the partial coverage of the SC $r_{\rm SC}$ is small, the lever arm is a factor of the order of $10^0$-$10^1$. This means that, for example, to change $\langle \phi\rangle$ by 1meV we need to apply a voltage to the gate of 1-10meV.
However, as $r_{\rm SC}$ increases, so does the lever arm and larger back gate potentials are needed to effectively deplete or fill the wire. This happens dramatically for the bottom-superlattice setup since the superlattice is placed between the back gate and the nanowire. Thus, for large $r_{\rm SC}$, the screening effect of the SC fingers is strong. By contrast, in the top-superlattice setup the lever arm converges to a finite small value corresponding to that of the continuously covered nanowire.

Since the electrostatic potential close to the SC fingers is fixed by the boundary condition $V_{\rm SC}$, large lever arms lead to large electrostatic variations along the wire. This can be detrimental for the stability of Majorana states, since these large variations lead in turn to the formation of localized states (as explained in Sec. \ref{Impact} of the main text).
The standard deviation $\sigma_{\phi}$ of the electrostatic potential along and across the wire shown in Figs. \ref{FigSI3-1}(e,f) (for both setups) gives an idea of the size of these potential variations. As pointed out before, for small $r_{\rm SC}$, when the lever arm is small as well, the variations are negligible. However, for larger $r_{\rm SC}$, the variations are larger, specially for the bottom-superlattice setup.

\begin{figure}[h]
\centering
\includegraphics{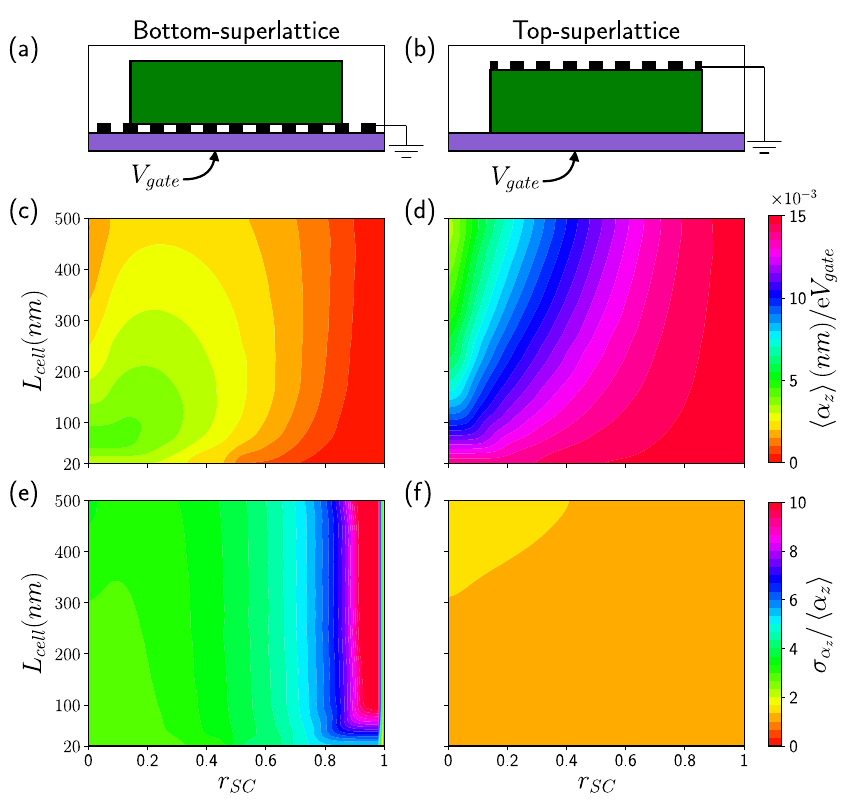}
\caption{\label{FigSI3-2} (Colour online) Similar analysis as in Fig. \ref{FigSI3-1} but for the spin-orbit coupling. Two setups are considered, bottom-superlattice (a) to the left and top-superlattice (b) to the right. (a,b) Sketches of both systems. (c,d) Variation of the spatially averaged $\alpha_{\rm z}$ Rashba coupling inside the wire due to the voltage applied to the back gate. (e,f) Dispersion of $\alpha_{\rm z}$ variation defined as $\sigma_{\alpha_z}=\sqrt{\left< \alpha_z^2\right>-(\left< \alpha_z\right>)^2}$.}
\end{figure}

Since the Rashba coupling depends locally on the electric field, the spin-orbit strength also depends on the superlattice parameters. The average value of $\alpha_z$ along the wire is shown in Figs. \ref{FigSI3-2}(c,d) for both superlattice setups. We only consider the contribution of the back gate potential (fixing $V_{\rm SC}=0$ and $\rho_{\rm surf}=0$).
For small $r_{\rm SC}$, the Rashba coupling is roughly 5meV$\cdot$nm when 1V is applied to the back gate (for both devices). As $r_{\rm SC}$ is increased, $\left<\alpha_z\right>$ decreases for the bottom-superlattice setup until it reaches zero, while it increases for the top-superlattice one until it reaches 15meV$\cdot$nm (when 1V is applied to the back gate), corresponding to the value of the mean Rashba coupling in the homogeneous nanowire. This qualitative difference is again due to the strong screening effect of the SC fingers in the bottom-superlattice setup.

For completeness, we show the dispersion of the $\alpha_z$ spin-orbit coupling variation along the wire in  Figs. \ref{FigSI3-2}(e,f). We find that the dispersion is constant in the top-superlattice setup (f) regardless of the superlattice parameters. However, the spin-orbit variations increase with $r_{\rm SC}$ in the bottom-superlattice one. 

\end{document}